\documentclass[a4paper,11pt]{report}
\usepackage{mdwlist,mathdots,dsfont}
\usepackage{mathrsfs,cite,marvosym,multirow,bbold}
\usepackage{bm,amssymb,slashed,amsmath,color,verbatim,graphicx,multicol,amsthm}
\usepackage[T1]{fontenc}
\usepackage{ae,aecompl}
\usepackage[font=sf, labelfont={sf,bf}, margin=1cm]{caption}
\usepackage{subfigure}
\usepackage[a4paper,top=3.5cm,right=3.1cm,left=3.1cm,bottom=3.5cm]{geometry}
\font\tenmsb=msbm10
\font\sevenmsb=msbm7
\font\fivemsb=msbm5
\newfam\msbfam
\textfont\msbfam=\tenmsb
\scriptfont\msbfam=\sevenmsb
\scriptscriptfont\msbfam=\fivemsb

\numberwithin{equation}{section}

\definecolor{IndianRed}{rgb}{0.8,0.36,0.36}
\definecolor{DarkGreen}{rgb}{0,0.5,0}
\definecolor{Gray}{rgb}{0.5,0.5,0.5}

\newcommand{\ir}[1]{\color{IndianRed}{#1}}

\newcommand{\be}{\begin{equation}}
\newcommand{\ee}{\end{equation}}
\newcommand{\bea}{\begin{eqnarray}}
\newcommand{\eea}{\end{eqnarray}}

\newcommand\bra[1]{\left\langle #1\right|}
\newcommand\ket[1]{\left|#1\right\rangle}
\newcommand{\dt}[2][L]{\det\nolimits_{#1}\left[ #2 \right ]}

\newcommand{\de}[2][]{\frac{\partial #1}{\partial #2}}
\newcommand{\stack}[2]{\begin{subarray}{c} #1 \\ #2 \end{subarray}}

\newcommand{\nn}{\nonumber}
\newcommand{\pic}[3]{\raisebox{#1}{\includegraphics[height=#2]{#3}}}

\DeclareMathOperator{\e}{\rm e}
\DeclareMathOperator{\dd}{\rm d}
\DeclareMathOperator{\sign}{sign}
\newcommand\ii{{\,\rm i}}
\long\def\symbolfootnote #1{\begingroup%
    \def\thefootnote{\fnsymbol{footnote}}\footnote[1]{#1}\endgroup}

\theoremstyle{plain}
\newtheorem{thm}{Theorem}[section]
\newtheorem{lm}[thm]{Lemma}
\newtheorem{prop}[thm]{Proposition}
\newtheorem*{cor}{Corollary}

\theoremstyle{definition}
\newtheorem{defn}{Definition}[section]
\newtheorem{conj}{Conjecture}[section]

\theoremstyle{remark}

\newcommand{\thmref}[1]{Theorem~\ref{thm:#1}}
\newcommand{\conjref}[1]{Conjecture~\ref{conj:#1}}
\newcommand{\lmref}[1]{Lemma~\ref{lm:#1}}
\newcommand{\propref}[1]{Proposition~\ref{prop:#1}}

\newcommand{\secref}[1]{Section~\ref{sec:#1}}
\newcommand{\apref}[1]{Appendix~\ref{ap:#1}}
\newcommand{\figref}[1]{Figure~\ref{fig:#1}}
\newcommand{\defref}[1]{Definition~\ref{def:#1}}
\newcommand{\chapref}[1]{Chapter~\ref{chap:#1}}

\usepackage{setspace}
\begin{document}

\begin{titlepage}
\title{
\vspace{3cm}
Finite size lattice results for the two-boundary Temperley--Lieb loop model}
\author{Anita Kristine Ponsaing
\vspace{1cm}\\
Submitted in total fulfilment of the requirements of\\
the degree of Doctor of Philosophy\\
\vspace{0.5cm}\\
June 2010\\
\vspace{1cm}\\
{\small\em Department of Mathematics and Statistics}\\
{\small\em The University of Melbourne}
\date{}
}
\maketitle
\end{titlepage}
\pagenumbering{roman}
\begin{abstract}
This thesis is concerned with aspects of the integrable Temperley--Lieb loop (TL($n$)) model on a vertically infinite lattice with two non-trivial boundaries. The TL($n$) model is central in the field of integrable lattice models, and different values of $n$ relate to different physical models. For instance, the point $n=0$ relates to critical dense polymers and a corresponding logarithmic conformal field theory. 
The point $n=1$ corresponds to critical bond percolation on the square lattice, and has connections with a combinatorial counting problem of alternating sign matrices and their symmetry classes. For general $n$, the TL($n$) model is closely related to the XXZ quantum spin chain and the $6$-vertex model.

We construct the transfer matrix of the model, which describes the weights of all the possible configurations of one row of the lattice. When $n=1$ the ground state eigenvector of this matrix can be interpreted as a probability distribution of the possible states of the system. Because of special properties exhibited by the transfer matrix at $n=1$, we can show that the eigenvector is a solution of the $q$-deformed Knizhnik--Zamolodchikov equation, and we use this fact to explicitly calculate some of the components of the eigenvector. In addition, recursive properties of this transfer matrix allow us to compute the normalisation of the eigenvector, and show that it is the product of four Weyl characters of the symplectic group. Previous work in this area has produced results for the TL($1$) loop model with periodic boundary conditions, two trivial boundaries and mixed (one trivial, one non-trivial) boundaries, but until recently little progress had been made on the case with two non-trivial boundaries. This boundary condition lends itself to calculations relating to horizontal percolation, which is not possible with the other boundary conditions.

One of these calculations is a type of correlation function that can be interpreted as the density of percolation cluster crossings between the two boundaries of the lattice. It is an example of a class of parafermionic observables recently introduced in an attempt to rigorously prove conformal invariance of the scaling limit of critical two-dimensional lattice models. It also corresponds to the spin current in the Chalker--Coddington model of the quantum spin Hall effect. We derive an exact expression for this correlation function, using properties of the transfer matrix of the TL($1$) model, and find that it can be expressed in terms of the same symplectic characters as the normalisation.

In order to better understand these solutions, we use Sklyanin's scheme to perform separation of variables on the symplectic character. We construct an invertible separating operator that transforms the multivariate character into a product of single variable polynomials. Analysing the asymptotics of these polynomials will lead, via the inverse transformation, to the asymptotic limit of the symplectic character, and thus to the asymptotic limit of the ground state normalisation and correlation function of the loop model. We construct the separating operator by viewing the symplectic characters as eigenfunctions of a quantum integrable system, and also explicitly construct the factorised Hamiltonian for this system.
\end{abstract}
\chapter*{Declaration}
\setcounter{page}{3}
This is to certify that:
\begin{itemize}
\item the thesis comprises only my original work towards the PhD except where indicated in the Preface,
\item due acknowledgement has been made in the text to all other material used,
\item the thesis is fewer than 100,000 words in length, exclusive of tables, maps, bibliographies and appendices.
\end{itemize}
\vspace{1cm}

\chapter*{Acknowledgments}
This thesis has benefited from the financial, moral, and academic support of many different parties. I was supported by a Melbourne Research Scholarship throughout the duration of my candidature, and I would also like to acknowledge the financial support of the Australian Research Council and the Research Support Scheme of the Department of Mathematics and Statistics for allowing me to travel to numerous conferences. I would particularly like to acknowledge the John Hodgson Memorial Fund, which supplied a scholarship allowing me to attend a trimester in Paris that greatly influenced my study.

I would like to thank Christian Krattenthaler and the Erwin Schr\"odinger Institute for Theoretical Physics in Vienna for their hospitality and financial support during the
Workshop on Combinatorics and Statistical Physics in May 2008, and especially the Institut Henri Poincar\'e in Paris and the organisers of the 2009 autumn trimester on Statistical Physics, Combinatorics and Probability, for their hard work organising the trimester, for the opportunity to attend, and again for the financial support given. For their hospitality during my visits I would also like to extend my thanks to Bernard Nienhuis and the Institute for Theoretical Physics at the University of Amsterdam, and Fabian Essler and the Rudolf Peierls Centre for Theoretical Physics in Oxford.

Thanks to Luigi Cantini, who noticed an error in our original draft of the paper that formed the bulk of Chapter 2; and Vladimir Mangazeev, who brought to our attention Sklyanin's separation of variables method, which is used in Chapter 4. For inspiring and useful mathematical discussions, I would like to thank John Cardy, Mihai Ciucu, Phillipe Di Francesco, Murray Elder, Fabian Essler, Peter Forrester, Cl\'ement Hongler, Bernard Nienhuis, Pavel Pyatov, Arun Ram, Stanislav Smirnov, Robert Weston, and Paul Zinn-Justin. A special thanks to Mark Sorrell, whose unfailing interest in my work encouraged and helped me to push past a number of mathematical hurdles. Thanks also to Caley Finn, Peter Forrester, and Anthony Mays for their efforts in proofreading this thesis. I realise it's not the most exciting way to spend a weekend, and I really appreciate it.

To all the friends I made in the department, thanks for making my time in Melbourne so enjoyable. Many have helped me with mathematical problems, and every single one has taught me something. I list here just a few: Wendy Baratta, Nick Beaton, Caley Finn, Steve Mc Ateer, Gus Schrader, Maria Tsarenko, and Michael Wheeler. I would like to thank Christopher Campbell for providing a sympathetic ear whenever I needed one. Special thanks to Keiichi Shigechi, who took the time to help a relative stranger through the first bout of PhD turbulence, and who has remained a close friend through thick and thin.

Thanks to my family, whose unconditional love and support, even from a thousand miles away, has somehow always managed to reach me. Thanks to Marianne and Noon, who made sure I knew I had someone nearby I could always turn to. Special thanks to Anthony Mays, for his infectious curiosity and pursuit of truth, and for all the help and support he has given me over the last three years.

Finally, my unreserved and heartfelt thanks to Jan de Gier. You always knew when I needed support and when I needed pushing. You have always treated me as a colleague; you listened to my opinions and allowed the teaching to go both ways. Your love of mathematics and physics, and your passion for research, have been inspiring, and you always had the time and patience for me when I didn't understand something. Not to mention the number of times you supported me when problems that had nothing to do with maths turned up. You have turned me from a student into a researcher. I'm lucky to have had you as a supervisor, and I feel privileged to be able to call you a friend.
\tableofcontents
\listoffigures
\addcontentsline{toc}{chapter}{List of Figures}
\newpage
\newpage
\chapter*{Introduction}
\pagenumbering{arabic}
\addcontentsline{toc}{chapter}{Introduction}
\label{chap:Intro}
Statistical mechanics is the study of systems composed of large numbers of small particles. The aim of the study is to investigate how global, collective effects arise from small-scale local interactions between the particles. The systems in question are too large to make accurate predictions on the level of individual particles, so the focus is instead on statistical quantities. Examples include the average density of particles and correlation functions between two chosen points. These statistical quantities can be used to calculate derived quantities such as heat capacity, energy densities and magnetisation.

Global quantities of a system are usually expressed in terms of the system's temperature, as well as external parameters such as a magnetic field. As these parameters pass certain values, the global quantities can be observed to change dramatically, for instance at the boiling point of water. These values are known as critical points. The location and nature of any critical points of a system are more examples of quantities of interest.

Near a critical point, global quantities can be approximated by a power law (say, $(T-T_c)^{\alpha}$, where $T_c$ is the critical temperature). The exponent $\alpha$ in this approximation is called a critical exponent, and indicates the strength of the divergence as the parameter passes the critical point. It is widely believed (and experimentally verified) that systems with the same dimensionality, symmetries, and range of interactions will have the same critical exponents --- a property known as universality.

A well-known statistical mechanical model is for magnetisation in a bar of iron. Each individual atom can be thought of as a small magnet, whose alignment affects the alignment of its neighbours. The strength of the effect depends on various factors, including the temperature of the bar and the external magnetic field. At low overall temperatures the atoms will prefer to keep their current alignment, but they will also try to align with their neighbours and the external field. This leads to the phenomenon of the bar having a spontaneous magnetisation if the external field is turned off. At high temperatures the particles are more likely to lose their alignment, so there is no spontaneous magnetisation. The temperature at which the spontaneous magnetisation disappears is the critical temperature, and much interesting physics happens near this point.

The local interactions of a system can be described by an energy function $E$. This can be variously interpreted as a function, a matrix, or more generally an operator. By acting on a possible state $s$ with the energy function, we get the energy $E(s)$ associated with that state. When the system is in thermodynamic equilibrium at a temperature $T$, the probability of observing a state decreases as the energy of that state increases. More precisely, the probability is
\[ P(s)=\frac1{Z}\e^{-\frac{E(s)}{kT}},\]
where $k$ is Boltzmann's constant (approx. $1.38\times 10^{-23} J/K$), and the factor $\e^{-E(s)/kT}$ is known as the Boltzmann weight of the state $s$. The normalisation $Z$ is defined as
\[Z=\sum_s \e^{-\frac{E(s)}{kT}},\]
and is known as the partition function (for a derivation of these expressions see \cite{Rushbrooke49}). The partition function holds all the information about the probabilities of all possible states, and as such is an important quantity, especially for calculating expected values of observable properties. The partition function is also closely related to the thermodynamic free energy $F$ of the system, by
\[ F=-kT\ln{Z},\qquad Z=\e^{-\frac{F}{kT}}. \]
The free energy is the amount of energy in the system that is available to do work.

The expected value of an observable $A$ is the weighted sum over the values of $A$ for all possible states, weighted by the probability of each state. In short,
\[ \langle A\rangle=\frac1{Z}\sum_s A(s)\e^{-\frac{E(s)}{kT}}. \]
Another important physical quantity is the correlation function, which describes how two spatially separated points affect each other's behaviour. For a quantity $A$, the correlation function is given by
\[ \langle A_1A_2 \rangle - \langle A_1\rangle\langle A_2\rangle,\]
where $A_1$ is the quantity evaluated at the first point, and $A_2$ is evaluated at the second.

A system is said to be solvable if it is possible to analytically calculate the physical quantities described above. In other words, the partition function (or equivalently, the free energy) must be calculable \cite{Bax82}. For physical systems, this is usually an impossible task, and as such, most of the work done in this field is either based on numerical approximations or simplified mathematical models. Nevertheless, these models can be quite powerful, as universality predicts that a model with the same dimensionality, symmetry and range of interactions as a physical system will also have exactly the same the critical exponents.

\section*{Lattice models}
Some of the most commonly used models in statistical mechanics are lattice models, where the particles are located at regular intervals, and interactions are defined between them. Calculating the partition function thus involves solving the combinatorial problem of finding all possible lattice configurations, and then summing the probabilities for each configuration; the latter is generally the more difficult task.

Even though these models do not reflect reality in full detail, by letting the number of lattice sites tend to infinity they can be used to obtain a good idea of the behaviour of a physical system. This infinite limit is known as the thermodynamic limit, or equivalently the asymptotic limit. In addition, as previously mentioned, universality predicts that an appropriately designed lattice model will have the same critical exponents as a physical system.

One of the most useful tools in the study of lattice models is the method of transfer matrices. Roughly speaking, a transfer matrix describes all possible configurations of one row of the lattice. If the height and width of the lattice are $M$ and $L$ respectively, it can also be shown that in the large $M$ limit, the partition function goes like
\[ Z_{LM}\sim\Lambda_L^M, \]
where $\Lambda_L$ is the largest eigenvalue of the transfer matrix (corresponding to the state with the lowest energy, or the ground state). The thermodynamic limit then corresponds to taking the limit as $L$ goes to infinity.

The transfer matrix can be constructed from operators called $R$-matrices that act locally. If it can be shown that the $R$-matrices satisfy the Yang--Baxter equation, then it is possible to construct an infinite family of commuting transfer matrices. This infinite family is parameterised by the so-called spectral parameter. The fact that the family of transfer matrices commute implies that they are simultaneously diagonalisable and share eigenvectors, and since each member of the family has a different value of the spectral parameter, these eigenvectors cannot depend on the spectral parameter. This fact provides access to a host of methods for finding the eigenvalues and eigenvectors of the transfer matrix, and often leads to the system being solvable.

The most famous of these methods is the Bethe ansatz method. This method works by guessing a form for the eigenvectors of the transfer matrix (called the ansatz). The ansatz depends on a set of parameters, which are fixed by constraint equations resulting from the transfer matrix eigenvalue equation. The Bethe ansatz method relies on the existence of a pseudovacuum --- an eigenvector of the transfer matrix for which there exists an annihilation operator.

Perhaps the most widely known lattice model is the Ising model, which is a model of a magnet, as described earlier. The Ising model was posed by Lenz in 1920, and solved for one dimension by Ising in 1925 \cite{Ising25}. In two dimensions the solution for zero external field was given by Onsager in 1944 \cite{Onsager44}. For non-zero external magnetic field, Zamolodchikov found a solution in 1989 for a field-theoretic version of the two-dimensional Ising model in the thermodynamic limit \cite{Zamolodchikov89}. In 1992, Warnaar, Nienhuis and Seaton found a solution for a two-dimensional model in the same universality class \cite{WarnaarNS92}.

The Ising model is typically defined on a square lattice. At each site is a molecule with a magnetic moment $\sigma_i$, which can take the value $+1$ (up) or $-1$ (down). These magnetic moments can also be regarded as spins. Each spin interacts only with its closest neighbours, and the interaction depends on the values of the spins and a coupling constant $\mathcal J$. The energy function of the Ising model is
\[E_{\text I}=\frac{\mathcal J}{2}\sum_{<ij>}\sigma_i\sigma_j-B\sum_i \sigma_i,\]
where the first sum is over all nearest neighbours, and $B$ is the strength of the external field. When $\mathcal J<0$, this system models a ferromagnet, since the energy is lower when adjacent spins are aligned. When $\mathcal J>0$ the system models an antiferromagnet. In two dimensions, the model exhibits spontaneous magnetisation below a critical temperature, but in one dimension no such behaviour occurs.

Naturally this is a highly idealised model, as real magnetic moments can point in more than two directions. Several generalisations of the Ising model have since been proposed, including the Potts model and the $n$-vector model.

The $Q$-state Potts model, or simply Potts model \cite{Tsarenko09,Bax82,Wu82}, has scalar spins that can take one of $Q$ values, and the interactions between nearest neighbours are given by Kronecker delta functions. This is a well-known and widely studied model in the field of two-dimensional lattice models. In its current form it was first written down by Potts in 1952 \cite{Potts52}. The energy function of the Potts model is
\[ E_{\text P}=-\mathcal J\sum_{<ij>}\delta(\sigma_i,\sigma_j), \]
and the partition function is then
\[ Z_{\text P}=\sum_{\sigma}\e^{\frac{\mathcal J}{kT}\sum_{<ij>}\delta(\sigma_i,\sigma_j)}, \]
where the sum is over all possible spin configurations. This can be rewritten as
\[ Z_{\text P}=\sum_{\sigma}\prod_{<ij>}(1+u\delta(\sigma_i,\sigma_j)), \]
with $u=\e^{\frac{\mathcal J}{kT}}-1$.

\begin{figure}[ht]
\begin{center}
\includegraphics[height=40pt]{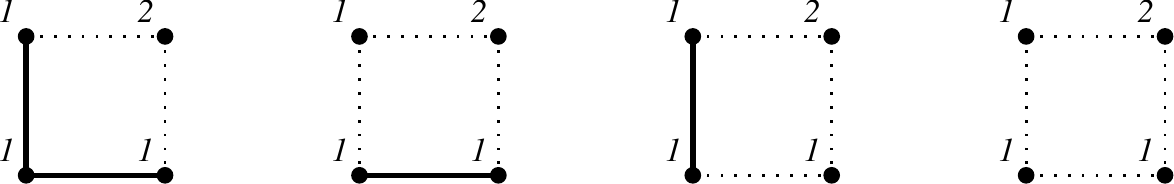}
\end{center}
\caption{Possible bond-graphs for the spin configuration ((1,2),(1,1)).}
\label{fig:bondgraphs}
\end{figure}

The Potts model can be reformulated as a loop model in the following way \cite{Bax82}. Consider the system of spins as a planar graph $\mathcal L$ where the spins sit on vertices and edges are drawn between neighbouring spins. On each edge, a bond may be drawn according to the following rules: if a pair of neighbouring vertices have different spin values, no bond is drawn; if the vertices share the same spin value, there is a choice of drawing a bond on the edge between them (with a weight of $u$), or leaving the edge blank (with weight $1$). We refer to an arrangement of bonds on $\mathcal L$ as a bond-graph. Then the expansion of each product in the partition function represents all the possible bond-graphs associated with a spin configuration. An example of this is shown in \figref{bondgraphs}.

By considering instead the spin configurations possible on a particular bond-graph, we can turn the partition function into a sum over bond-graphs. Let $C$ be the number of connected components of the bond-graph (clusters of bonds, including single vertices). All the vertices in a connected component must share a spin value. This spin value could be any of the $Q$, and since each connected component is independent of the others, there are $Q^C$ ways to choose these values for the entire bond-graph. A bond-graph with $l$ bonds has a weight of $u^l$. Then the partition function is a sum over all the possible bond-graphs $G$ on $\mathcal L$,
\[ Z_{\text P}=\sum_G Q^Cu^l. \]
The connection to the loop model is made by drawing loops around the bonds in the following two ways:
\begin{center}
\includegraphics[height=20pt]{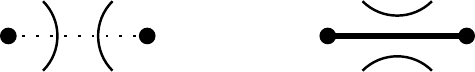}
\end{center}
Every cluster has $Q$ possible spin values, which leads to every closed loop in the loop expansion having a weight of $\sqrt{Q}$.

At $Q=1$ the loop model is equivalent to the two-dimensional square lattice bond percolation model \cite{PearceRZ06,SaleurD87}. In this model, bonds are drawn between neighbouring sites of the lattice with probability $p$. On an infinite lattice, percolation occurs when there is an infinite sized cluster of bonds. This model has a second order phase transition (that is, the first derivative of the free energy is continuous but the second is not) at critical probability $p=1/2$, above which percolation occurs with probability $1$.

By orienting the loops, the Potts loop model can be generalised to a directed loop model, which can itself be mapped to the six vertex model. These mappings are explained in more detail in Chapter 12 of \cite{Bax82}, and in \cite{BaxKW76}.

The six vertex model is another classical two-dimensional lattice model, also known as the square ice model. On a square lattice, oxygen atoms are placed at each vertex, and on the edges between them are placed hydrogen atoms. A single water molecule consists of one oxygen bonded with two hydrogens, and correspondingly the `ice rule' for the lattice model states that of the four hydrogens surrounding an oxygen, two will be near the oxygen and two will be distant from it. The ice rule leads to six possible configurations around a vertex:
\begin{center}
\includegraphics[height=100pt]{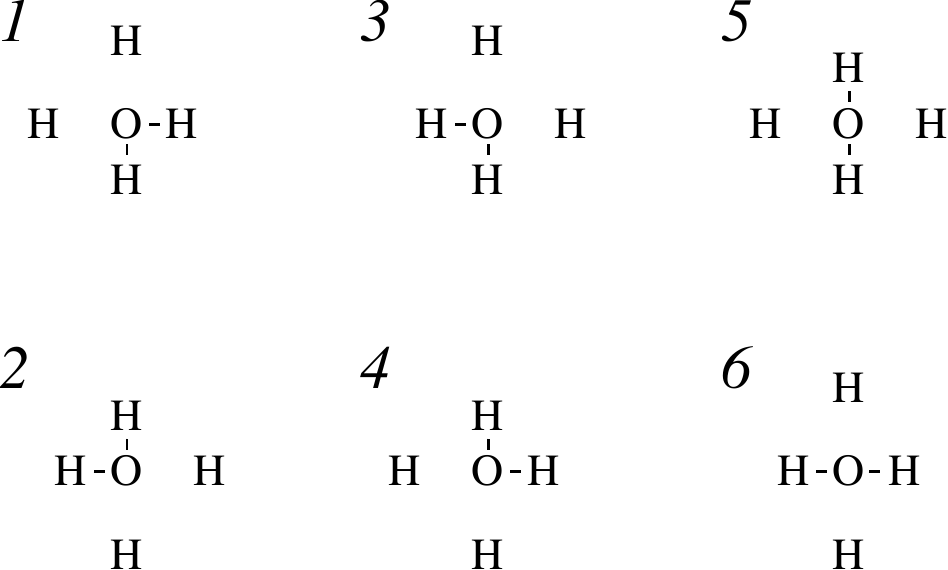}
\end{center}

The six possibilities are usually drawn as arrow configurations, where each arrow corresponds to a hydrogen atom and points to the nearest oxygen.
\begin{figure}[hb]
\begin{center}
\includegraphics[height=120pt]{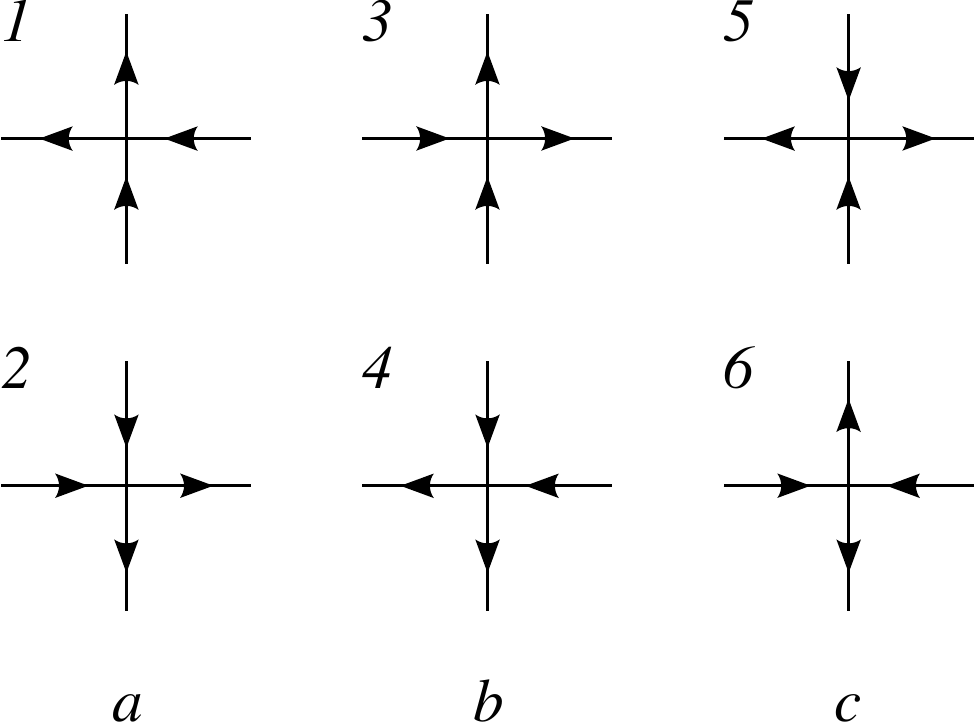}
\end{center}
\caption[The $6$ vertex configurations]{The $6$ vertex configurations with associated Boltzmann weights $a$, $b$, and $c$.}
\label{fig:6v}
\end{figure}
Each vertex configuration has a Boltzmann weight associated with it as shown in \figref{6v}. The Boltzmann weights $a$, $b$, and $c$ appear in the partition function of the model,
\[ Z_{6\text v}=\sum_{\text{conf.}} a^{m_a} b^{m_b}c^{m_c}, \]
where the sum is over all possible configurations of the lattice and $m_a$, $m_b$ and $m_c$ are the numbers of vertices of types $1$ \& $2$, $3$ \& $4$, and $5$ \& $6$ respectively.

The six vertex model has strong connections to combinatorics. When defined on a finite $M\times M$ grid, with `domain wall' boundary conditions --- outward-pointing arrows at the top and bottom, and inward-pointing arrows at the sides --- the possible configurations of the lattice are in bijection with $M\times M$ alternating sign matrices (ASMs). The bijection is made by replacing all vertices of type $1$, $2$, $3$, and $4$ with a $0$, vertices of type $5$ with a $-1$ and type $6$ with a $1$.
\[ \pic{-48pt}{100pt}{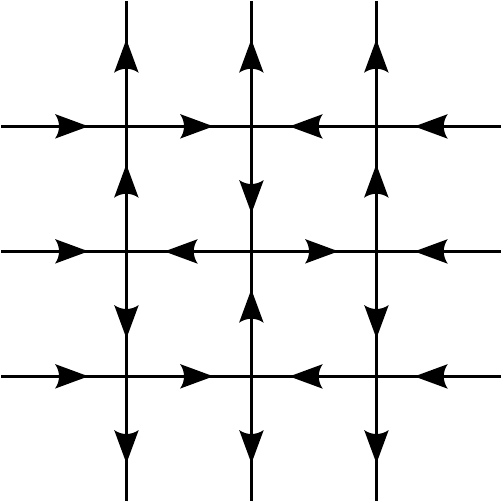}\qquad\rightarrow\qquad \pic{-28pt}{60pt}{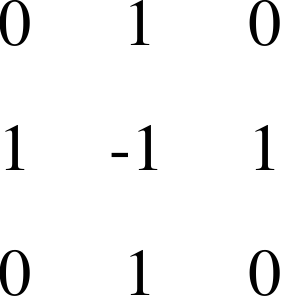} \]
A similar bijection can also be made with fully packed loop diagrams (FPLs). For a thorough review see \cite{Propp01}.

The six vertex model was solved (that is, the partition function was calculated) in 1967 by Lieb \cite{Lieb67a,Lieb67b} and Sutherland \cite{Sutherland67}. Lieb also found that the transfer matrix of the six vertex model shares eigenvectors with the Hamiltonian of the one-dimensional quantum XXZ spin chain, which was solved by Yang and Yang using the Bethe ansatz in the previous year \cite{YangY66b,YangY66a}.

The XXZ model is a quantum generalisation of the Ising model. Its Hamiltonian is a sum over nearest neighbours,
\[ H=\mathcal J\sum_{<ij>}\left(\sigma_i^x\sigma_j^x+\sigma_i^y\sigma_j^y+\Delta\sigma_i^z\sigma_j^z\right), \]
where $\sigma^x$, $\sigma^y$, and $\sigma^z$ are spin operators represented by the Pauli matrices, and $\Delta$ is known as the anisotropy parameter. When $\Delta=1$, the system becomes the Heisenberg model, and the value $\Delta=0$ corresponds to the free fermion point, where the model is related to the Ising model.

The eigenvectors of the six vertex transfer matrix only depend on the Boltzmann weights in a certain combination, which, as it turns out, exactly corresponds to the XXZ anisotropy parameter,
\[ \Delta=\frac{a^2+b^2-c^2}{2ab}. \]
In fact, the transfer matrix of the six vertex model and the Hamiltonian of the one-dimensional quantum XXZ spin chain are related through the Taylor expansion of the transfer matrix,
\[ T_{6\text v}(\mu)=T_{6\text v}(0)\left(1+\mu\ H_\text{XXZ}+\ldots\right),\]
where $\mu$ is the spectral parameter.

This relationship is an example of a common occurrence in classical two-dimensional lattice models, where the logarithmic derivative of the transfer matrix taken at $\mu=0$ corresponds to the Hamiltonian of a related one-dimensional quantum system. The Hamiltonian is usually a local operator, unlike the transfer matrix, so it is often easier to deal with, however it can also contain less information.

Another generalisation of the Ising model is the $n$-vector model (also known as the O($n$) model), introduced by Stanley in 1968 \cite{Stanley68}. In this model, each spin is a unit vector with $n$ components. The energy function of the model is
\[ E_{\text O(n)}=-\mathcal J\sum_{<ij>}\sigma_i\cdot\sigma_j. \]
The name O($n$) refers to the fact that the spins have the symmetry of the orthogonal group. The limit $n\rightarrow 0$ is related to the theory of self-avoiding walks, which has an application in the study of polymers.

The $n$-vector model, similarly to the Potts model, has a reformulation as a loop model (see \cite{DubailJS09,Nien90,BloteN89,BloteN94,KunzW88}), and this allows for a generalisation such that $n$ need not be a positive integer. The O($n$) loop model on a square lattice has nine possible states for each lattice face:
\begin{center}
\includegraphics[height=100pt]{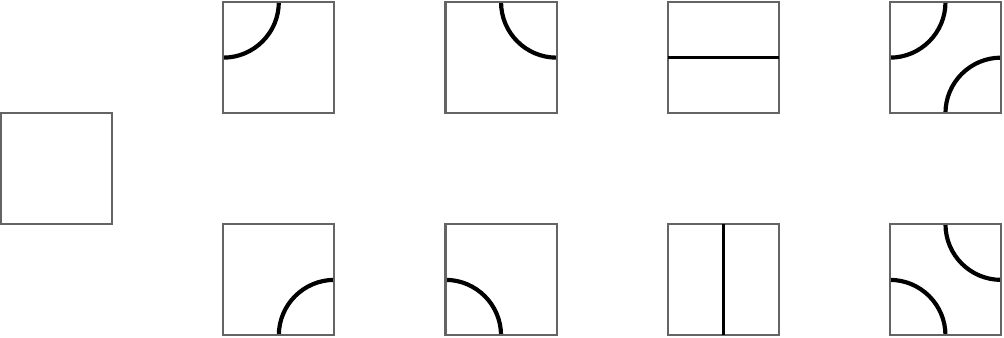}
\end{center}
The loops decorating the faces (or plaquettes) correspond to interactions between spins (the spins are located on the centres of the edges). Each plaquette has a weight associated with it, and various results have been obtained for systems with different restrictions on the weights.

The completely packed version \cite{MitraN04}, where the weights of all but the last two pictures are set to zero, is the one this thesis will be focusing on. The model has a rich underlying mathematical structure, best understood in terms of the Temperley--Lieb algebra. For this reason we will refer to the completely packed O($n$) loop model as the Temperley--Lieb loop model, or the TL($n$) model. This model is related to the $Q$-state Potts model on the square lattice by $\sqrt Q=n$.

\section*{The Temperley--Lieb loop model}
It would seem natural to define the Temperley--Lieb loop model in terms of its origins as the $n$-vector model. However the connection is mathematically involved \cite{KunzW88,DubailJS09,BloteN94,DomanyMNS81} and not altogether intuitive. Further, in recent years the TL($n$) model has garnered a significant amount of interest quite separate from the spin model formulation. For these reasons we will define it directly as a loop model.

The TL($n$) loop model is defined on a horizontally finite, vertically semi-infinite square lattice, where each face of the lattice has loops drawn on it in one of two orientations, as described earlier. We choose the infinite direction to be downwards, but this is simply a convention. Weights are then given to closed loops, and we are interested in the remaining connectivities at the top of the lattice. The lattice may be wrapped around a cylinder to form the periodic TL($n$) model, or placed on a finite width strip with various types of boundary conditions at the left and right. If loops are allowed to end on a boundary, we call the boundary open (or non-diagonal), otherwise the boundary is closed (or diagonal). In addition to the periodic model, there are three notable versions of the model; namely the zero-boundary model with closed boundaries at the left and right, the one-boundary model with one closed and one open boundary, and the two-boundary model with two open boundaries. We are interested in the latter case.

The transfer matrix of the TL($n$) model describes all the possible configurations on one row of the semi-infinite lattice, and one can add rows to the lattice by acting with the transfer matrix at the top of the semi-infinite strip. When $n=1$ the largest eigenvalue of the transfer matrix is $1$, so the partition function is trivial.\symbolfootnote{This fact is in agreement with the triviality of the Potts model at $Q=1$, when every spin has the same value. There exists a sypersymmetry interpretation of this point, see \cite{FendR02}.} These properties allow us to calculate other statistical quantities of interest for the two-boundary model with $n=1$, such as the probability of observing a certain configuration at the top of the lattice, and a correlation function between two spatially separated points.

The TL($n$) model, at special values of $n$, is closely related to a number of different statistical mechanical models. At $n=1$ there is an easy mapping to the two-dimensional bond percolation problem, which is made by drawing hulls around the percolation clusters as pictured:
\begin{center}
\includegraphics[height=100pt]{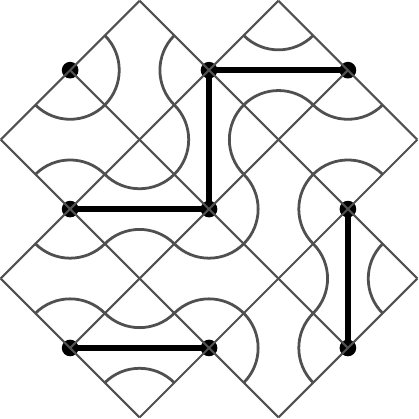}
\end{center}
Percolation of the bonds then corresponds to percolation of the hulls, or loops. Also at $n=1$, the action of the transfer matrix on the lattice of the TL($1$) model can be interpreted as the evolution of a stochastic process (the stochastic Raise and Peel model \cite{PearceRdGN02}), and the ground state eigenvector of the transfer matrix is the steady state of this process. In addition, the TL($n$) model is closely related to the quantum XXZ spin chain in one dimension, as the Hamiltonians of the two models correspond to different representations of the underlying Temperley--Lieb algebra. The anisotropy parameter $\Delta$ in the XXZ model is related to $n$ by $n=-2\Delta$.

The ground state of the TL($1$) loop model with periodic boundary conditions has also been connected with the combinatorial counting problems of alternating sign matrices, fully packed loops and plane partitions. This connection has been known as the Razumov--Stroganov (RS) conjecture \cite{RazStr01,RazStr01Tw,RazStr04,RazStr05,DFZJ05gen}, but was recently proved using combinatorial means by Cantini and Sportiello in \cite{CantiniS11}. The original conjecture also sparked a host of related conjectures, largely concerning different boundary conditions or topologies on the TL($1$) model and different boundary conditions or symmetry requirements on the counting problems, which at the time of writing remain unproved \cite{BatdGN01,MitraNdGB04,dG05}. Notable by its absence is a conjecture relating to the ground state of the TL($1$) loop model with two distinct boundaries, which as yet has no connection to any similar combinatorial problem, though if the boundaries are identified a connection can be made \cite{dGR04}.

A key element in the solutions of TL($n$) model with different boundary conditions is the $q$-deformed Knizhnik--Zamolodchikov ($q$KZ) equation \cite{Cantini09,DF05,DF05Bound,DF07,dGP07,ZJ07,JimboM95,JimboM96,Pasq06,DFZJ05gen}. With the relationship $n=-(q+q^{-1})$, it can be shown that the ground state of the TL($1$) model satisfies the $q$KZ equation for $q=\e^{2\pi\ii/3}$. The version of the $q$KZ equation used depends on the boundary conditions. This connection provides a tool set for solving for the ground state, which would otherwise be inaccessible.

Recently, a further connection has been made between the TL($1$) loop model and the quantum spin Hall effect \cite{GruzbergLR99,Cardy00,KasaP07}. In particular the $2$-boundary version is closely connected to this effect, since a type of correlation function between the two boundaries (see \chapref{Flow}) corresponds to the spin current in a model of the quantum spin Hall effect.

\section*{Layout of the thesis}

This thesis is concerned with properties of the TL($n$) loop model with two open boundaries, at the special point $n=1$. Each horizontal position $i$ on the lattice has a variable $z_i$ associated with it, and the left and right boundaries have associated variables $\zeta_0$ and $\zeta_L$ respectively. In addition, the transfer matrix $T$ depends on the spectral parameter $w$, and since it can be shown that $[T(w),T(u)]=0$, the eigenvectors do not depend on this parameter.

\chapref{On} defines the TL($n$) loop model with general parameters, first by defining the Temperley--Lieb algebra and then describing the lattice model. The transfer matrix and Hamiltonian are defined, and the specialisation to $n=1$ is described.

The main results of \chapref{2BSoln} are the exact calculation of key components of the ground state eigenvector of the 2-boundary transfer matrix for finite system size $L$, as well as the normalisation of the eigenvector. These quantities are expressed in terms of the polynomial character of the symplectic group, which is the Schur polynomial for the root system of type $C$. The reason why the remaining components of the eigenvector become harder and harder to specify as the system size increases is discussed.

Another quantity that is exactly calculated for finite system size, in \chapref{Flow}, is the boundary-to-boundary correlation function, which is equivalent to the average density of percolating clusters between two points in a two-dimensional percolation model, defined on an infinite lattice of finite width. This quantity can also be interpreted as the spin current in the Chalker--Coddington model of the quantum spin Hall effect. The correlation function is expressed in terms of the same symplectic characters as the normalisation.

A key aspect of any statistical mechanical model is the thermodynamic limit. The results in \chapref{2BSoln} and \chapref{Flow} are obtained exactly for finite system sizes, which is usually not possible for critical systems. Generally it is only possible to derive approximate results in the thermodynamic limit. Our calculations thus allow for the calculation of exact thermodynamic quantities, by taking the limit as $L\rightarrow\infty$. It will then be possible to compare our results to the previously calculated non-rigorous continuum limit \cite{Cardy00}.

In order to obtain the thermodynamic limit, we need to consider the large $L$ asymptotics of the symplectic character that appears in all the key results. In particular we are interested in the effect the two boundaries have in the asymptotic limit. It is therefore desirable to analyse the asymptotics of the symplectic character when all its arguments are set to $1$, except for the two variables that correspond to the boundaries. In \chapref{Branching}, a separation of variables method initiated by Sklyanin \cite{Skl88} is extended to the symplectic character. It is hoped that an asymptotic approximation for the separated polynomial will be easier to find than the symplectic character itself, and by reversing the process of separation the result can be used to calculate the thermodynamic limit of the physical quantities associated with the TL($n$) model. At present, however, this is an open problem.

\newpage
\chapter{The Temperley--Lieb loop model}
\label{chap:On}

Sklyanin's transfer matrix \cite{Skl88} of the TL($n$) model, and consequently the Hamiltonian, can be expressed in terms of algebraic generators satisfying a Temperley--Lieb algebra \cite{TemperleyL71}, see for example \cite{DF05,PearceRZ06}. The imposed boundary conditions decide which particular version of the Temperley--Lieb (TL) algebra is needed. In this chapter we will introduce the TL($n$) model on a strip with open boundaries on both sides, which can be described in terms of the two-boundary Temperley--Lieb (2BTL) algebra \cite{dGN09}. Models with two reflecting or diagonal boundaries, as well as with mixed boundaries were studied in \cite{DF05,ZJ07,dGP07}. The periodic version was considered in \cite{MitraN04,RazStr04,RazStr01}.

\section{The Temperley--Lieb algebra}
\label{sec:TLA}
The Temperley--Lieb algebra is an $sl_2$ quotient of the Hecke algebra (see \apref{hecke}). The different types of the Hecke algebra lead to different boundary conditions of the TL algebra. The 2BTL algebra comes from the Hecke algebra of type $BC$ \cite{Ernst09,dGN09}, see also \cite{Daugherty10}.

One of two distinguished representations of the 2BTL algebra \cite{dGN09} is in a space of connectivities or link patterns, described in \secref{linkp}. This representation is relevant for the TL($n$) model with open boundaries \cite{MitraNdGB04,JacobS08}, and we will use it to illustrate the generators and relations that define the TL algebras.

We will give brief definitions of the TL algebra with the four most common boundary conditions, before describing the two-boundary version in more detail. We then use the 2BTL algebra to define the TL($n$) loop model for general $n$, and in \secref{spec} discuss the specialisation $n=1$ at which we obtain physical results for the system.

\subsection*{Trivial boundaries}
\begin{defn}
\label{def:0BTL}
The best known version of the TL algebra is generated by elements $e_i$, $1\leq i\leq L-1$, which satisfy relations
\begin{align*}
 e_i^2&=-2\cos\gamma\ e_i, &&\hspace{-3cm} \forall i,\\
 e_ie_{i+1}e_i&=e_i,&&\hspace{-3cm}  1\leq i\leq L-2,\\
 e_ie_{i-1}e_i&=e_i,&&\hspace{-3cm}  2\leq i\leq L-1,\\
 e_ie_j&=e_je_i,&&\hspace{-3cm}  |i-j|>1,
\end{align*}
where $L$ is the system size and $\gamma$ is a complex parameter, related to the parameter $n$ by $n=-2\cos\gamma$.
\end{defn}
In the link pattern representation, the generators look like strings across a strip,
\[e_i=\quad\raisebox{-10pt}{\includegraphics[width=120pt]{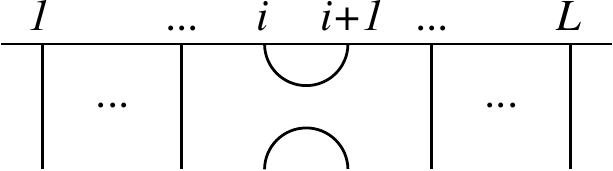}}\;,\]
and multiplication in the algebra from the left corresponds to vertical concatenation of the pictures from the top. The relations can be put in terms of two rules: firstly, from the rule $e_i^2=-2\cos{\gamma}\ e_i$, closed loops are removed and replaced with a factor of $-2\cos\gamma$,
\[ \raisebox{-21pt}{\includegraphics[width=120pt]{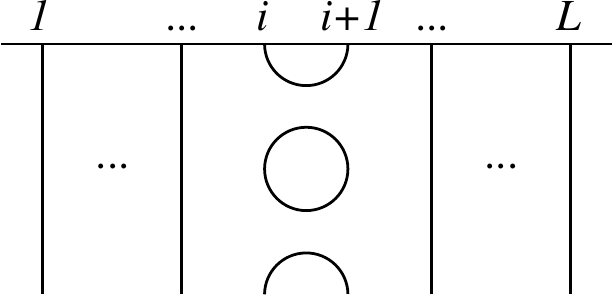}}\quad=-2\cos{\gamma}\quad\raisebox{-10pt}{\includegraphics[width=120pt]{TLei}}\quad,\]
and secondly, from the rule $e_ie_{i\pm 1}e_i=e_i$, strings are pulled tight,
\[\raisebox{-35pt}{\includegraphics[width=140pt]{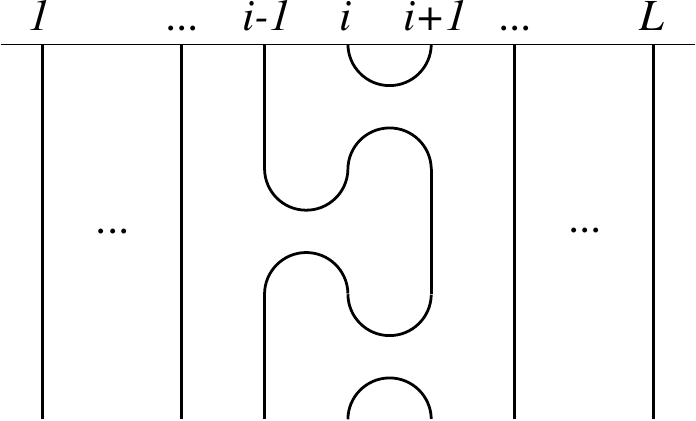}}\quad=\quad\raisebox{-10pt}{\includegraphics[width=120pt]{TLei}}\ . \]
\subsection*{Periodic boundaries}
\begin{defn}
\label{def:PTL}
The periodic version has the generators and relations from \defref{0BTL}, as well as an extra generator $e_L$ that satisfies the relations \cite{MartinS94,Jones99}
\begin{align*}
e_L^2&=-2\cos\gamma\ e_L,\\
e_ke_Le_k&=e_k,\\
e_Le_ke_L&=e_L,\raisebox{0pt}[0pt][0pt]{\raisebox{0.6em}{\ensuremath{\biggr\}\qquad\;\ k=1,L-1,}}}\\
e_Le_i&=e_ie_L,\ \,\qquad 2\leq i\leq L-1.
\end{align*}
This algebra is infinite-dimensional, so when $L$ is even an additional relation can be imposed,
\be
\label{eq:Pquo}
(e_1e_3\ldots e_{L-1})\ (e_2e_4\ldots e_L)\ (e_1e_3\ldots e_{L-1})=\eta\ (e_1e_3\ldots e_{L-1}).
\ee
This relation produces a finite-dimensional quotient of the periodic TL algebra.
\end{defn}
The pictorial representation is similar to the trivial boundary version, but this time the strip closes on itself to form the surface of a cylinder,
\[e_i=\ \raisebox{-42pt}{\includegraphics[width=100pt]{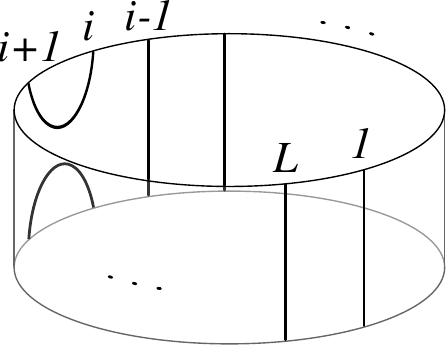}}\;.\]
In this representation, the relation $e_ie_{i-1}e_i=e_i$ looks like
\[\includegraphics[width=100pt]{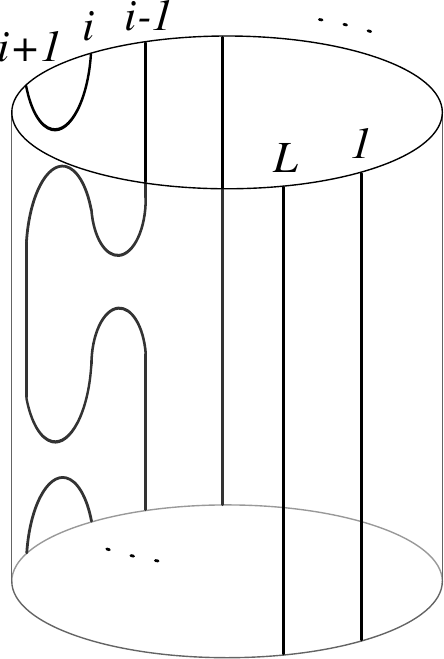}\raisebox{70pt}{\quad =\quad}\raisebox{35pt}{\includegraphics[width=100pt]{PTLei}}\raisebox{70pt}{\quad .}\]
The quotient relation \eqref{eq:Pquo} describes the weight given to non-contractible loops,
\[ \raisebox{-70pt}{\includegraphics[width=100pt]{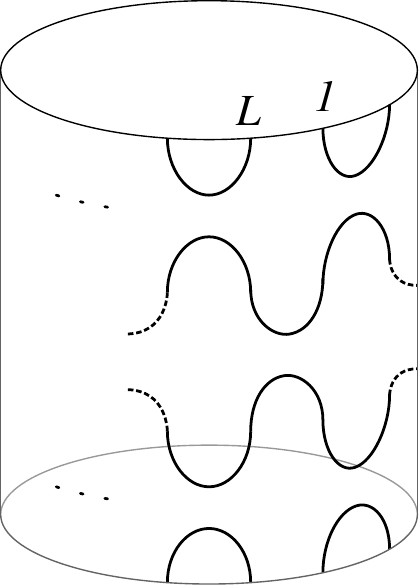}} \quad=\eta\quad\raisebox{-35pt}{\includegraphics[width=100pt]{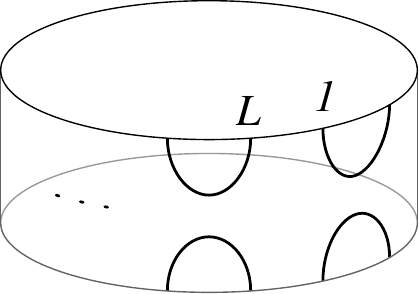}}\quad . \]
\subsection*{One boundary}
\begin{defn}
\label{def:1BTL}
The one-boundary Temperley--Lieb (1BTL) algebra, otherwise known as the blob algebra \cite{MartinS94}, also has the generators and relations of the trivial boundary version \defref{0BTL}, along with a left boundary generator $e_0$, depicted as
\[
e_0=\ \pic{-9pt}{30pt}{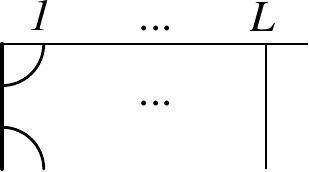}\ ,
\]
which satisfies
\[e_1e_0e_1=e_0,\qquad e_0^2=\frac{-\sin\omega_0}{\sin(\omega_0+\gamma)} e_0,\]
where $\omega_0$ is an additional complex parameter. Note that we do not impose $e_0e_1e_0=e_0$.
\end{defn}
Pictorially the first relation is
\[\pic{-35pt}{80pt}{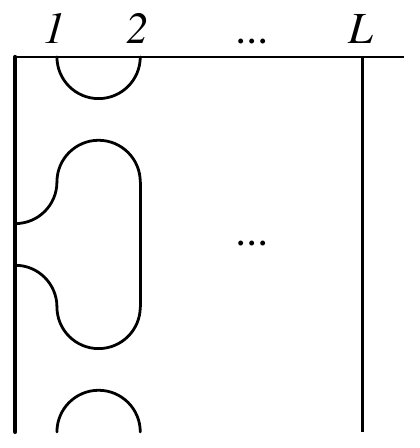}\quad =\quad\pic{-35pt}{80pt}{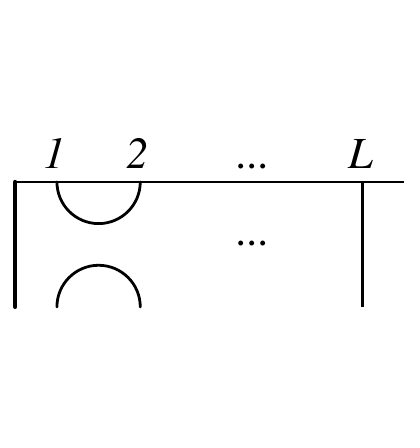}\quad,\]
and the second is
\[ \pic{-24pt}{60pt}{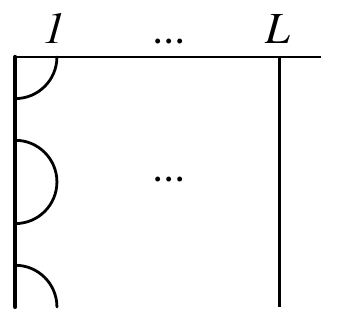}\quad=\frac{-\sin\omega_0}{\sin(\omega_0+\gamma)}\quad\pic{-24pt}{60pt}{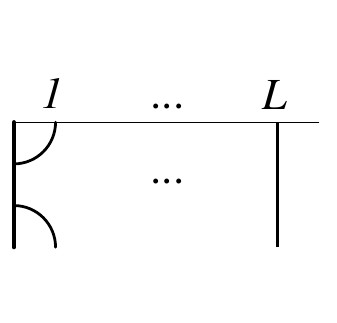}\quad.\]

Both of these relations involve a closed loop at the left boundary. However, only the second relation produces a non-trivial factor. To make clear the distinction, we introduce the notion of parity by supposing the generators are shaded in the following way:
\[ e_i=\quad\pic{-4.3mm}{9mm}{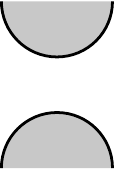}\quad,\quad i\text{ even},\qquad e_i=\quad\pic{-4.3mm}{9mm}{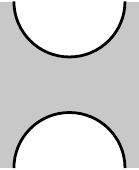}\quad,\quad i\text{ odd},\]
and $e_0$ has the same shading as the even case. Then a loop that has both ends connected to the left boundary produces a non-trivial factor iff the inside of the loop is shaded.

\subsection*{Two boundaries}
Finally, the version we will be working with is the 2BTL algebra \cite{DubailJS09b}, which has the generators of the one-boundary case above, as well as a right boundary generator $e_L$ depicted as
\[
e_L=\ \pic{-9pt}{30pt}{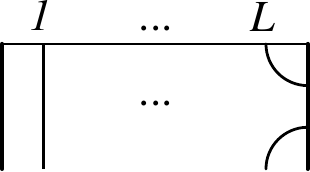}\ ,
\]
which satisfies similar relations to $e_0$,
\[e_{L-1}e_Le_{L-1}=e_L,\qquad e_L^2=\frac{-\sin\omega_L}{\sin(\omega_L+\gamma)} e_L.\]

\begin{defn}
\label{def:2BTL}
Written all together, the algebraic relations for the 2BTL are
\begin{align}
e_i^2 &=-2\cos{\gamma}\ e_i, &&\hspace{-2cm} 1\leq i\leq L-1,\nn\\
e_i^2 &=\frac{-\sin\omega_i}{\sin(\omega_i+\gamma)}e_i,&&\hspace{-2cm}i=0,L,\label{eq:ei2a}\\
e_ie_{i\pm1}e_i &= e_i, &&\hspace{-2cm}1\leq i\leq L-1,\nn
\end{align}
along with the idempotent relations \eqref{eq:DoubleQuotient} described in the next section.
\end{defn}
Loops connected to the left boundary behave as in the 1BTL, but at the right boundary the parity of $L$ must also be taken into account. If $L$ is even, a loop that has both ends connected to the right boundary produces a non-trivial factor iff the inside of the loop is shaded, but if $L$ is odd the factor appears iff the inside of the loop is \emph{not} shaded.

\subsection{The idempotent relations}
\begin{figure}[ht]
\centerline{
\pic{0pt}{90pt}{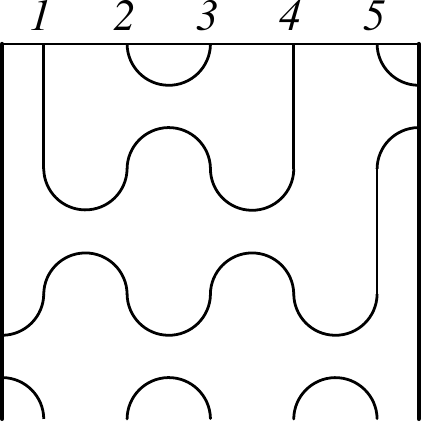}
}
\caption[The arrangement $e_2e_5e_1e_3e_0e_2e_4$]{The arrangement $e_2e_5e_1e_3e_0e_2e_4$ when $L=5$ has a loop stretching from the left boundary to the right boundary.}
\label{fig:line}
\end{figure}

The versions of the TL algebra with 0 and 1 boundaries are both finite-dimensional. However, the periodic and 2-boundary versions are not, since it is possible to form a string winding all the way around the cylinder (for the periodic case) or stretching from the left boundary to the right (for the two-boundary case), as in \figref{line}.

It was shown in \cite{dGN09} that all finite-dimensional irreducible representations of the 2BTL satisfy two additional relations, which we will describe now.
\begin{defn}
We first define two (unnormalised) idempotents $I_1$ and $I_2$ as follows:
\label{def:Idem}
\begin{align}
\label{eq:Idempotents}
& I_1=\left\{
\begin{array}{ll}
e_1e_3\cdots e_{L-1}, & \quad L \text{ even},\\
e_1e_3\cdots e_L, & \quad L \text{ odd},
\end{array}\right.
&& I_2=\left\{
\begin{array}{ll}
e_0 e_2\cdots e_L, & \quad L \text{ even},\\
e_0 e_2\cdots e_{L-1}, & \quad L \text{ odd}.\\
\end{array}\right.
\end{align}
For example, when $L=6$ the idempotents are
\[ I_1=\quad\pic{-11pt}{35pt}{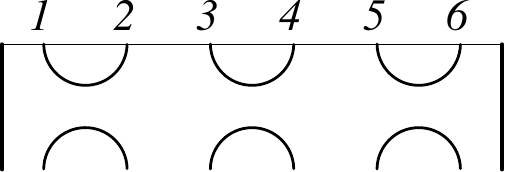}\quad,\qquad I_2=\quad\pic{-11pt}{35pt}{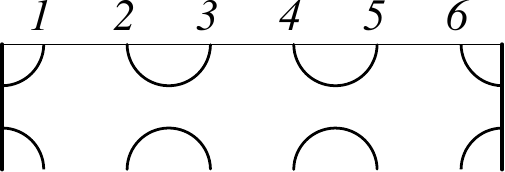}\quad.\]
\end{defn}

The double quotient of the 2BTL algebra has the additional relations:
\begin{align}
&I_1 I_2 I_1 = b I_1, && I_2 I_1 I_2 = b I_2,
\label{eq:DoubleQuotient}
\end{align}
where $b$ is an additional parameter. In our pictorial representation, these relations have the effect of removing pairs of loops stretching from the left boundary to the right and replacing them with a factor of $b$.

\subsection{Link pattern space}
\label{sec:linkp}
\begin{figure}[ht]
\begin{center}
\includegraphics[height=50pt]{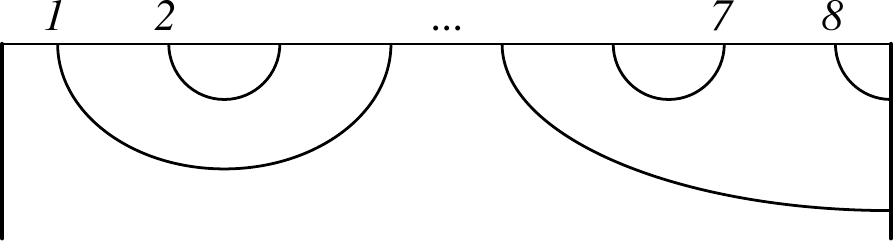}
\end{center}
\caption{A link pattern for $L=8$.}
\label{fig:linkpat}
\end{figure}

Here we will define the space ${\rm LP}_L$ spanned by link patterns (sometimes called connectivities) in terms of the loop representation of the 2BTL generators above. This space is equivalent to the space spanned by (a variant of) anchored cross paths \cite{Pyat04}. The space ${\rm LP}_L$ forms the state space of the TL($n$) loop model \cite{ShiU07}. An example of a link pattern is given in \figref{linkpat}.
\begin{defn}
A link pattern is a non-crossing matching of the integers $0,1,\ldots,L+1$. The matching between the integers $1,\ldots, L$ is pairwise, whereas $0$ and $L+1$ may be matched with, or connected to, an arbitrary number of other integers. The integers $0$ and $L+1$ are respectively referred to as the left and right boundary.
\end{defn}

We can express elements of ${\rm LP}_L$ as words in the generators of the 2BTL. We choose one of the idempotents to be the shortest word, and act with any combination of the 2BTL generators. Then by the relations \eqref{eq:ei2a} and \eqref{eq:DoubleQuotient}, the resulting picture will reduce to one of the link patterns of size $L$, multiplied by the weight as introduced by the relations. For example, with $L=2$ we define the shortest word to be $I_1=e_1$, and acting with the combination $e_0e_1$ produces
\[e_0e_1I_1=-2\cos\gamma\ e_0I_1.\]
In fact, the four basis elements of ${\rm LP}_2$ are $I_1$, $e_0I_1$, $e_2I_1$ and $e_0e_2I_1$, and these are represented respectively by the pictures
\[\includegraphics[width=220pt]{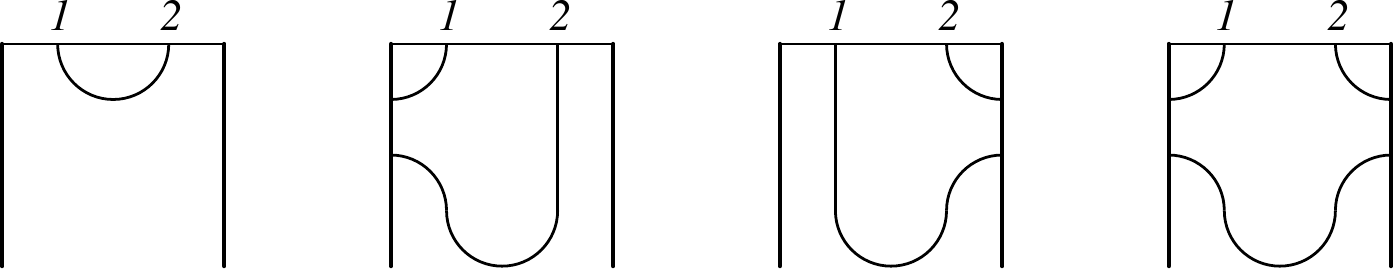}\ \raisebox{19pt}{\ .}\]

\begin{figure}[ht]
\begin{center}
\includegraphics[width=220pt]{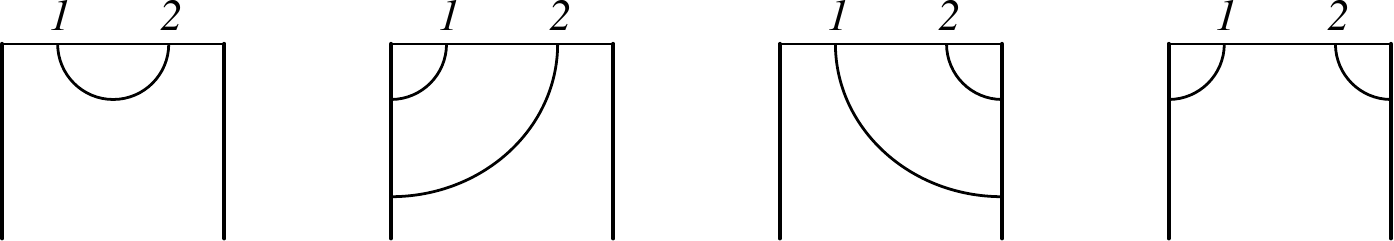}
\end{center}
\caption[The link patterns for $L=2$]{The link patterns for $L=2$, corresponding to $I_1$, $e_0I_1$, $e_2I_1$ and $e_0e_2I_1$ respectively.}
\label{fig:lp2}
\end{figure}

Note the line connecting the two boundaries in the final picture. If we had defined the shortest word to be $I_2=e_0e_2$ instead of $I_1$, a line would appear on the first three pictures and not the final one. Using the word representation of link patterns and starting from $I_1$, the link pattern with $1$ connected to the left boundary and $2$ to the right will always have one of these lines, and the others never will. Because of this, and because it is always possible to find out whether a picture should have a line, we omit single lines connecting the two boundaries when referring to a link pattern. In this way, the above pictures can be represented as the link patterns in \figref{lp2}.

We use a shorthand notation for the link patterns, in terms of a sequence of opening `(' and closing `)' parentheses. The $i$th bracket in the sequence refers to whether the $i$th site is connected to some place to the right of it (opening bracket) or to the left (closing bracket). Since the loops are non-crossing and every site from $1$ to $L$ must be connected, this provides a unique labelling. The above link patterns for $L=2$ are thus given respectively by
\[ (),\quad )),\quad ((,\quad )(. \]

Because we can independently place an opening or closing parenthesis at each site, the dimension of the space ${\rm LP}_L$ of link patterns of size $L$ for the two-boundary Temperley--Lieb algebra is
\[ \dim {\rm LP}_L = 2^L. \]

\subsection{Path representation}
\label{sec:path}

\begin{figure}[ht]
\begin{center}
\subfigure[$e_0$]{
\pic{0pt}{70pt}{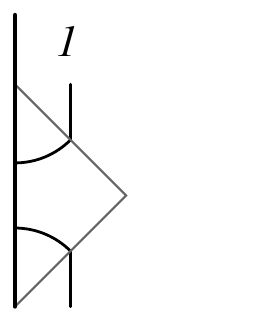}
}
\subfigure[$e_i$]{
\pic{0pt}{70pt}{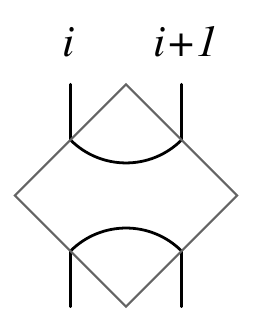}
}
\subfigure[$e_L$]{
\pic{0pt}{70pt}{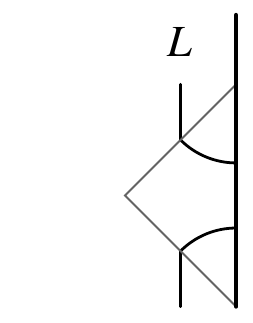}
}
\end{center}
\caption{Tiles for the 2BTL generators.}
\label{fig:pathdef}
\end{figure}

We now present a representation of the link pattern space using the graphical depiction of $e_i$ as a tilted square tile decorated with small loop segments; and of $e_0$ and $e_L$ as similarly decorated half-tiles lying against the boundaries of the picture, as shown in \figref{pathdef}.

\begin{figure}[ht]
\begin{center}
\pic{0pt}{40pt}{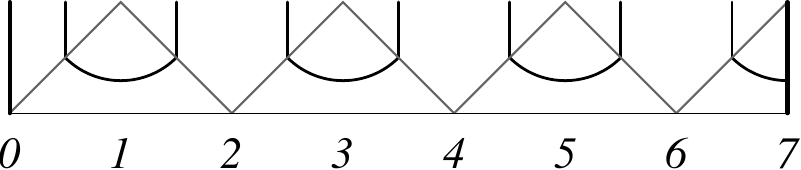}
\end{center}
\caption[The idempotent $I_1$ for $L=7$]{The idempotent $I_1$ for $L=7$. The label $i$ on the x-axis refers to the centre of the tile $e_i$.}
\label{fig:pathidem}
\end{figure}

The shortest word, given by one of the two idempotents, is depicted as a row of tilted half-tiles, which lie along the bottom of the picture, as in \figref{pathidem}.

Multiplication in the 2BTL algebra corresponds to vertical concatenation of the tiles. As an example, below are shown the algebraic relations $e_i^2=-2\cos\gamma\ e_i$ and $e_ie_{i-1}e_i=e_i$. The other relations are similar.

\[\pic{-33pt}{80pt}{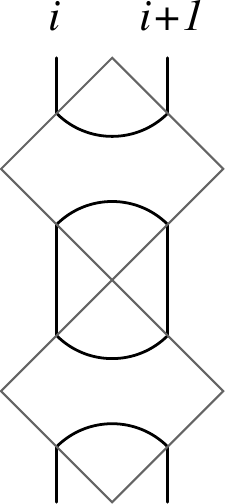}\quad=-2\cos{\gamma}\quad\pic{-33pt}{80pt}{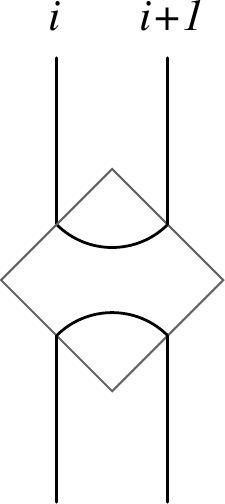}\quad,\qquad\pic{-33pt}{80pt}{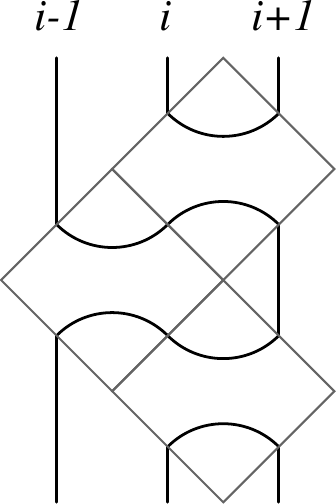}\quad =\quad\pic{-33pt}{80pt}{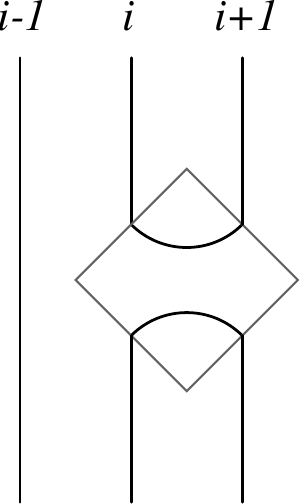}\quad.\]
\vspace{0.1cm}

\begin{figure}[ht]
\centerline{
\begin{picture}(200,200)
\put(0,0){\includegraphics[width=200pt]{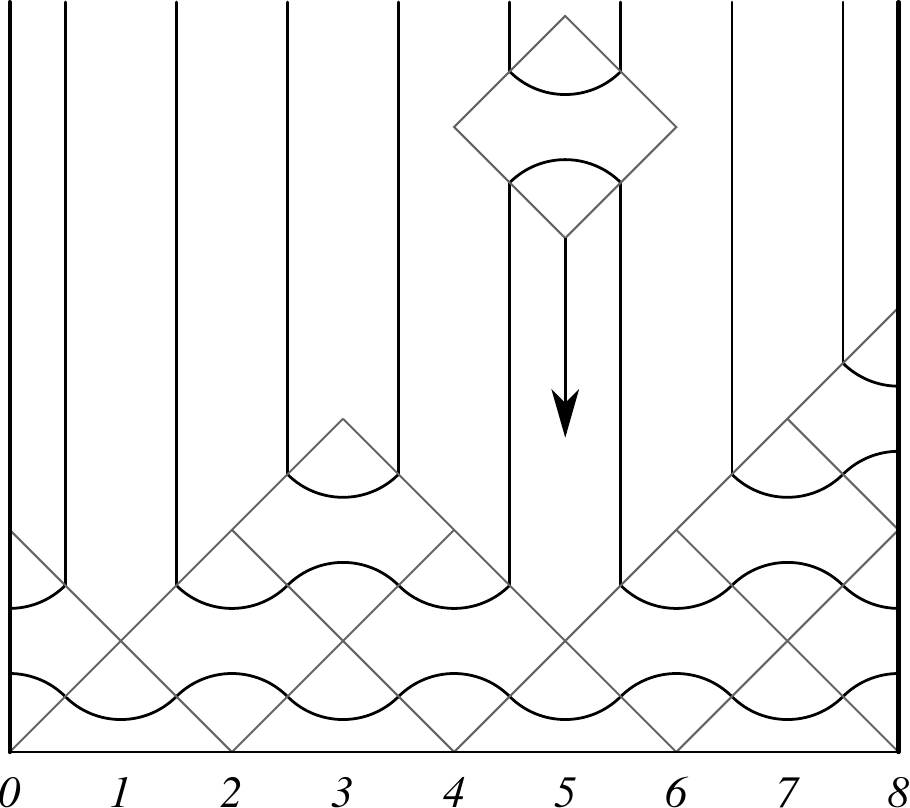}}
\end{picture}}
\caption[Path representation of the action of $e_5$ on the link pattern $)((){\ir )(}(($]{Path representation of the action of $e_5$ on the link pattern $)((){\ir )(}(($. The result will be the link pattern $)((){\ir ()}(($.}
\label{fig:e5onpath}
\end{figure}

Because of these relations, each link pattern corresponds uniquely to a path, which is traced out by the top of the tiles. A path is defined in this representation as beginning at the left boundary, ending at the right, and taking one step either up or down for each step to the right. The shorthand notation defined in the previous section works with the path representation as well, where steps down and up are symbolised by `$)$' and `$($' respectively. It is easily seen (as in \figref{e5onpath}) that when the path steps up, the loop in the link pattern connects somewhere to the right, and the opposite is true for down steps.

\subsection{Spin chain representation}
The 2BTL algebra has another distinguished representation, also of dimension $2^L$ \cite{MitraNdGB04}. This representation is in the tensor product space $(\mathbb{C}^2)^{\otimes L}$, giving rise to the quantum XXZ spin chain with non-diagonal boundary conditions on both sides. We define the Pauli spin matrices,
\begin{align*}
\sigma^{\rm x}&=\left[\begin{array}{cc}
0 & 1\\
1 & 0
\end{array}\right],\\
\sigma^{\rm y}&=\left[\begin{array}{cc}
 0 & -\ii\\
 \ii & 0 \end{array}\right],\\
\sigma^{\rm z}&=\left[\begin{array}{cc}
 1 & 0\\
 0 & -1 \end{array}\right],
\end{align*}
and will use the notation
\[ \sigma^{\rm x,y,z}_i =1^{\otimes i-1}\otimes \sigma^{\rm x,y,z}\otimes 1^{\otimes L-i}. \]
In the spin chain representation the Temperley--Lieb generators take the form \cite{dGNPR05}
\begin{align*}
e_i &\mapsto \frac12\left( \sigma_i^{\rm x}\sigma_{i+1}^{\rm x} + \sigma_i^{\rm y}\sigma_{i+1}^{\rm y} + \cos\gamma \left( \sigma_i^{\rm z}\sigma_{i+1}^{\rm z}-1\right) - \ii\sin\gamma  \left( \sigma_i^{\rm z}- \sigma_{i+1}^{\rm z}\right)\right),\\
e_0 &\mapsto \frac{1}{2\sin(\omega_0+\gamma)}\left( \cos\theta_1\ \sigma_1^{\rm x} + \sin\theta_1\ \sigma_1^{\rm y} +\ii \cos\omega_0\ \sigma_1^{\rm z} - \sin\omega_0\right),\\
e_L &\mapsto \frac{1}{2\sin(\omega_L+\gamma)}\left( \cos\theta_2\ \sigma_L^{\rm x} + \sin\theta_2\ \sigma_L^{\rm y} -\ii \cos\omega_L\ \sigma_L^{\rm z} - \sin\omega_L\right),
\end{align*}
where $\theta_1$ and $\theta_2$ are related to the idempotent parameter $b$ by
\[
b=
\left\{
\begin{array}{ll}
\frac{\textstyle \cos(\theta_1-\theta_2)+\cos(\gamma+\omega_0+\omega_L)}{\textstyle 2\sin(\gamma+\omega_0)\sin(\gamma+\omega_L)},&\qquad L\;\text{even},\\[3mm]
\frac{\textstyle \cos(\theta_1-\theta_2)+\cos(\omega_0-\omega_L)}{\textstyle 2\sin(\gamma+\omega_0)\sin(\gamma+\omega_L)},&\qquad L\;\text{odd}.
\end{array}
\right.
\]
Note that $b$ only depends on the difference of these two parameters. Due to the rotational symmetry in the spin $x$-$y$ plane, the extra freedom ($\theta_1+\theta_2$) can be interpreted as a free gauge parameter on which the algebra does not depend.

\section{The TL($n$) lattice}
\label{sec:lattice}
\begin{figure}[ht]
\begin{center}
\includegraphics[width=200pt]{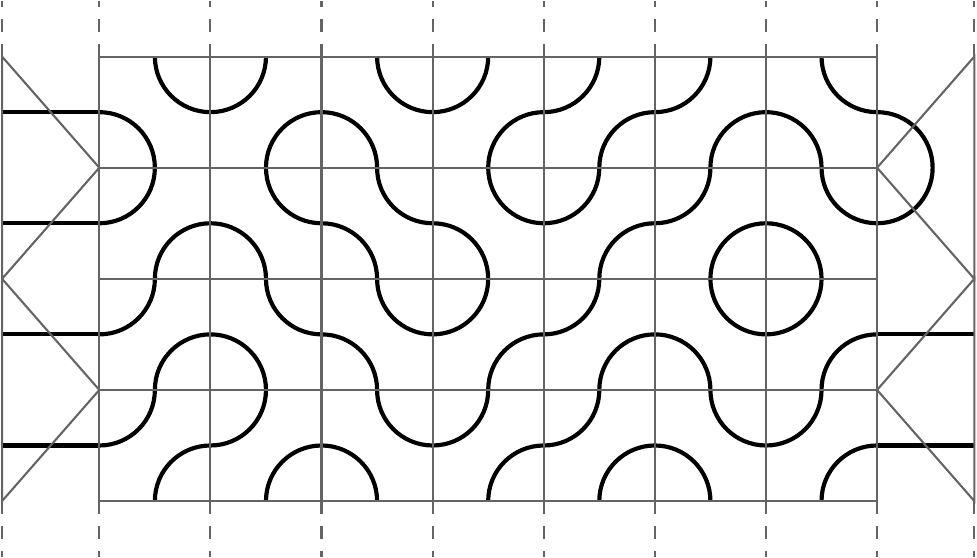}
\end{center}
\caption{One of the configurations of four rows of the lattice for $L=7$.}
\label{fig:lattice}
\end{figure}

The two-boundary TL($n$) loop model is defined on a lattice of width $L$, vertically infinite with a boundary on the left and right hand sides. Each face of the lattice is decorated with loops, which can end on the boundary or close back on themselves. Each bulk (square) face has two possible configurations of loops,
\[\pic{-8pt}{20pt}{R1}\quad \text{and}\quad \pic{-8pt}{20pt}{R2}\quad.\]
The boundary (triangle) faces also have two possible configurations, one joining the loops on two adjacent rows of the lattice, the other connecting them both to the boundary. At the left boundary the possibilities are
\[\pic{-18pt}{40pt}{K01}\qquad \text{and}\qquad \pic{-18pt}{40pt}{K02}\quad,\]
and at the right boundary the above pictures are reflected. Because of the action of the boundary faces, it is easy to see that the rows of the lattice come in pairs, as illustrated in \figref{lattice}.

\begin{figure}[ht]
\begin{center}
\includegraphics[width=200pt]{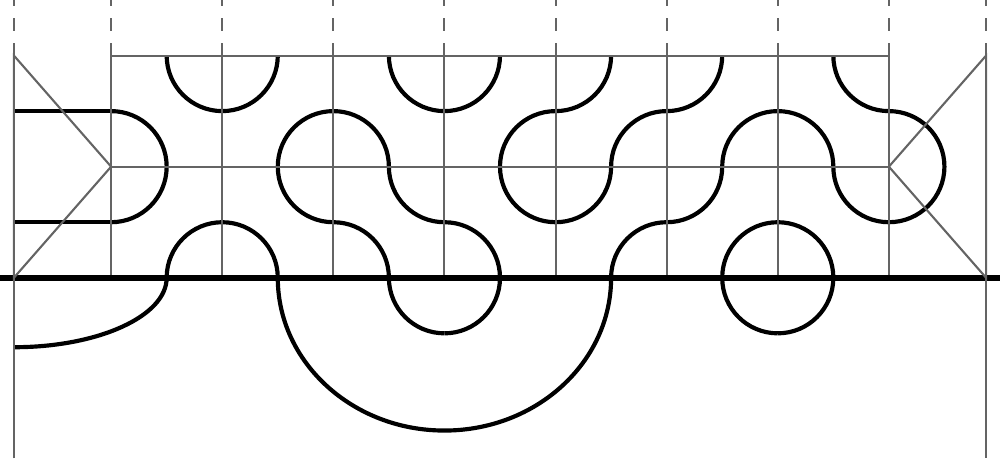}
\end{center}
\caption[Link pattern produced by \figref{lattice}]{The link pattern below the line is produced by the bottom two rows of the lattice in \figref{lattice}.}
\label{fig:tmlp}
\end{figure}

One of the important aspects of this model is how the loops connect different parts of the lattice. If we choose a configuration for each of the faces, and cut the lattice along a horizontal line between two pairs of rows, it is clear that below the line there are a collection of open loop segments, which connect different horizontal positions (see \figref{tmlp}).
In the example, position 1 is connected to the left boundary, position 2 to 5, 3 to 4, and 6 to 7. This is a clear example of a link pattern as defined in \secref{linkp}. As in the 2BTL algebra, we can ignore the specific paths the loops take and replace closed loops and loops that have both ends connected to a boundary by their respective weights. Once this is done the only thing left to consider is the remaining link pattern, which we call $\alpha$. Every configuration of the semi-infinite lattice below the horizontal line has one of the $2^L$ link patterns at the top.

The two rows of the lattice above the link pattern can have one of $2^{2L+2}$ possible configurations, and each one will map a link pattern into another. As an example, the configuration in \figref{tmlp} produces the link pattern
\[\includegraphics[width=200pt]{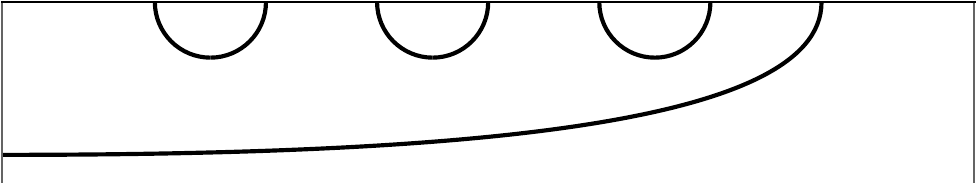}\ \raisebox{15pt}{,}\]
and introduces the weight $4\cos^2\gamma$ from the two closed loops.\symbolfootnote{The loop at the left boundary does not contribute a weight, as it only involves one left boundary tile, and is therefore a loop produced by the rule $e_1e_0e_1=e_1$. At the end of this chapter all the weights will be set to $1$ and this technicality will no longer be important.} We represent the two lattice rows as a matrix $T$ (called the transfer matrix), with the $ij$th entry being the sum of weights produced by all the configurations that map the $j$th link pattern to the $i$th (where there is some ordering on the link patterns).

A vector with a basis in the space of link patterns can be written as
\[ \ket\Psi=\sum_{\alpha\in\text{LP}_L} \psi_\alpha\ket\alpha. \]
If this vector is the unique `ground state' eigenvector of $T$,
\[ T\ket\Psi=\Lambda_{\text{max}}\ket\Psi,\]
where $\Lambda_{\text{max}}$ is the maximum eigenvalue, then repeated action of $T$ on some initial state $\ket{\text{in}}$ produces
\[ T^k\ket{\text{in}} \sim \Lambda_{\text{max}}^k\ket\Psi,\qquad \text{as}\qquad k\rightarrow\infty. \]
This process can be seen as building up a semi-infinite lattice with infinite copies of the transfer matrix. The eigenvector then expresses the relative weights of all possible link patterns of size $L$, produced by the configurations on the lattice.

As shown by Di Francesco and Zinn-Justin for other types of boundary conditions \cite{DF05,ZJ07}, it is possible to derive exact closed form expressions for certain properties of the ground state eigenvector for finite system sizes. To achieve this one needs to generalise the model just described, by considering an inhomogeneous version of the transfer matrix. This is done in the next section.

\subsection{Baxterisation}
\label{sec:bax}

In order to define the transfer matrix we will first introduce the operators $\check{R}$ and $\check{K}$, as well as their unchecked versions, to be defined shortly. We furthermore list some useful properties that we will need in later calculations. Throughout the following we will use the notation $[z]$ for
\[ [z] = z-z^{-1}, \]
and define
\[
q=\e^{\ii\gamma}. \]

\begin{defn}
\label{def:RKcheck}
The Baxterised elements $\check R_i(z)$, and the boundary Baxterised elements $\check K_0(z,\zeta)$ and $\check K_L(z,\zeta)$ of the Temperley--Lieb algebras are defined as
\be
\label{eq:R_z}
\begin{split}
\check R_i(z)&=\,\frac1{\kappa(z)}\frac{[q/z]-[z]\,e_i}{[qz]}\,,\\
\check K_i(z,\zeta)& =\,\frac{k(z,\zeta,\omega_i)+[q\e^{\ii \omega_i}][z^2]e_i}{k(1/z,\zeta,\omega_i)},\quad i=0,L,\\
\end{split}
\ee
where the parameter $z$ is called the spectral parameter. Each boundary element can also be equipped with an additional free parameter $\zeta$.
\end{defn}
The weight $k(z,\zeta,\omega_i)$ here is given by
\be
k(z,\zeta,\omega)=[z\e^{\ii\omega/2}/\zeta][z\e^{\ii\omega/2}\zeta],
\label{eq:kdef}
\ee
and $\kappa(z)$ is defined in terms of $q$-Pochhammer symbols \cite{JimboM95},
\be
\label{eq:kappa}
\kappa(z)=\left\{
\begin{array}{ll}
z\ \dfrac{(q^6z^2;q^4)_\infty(q^4z^{-2};q^4)_\infty}{(q^6z^{-2};q^4)_\infty(q^4z^2;q^4)_\infty}, & |q|<1,\\[4mm]
\dfrac{1}{z}\ \dfrac{(q^{-6}z^{-2};q^{-4})_\infty(q^{-4}z^2;q^{-4})_\infty}{(q^{-6}z^2;q^{-4})_\infty(q^{-4}z^{-2};q^{-4})_\infty}, & |q|>1,
\end{array}
\right.\\
\ee
where
\[ (a;b)_\infty=\prod_{n=0}^\infty(1-ab^n). \]
This normalisation factor satisfies the functional relations
\be
\begin{split}
\label{eq:kapparel1}
\kappa(z)\kappa(1/z)&=1,\\
\frac{\kappa(-q/z)}{\kappa(z)}&=\frac{-[qz]}{[q^2/z]},
\end{split}
\ee
and has the special value
\be
\label{eq:kapparel2}
\kappa(1)=1.
\ee

Note that the definition \eqref{eq:kappa} for $\kappa$ is non-analytic across the unit circle $|q|=1$. For these values of $q$ there is an alternate definition of $\kappa$, described in \cite{JimboM96}. At the special point $q=\e^{2\pi\ii/3}$, we can take $\kappa\equiv 1$, as this satisfies the functional relations \eqref{eq:kapparel1} and \eqref{eq:kapparel2}. We note, however, that this choice for $\kappa$ is not the limit of either function in \eqref{eq:kappa} as $q\rightarrow\e^{2\pi\ii/3}$.

\begin{prop}
\label{prop:RKcheckrelns}
The Baxterised elements obey the usual Yang--Baxter and reflection equations with spectral parameters:
\begin{align}
\label{eq:RKcheckYbeReflect}\check R_i(z)\check R_{i+1}(zw)\check R_i(w) &= \check R_{i+1}(w)\check R_i(zw)\check R_{i+1}(z),\nn\\
\check K_0(z,\zeta)\check R_{1}(zw)\check K_0(w,\zeta)\check R_1(w/z) &= \check R_{1}(w/z)\check K_0(w,\zeta)\check R_{1}(zw)\check K_0(z,\zeta),\\
\check K_L(z,\zeta)\check R_{L-1}(zw)\check K_L(w,\zeta)\check R_{L-1}(w/z) &= \check R_{L-1}(w/z)\check K_L(w,\zeta)\check R_{L-1}(zw)\check K_L(z,\zeta).\nn
\end{align}
They furthermore satisfy the unitarity relations
\be
\label{eq:RKcheckId}
\begin{split}
\check R_i(z)\check R_i(1/z)&=1,\\
\check K_i(z,\zeta)\check K_i(1/z,\zeta)&=1,\quad i=0,L.
\end{split}
\ee
\end{prop}
\begin{proof}
These relations can be easily checked using the algebraic rules \eqref{eq:ei2a}, or using a graphical notation like the one in \figref{e5onpath}.
\end{proof}

The above Baxterised elements are special cases of $R$-matrices, which can be defined more generally using the Hecke algebra. This is explained in more detail in \apref{hecke}.

We now introduce a graphical version of the Baxterised elements, using the planar Temperley--Lieb--Jones algebra \cite{Jones99}, which we will be able to use in a more general context than \figref{e5onpath}.

\begin{defn}
\label{def:RK}
We define the $R$-operator $R(w,z)$ to be the following linear combination of pictures:
\[
R(w,z)=\frac1{\kappa(w/z)}\Biggl(\frac{[qz/w]}{[qw/z]}\quad\raisebox{-10pt}{\includegraphics[height=25pt]{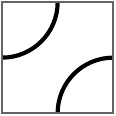}}\;+\;\frac{[z/w]}{[qw/z]}\quad\raisebox{-10pt}{\includegraphics[height=25pt]{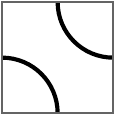}}\Biggr),
\]
and graphically abbreviate $R(w,z)$ by
\[
R(w,z) =\raisebox{-25pt}{\includegraphics[height=50pt]{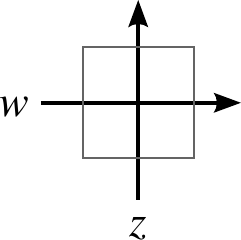}}.
\]
Note that we can use this picture in any orientation, as the arrows uniquely determine how the spectral parameters $z$ and $w$ enter in $R$. It is worthwhile to point out that $R(cw,cz)=R(w,z)$, for any constant $c$. We also define the boundary $K$-operators by
\begin{align*}
K_0(w)&=\frac{k(qw,\zeta_0,\omega_0)}{k(1/qw,\zeta_0,\omega_0)}\quad\raisebox{-18pt}{\includegraphics[height=40pt]{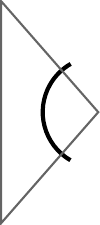}}\;+\frac{[q\e^{\ii \omega_0}][(qw)^2]}{k(1/qw,\zeta_0,\omega_0)}\quad\raisebox{-18pt}{\includegraphics[height=40pt]{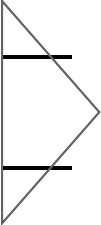}}\\
&=\raisebox{-26pt}{\includegraphics[height=50pt]{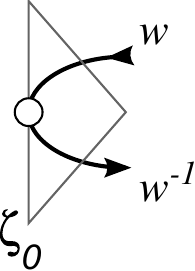}}\ ,\\
K_L(w)&=\frac{k(w,\zeta_L,\omega_L)}{k(1/w,\zeta_L,\omega_L)}\quad\raisebox{-18pt}{\includegraphics[height=40pt]{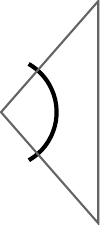}}\;+\frac{[q\e^{\ii \omega_L}][w^2]}{k(1/w,\zeta_L,\omega_L)}\quad\raisebox{-18pt}{\includegraphics[height=40pt]{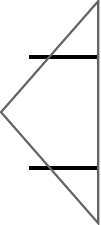}}\\
&=\raisebox{-26pt}{\includegraphics[height=50pt]{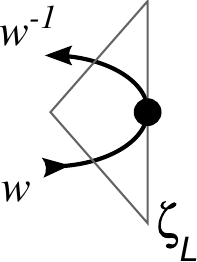}}.
\end{align*}
\end{defn}
The Baxterised elements $R$, $K_0$ and $K_L$ will be used to define the transfer matrix of the system. We can also write $K_0$ as
\be
\label{eq:KK}
\pic{-26pt}{50pt}{K0}\ =\pic{-26pt}{50pt}{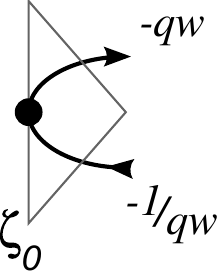}\ ,
\ee
where
\[\raisebox{-26pt}{\includegraphics[height=50pt]{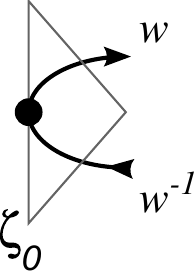}}=\frac{k(w,\zeta_0,\omega_0)}{k(1/w,\zeta_0,\omega_0)}\quad\raisebox{-18pt}{\includegraphics[height=40pt]{K01}}\;+\frac{[q\e^{\ii \omega_0}][w^2]}{k(1/w,\zeta_0,\omega_0)}\quad\raisebox{-18pt}{\includegraphics[height=40pt]{K02}}\ ,\]
which will be useful for defining the relations satisfied by $K_0$.

\begin{prop}
\label{prop:RKrelns}
The unitarity relations \eqref{eq:RKcheckId} for $R$  and $K$ can be graphically depicted as
\be
\label{eq:RId}
\raisebox{-15pt}{\includegraphics[width=80pt]{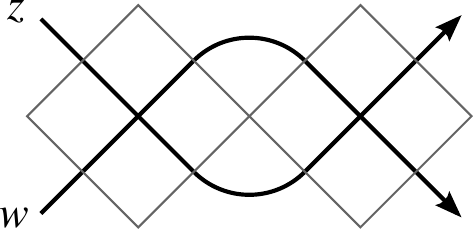}}\quad=\quad\raisebox{-15pt}{\includegraphics[width=80pt]{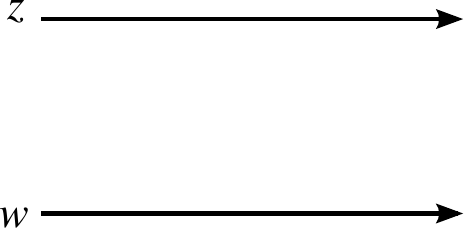}}\;,
\ee
and
\[
\raisebox{-55pt}{\includegraphics[height=120pt]{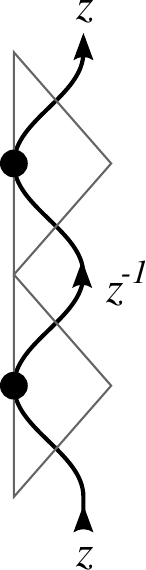}}\quad=
\quad\raisebox{-55pt}{\includegraphics[height=120pt]{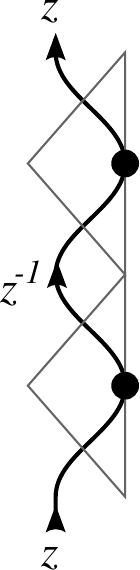}}\quad=\quad\raisebox{-55pt}{\includegraphics[height=120pt]{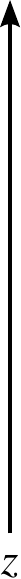}}\;.
\]
In addition, the Yang--Baxter and reflection equations \eqref{eq:RKcheckYbeReflect} can be written as
\be
\label{eq:YBE}
\raisebox{-45pt}{\includegraphics[height=80pt]{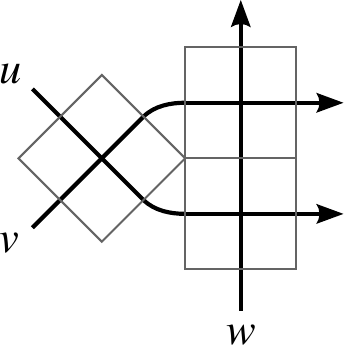}}\quad=\quad\raisebox{-45pt}{\includegraphics[height=80pt]{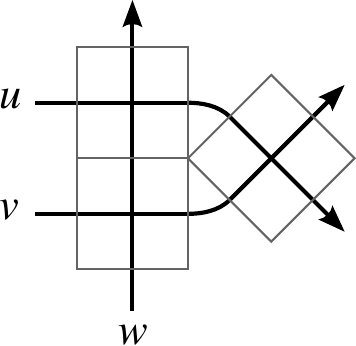}}\;,
\ee
and
\[
\raisebox{-55pt}{\includegraphics[height=120pt]{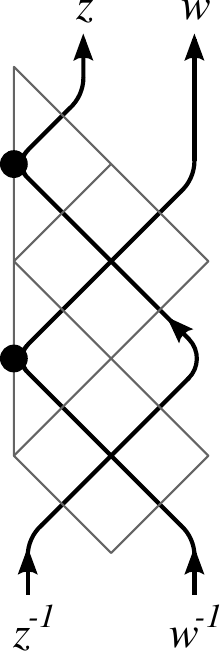}}\quad=\quad\raisebox{-55pt}{\includegraphics[height=120pt]{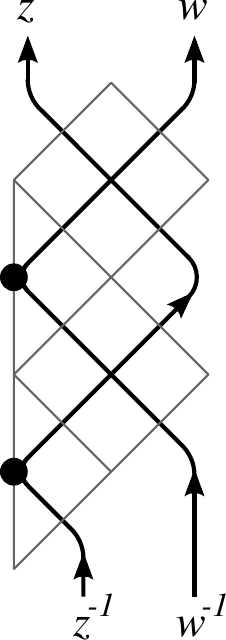}},\qquad\qquad\qquad\raisebox{-55pt}{\includegraphics[height=120pt]{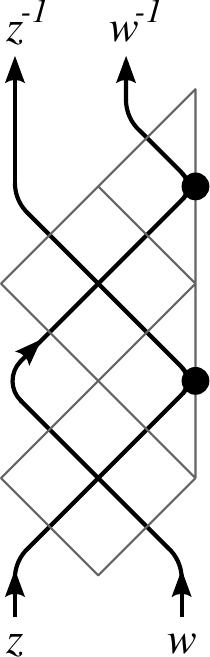}}\quad=\quad\raisebox{-55pt}{\includegraphics[height=120pt]{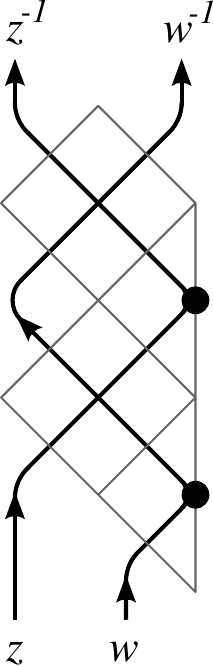}}\;.
\]
We furthermore note the crossing relation satisfied by $R$, i.e., $R(z,w)=R(-qw,z)$, which graphically reads \cite{JimboM95}
\be
\raisebox{-25pt}{\includegraphics[height=50pt]{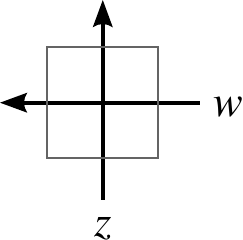}}\quad=\quad\raisebox{-25pt}{\includegraphics[height=50pt]{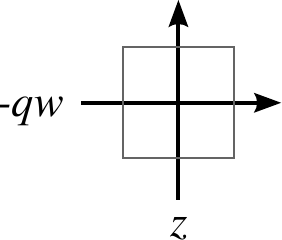}}\;.
\label{eq:crossing}
\ee
\end{prop}
The above relations are straightforward to prove from the definitions, by considering two loops to be the same if they have the same connectivities, and replacing closed loops with a factor of $-(q+q^{-1})$. The unitarity relation for $R$ \eqref{eq:RId} is proved here as an example.
\begin{proof}
The LHS of \eqref{eq:RId} produces four pictures,
\[\includegraphics[height=40pt]{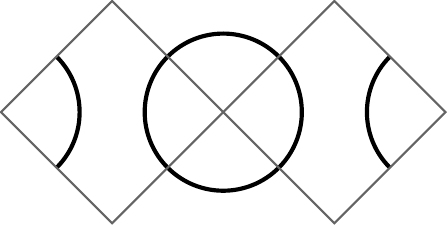}\raisebox{18pt}{\ ,}\qquad\includegraphics[height=40pt]{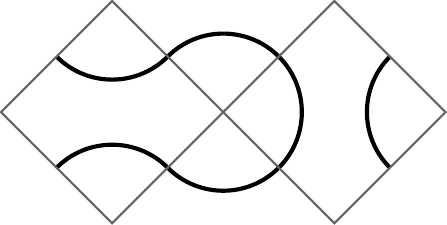}\raisebox{18pt}{\ ,}\qquad\includegraphics[height=40pt]{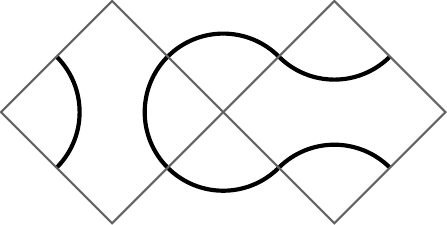}\raisebox{18pt}{\ ,}\qquad\includegraphics[height=40pt]{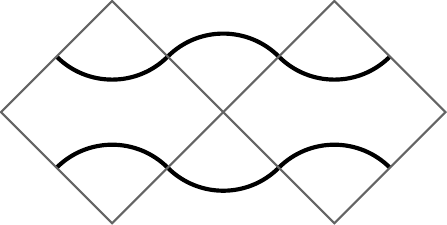}\raisebox{18pt}{\ .}\]
The first three of these have the same connectivity, so they can be grouped together with coefficient
\[\frac1{\kappa(w/z)\kappa(z/w)}\left(\frac{-(q+q^{-1})[w/z][z/w]+[qw/z][z/w]+[w/z][qz/w]}{[qz/w][qw/z]}\right)=0.\]
This leaves only the fourth picture, which is equal to the identity.
\end{proof}

Finally, we will also define slightly different versions of $K_0$ and $K_L$, which will be useful in proving the commutativity of the transfer matrix,
\be
\label{eq:Knew}
\pic{-26pt}{50pt}{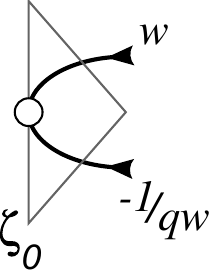}\quad=\quad\pic{-26pt}{50pt}{K0}\quad,\qquad\qquad\qquad\pic{-26pt}{50pt}{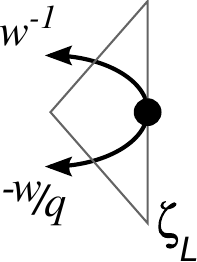}\quad=\quad\pic{-26pt}{50pt}{KL}\quad.
\ee
These $K$-matrices satisfy the following versions of the reflection equations,
\be
\label{eq:reflect}
\pic{-55pt}{120pt}{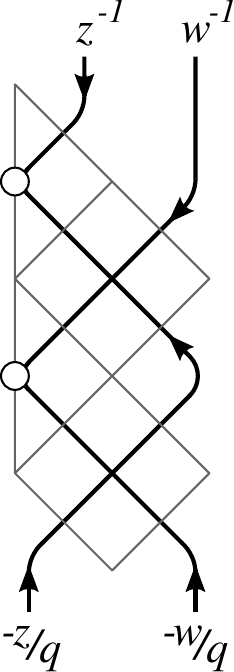}\quad=\quad\pic{-55pt}{120pt}{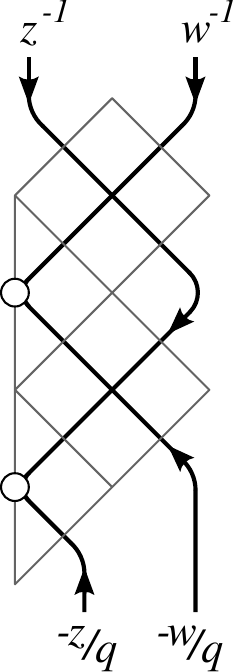}\quad,\qquad\qquad\qquad\pic{-55pt}{120pt}{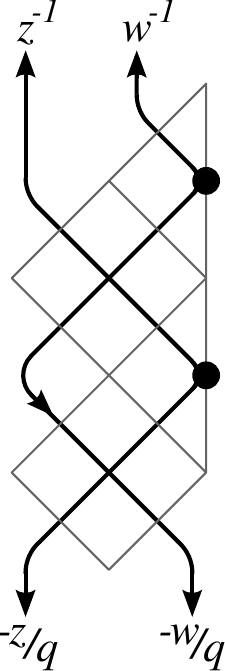}\quad=\quad\pic{-55pt}{120pt}{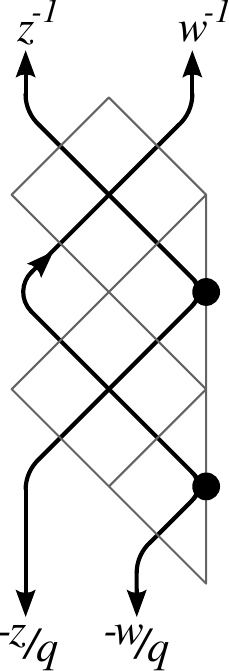}\quad.
\ee
These are easily proved using the crossing relation \eqref{eq:crossing}.

\subsection{Transfer matrix}
\label{sec:transfermatrix}

\begin{defn}
\label{def:transmat}
Using the pictorial versions of the $R$ and $K$-matrices in \defref{RK} we define Sklyanin's double row transfer matrix $T_L(w)=T_L(w;z_1,\ldots,z_L;\zeta_0,\zeta_L)$ pictorially in the following way (see \cite{Skl88,DF05,PearceRZ06}),
\[
T_L(w) = \raisebox{-40pt}{\includegraphics[height=80pt]{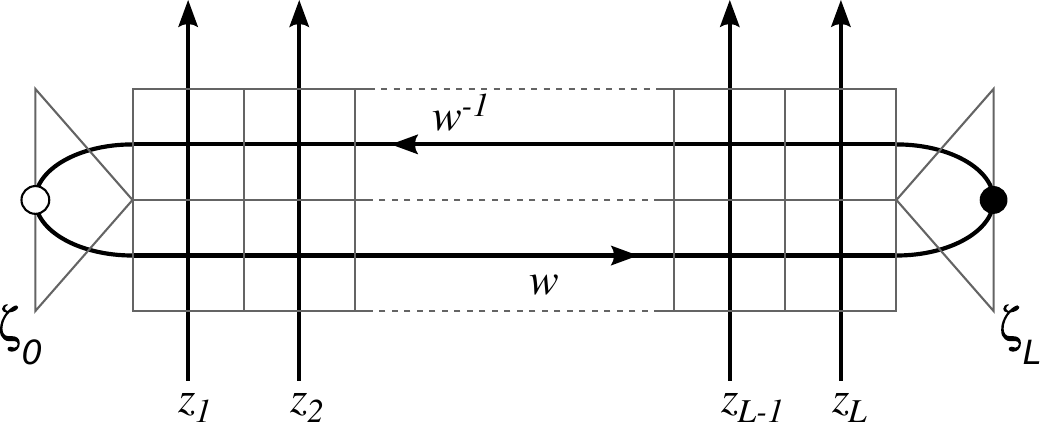}}\;.
\]
\end{defn}
This can also be written as
\[ T_L(w)=Tr_w\left(K_0(w^{-1})\ R(w,z_1)\ldots R(w,z_L)\ K_L(w)\ R(z_L,w^{-1})\ldots R(z_1,w^{-1})\right),\]
where, in terms of pictures, the trace means that we join the two ends of the line with the parameter $w$ attached to it, as shown in the above diagram.
\begin{prop}
All the possible values of $w$ give us a commuting family of transfer matrices (see for example \cite{Skl88}), i.e.,
\[
[T_L(v),T_L(w)]=0,
\]
and hence $T(w)$ defines an integrable lattice model.
\end{prop}
\begin{proof}
First, note that with the notation \eqref{eq:Knew} and the crossing relation, $T(w)T(v)$ may be depicted
\[\pic{0pt}{130pt}{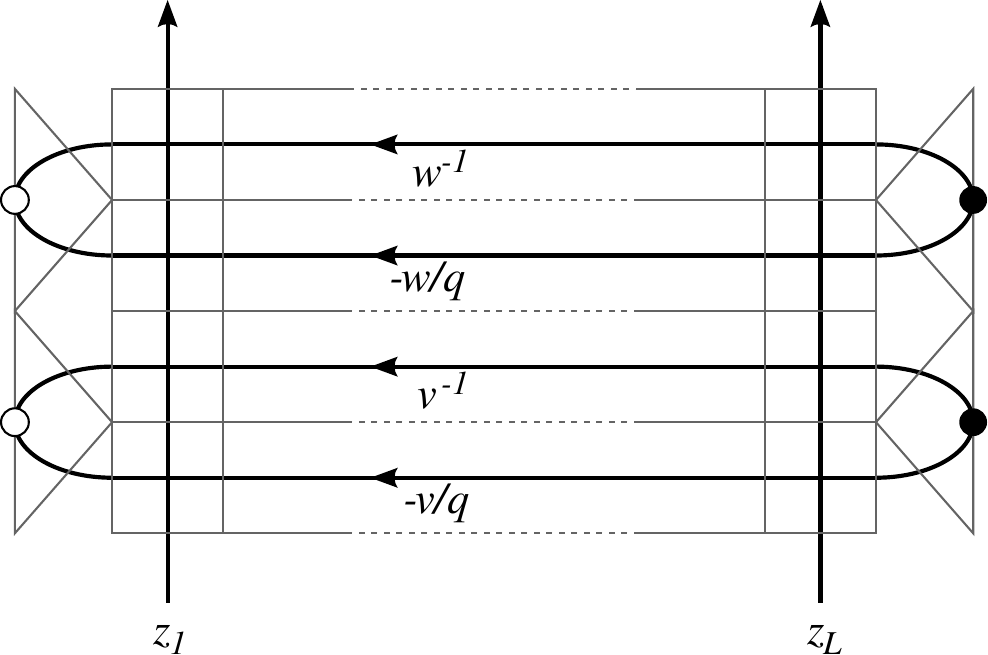}\quad\raisebox{60pt}{.}\]
Using the unitarity relation \eqref{eq:RId} twice, and repeated application of the Yang-Baxter equation \eqref{eq:YBE}, the above becomes
\[\pic{0pt}{170pt}{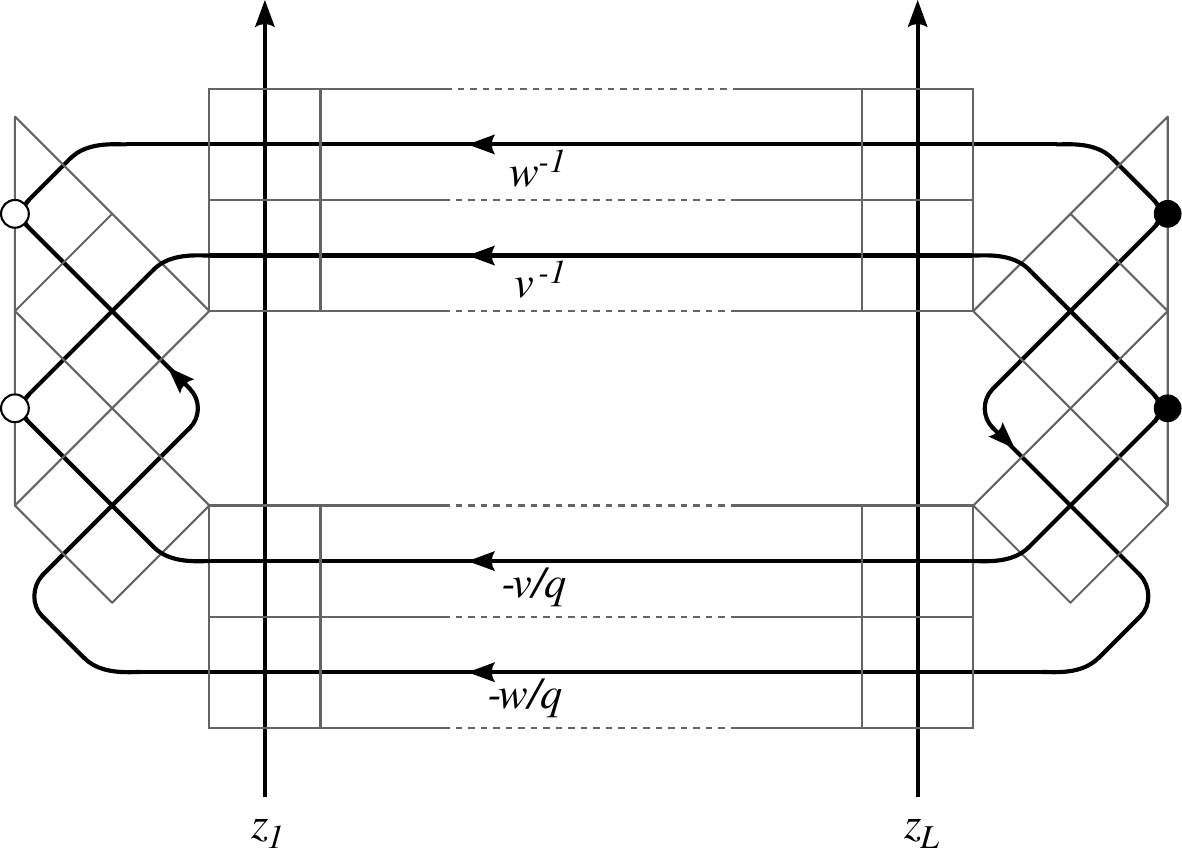}\quad\raisebox{80pt}{.}\]
Now, at both boundaries, the reflection equations \eqref{eq:reflect} can be applied, resulting in
\[\pic{0pt}{170pt}{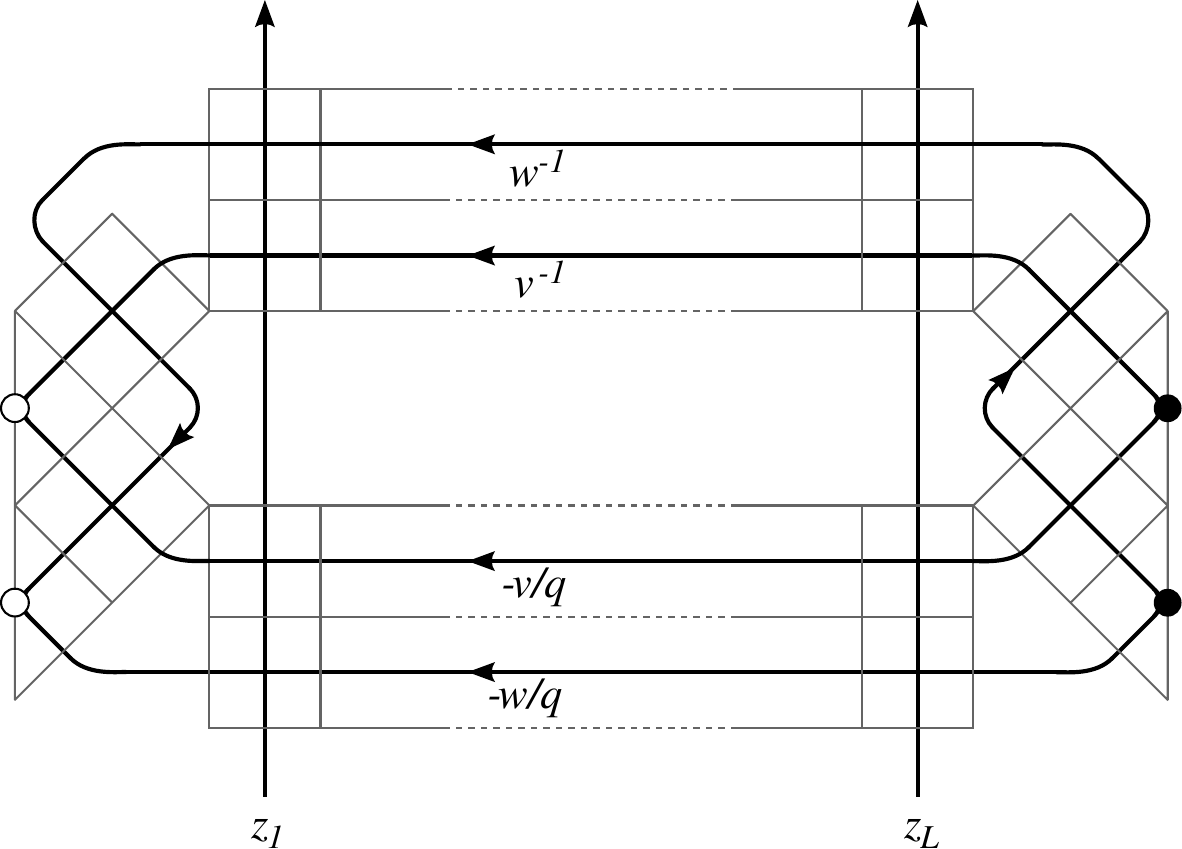}\quad\raisebox{80pt}{,}\]
and the process involving the Yang--Baxter equation and unitarity relations can be reversed, to finally produce
\[\pic{0pt}{130pt}{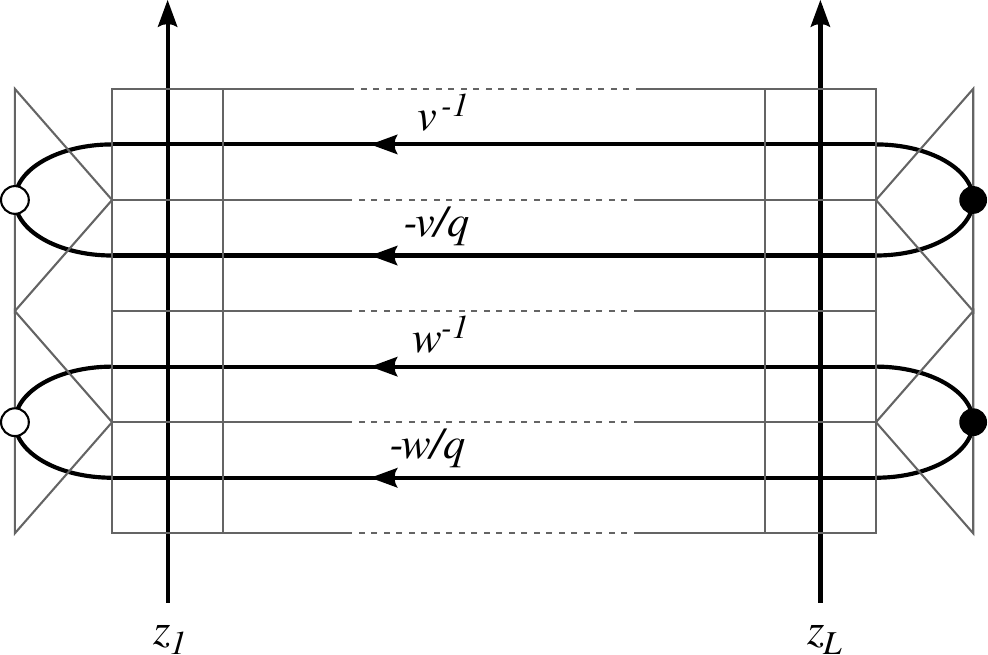}\quad\raisebox{60pt}{,}\]
which is the graphical depiction of $T(v)T(w)$.
\end{proof}
As a consequence of this commutativity, the eigenvectors of $T_L$ do not depend on the spectral parameter $w$, but only on $z_1,\ldots,z_L$,
\[T_L(w;z_1,\ldots,z_L)\ket{\Psi(z_1,\ldots,z_L)}\propto\ket{\Psi(z_1,\ldots,z_L)}.\]

Following \cite{DFZJ05,DF05}, we note that the Yang--Baxter and reflection equations \eqref{eq:RKcheckYbeReflect} also immediately imply the following interlacing conditions of the transfer matrix with $\check R_i$,  $\check K_0$ and  $\check K_L$.
\begin{prop}
\label{prop:interlace}
\begin{align}
\check R_i(z_i/z_{i+1})T_L(w;z_1,\ldots,z_L) &= T_L(w;z_1,\ldots,z_{i+1},z_i,\ldots,z_L)\check R_i(z_i/z_{i+1}),\nn\\
\label{eq:interlace}\check K_0(z_1^{-1},\zeta_0)T_L(w;z_1,\ldots,z_L) &= T_L(w;z_1^{-1},z_2,\ldots,z_L)\check K_0(z_1^{-1},\zeta_0),\\
\check K_L(z_L,\zeta_L)T_L(w;z_1,\ldots,z_L) &= T_L(w;z_1,\ldots,z_{L-1},z_L^{-1})\check K_L(z_L,\zeta_L).\nn
\end{align}
\end{prop}
Pictorially, the first relation is
\[ \pic{3pt}{90pt}{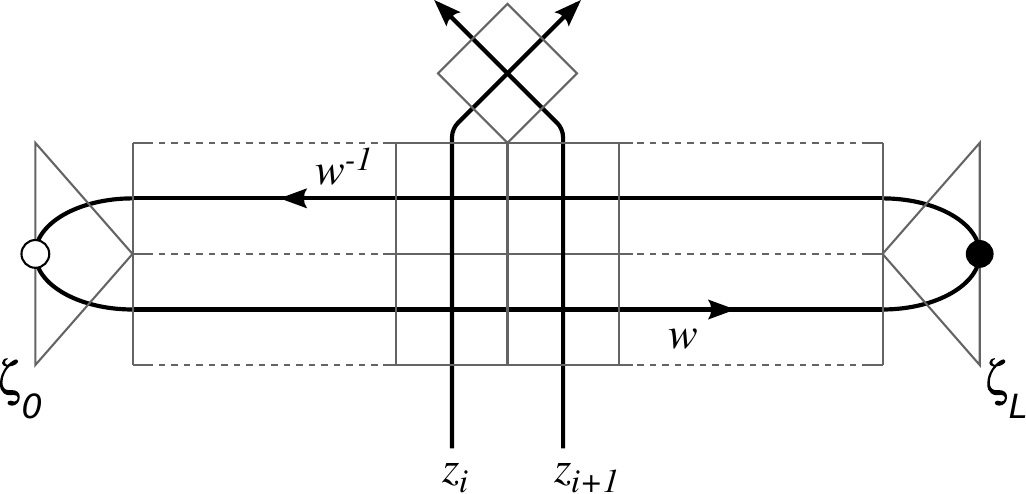}\raisebox{43pt}{\quad=\quad}\pic{-5pt}{90pt}{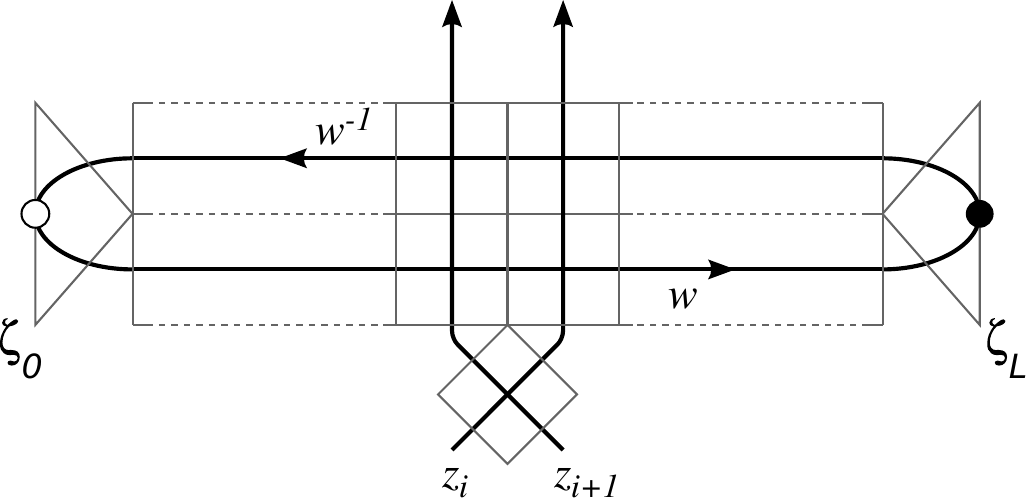}\raisebox{43pt}{\quad.} \]
The other relations have similar depictions.

\subsection{Hamiltonian}
\label{sec:Ham}
\begin{defn}
\label{def:Ham}
The Hamiltonian of the TL($n$) loop model with open boundaries is defined in terms of the logarithmic derivative of the transfer matrix with respect to $w$ at the point $z_1=\ldots=z_L=w=1$;
\begin{align*}
H&=\frac{[q]}{4}\left.\de{w}\right|_{w=1}\log{T(w;1,\ldots,1)}\\
&=\frac{[q]}{4}\left.\left(\de[\check K_0(w)]{w}+\de[\check K_L(w)]{w}+2\sum_{j=1}^{L-1}\de[\check R_j(w)]{w}\right)\right|_{w=1}+\text{ const.}
\end{align*}
\end{defn}
This can be calculated and expressed as the following operator,
\be
\label{eq:ham}
H = c_0\left(\frac{-\sin{\omega_0}}{\sin{(\omega_0+\gamma)}}-e_0\right) + c_L\left(\frac{-\sin{\omega_L}}{\sin{(\omega_L+\gamma)}}-e_L\right) + \sum_{j=1}^{L-1} (-2\cos{\gamma}-e_j)
+\text{ const.},
\ee
where $c_0,c_L$ are
\[
c_i=\frac{4\sin{\gamma}\sin{(\omega_i+\gamma)}}{k(1,\zeta_i,\omega_i)}.
\]

\section{Specialisation $n=1$}
\label{sec:spec}

In the next chapter we will introduce the $q$-deformed Knizhnik--Zamolodchikov equation, which depends on $q$ as well as an additional parameter $s$. At $s^4q^3=1$ the solution of this equation is polynomial \cite{KasaT07,KasaP07}, and in \secref{qKZ} we will use the interlacing conditions \eqref{eq:interlace} to show that at $q^3=s^4=1$ the ground state eigenvector of the TL($n$) loop model is a solution of the $q$KZ equation.

The specialisation $q^3=1$ is equivalent to setting $n=1$ or $\gamma=2\pi/3$, and this also corresponds to anisotropy of the XXZ spin chain, $\Delta:=\cos\gamma=-1/2$. In order to construct a special representation we will also set $\omega_0=\omega_L=-2\gamma$ and $b=1$. At this point we have several simplifications, the first of which is that $k$ as given in \eqref{eq:kdef} becomes
\be
\label{eq:k2}
k(z,\zeta,\omega)=[z/q\zeta][z\zeta/q]=:k(z,\zeta).
\ee

As stated in \secref{bax}, when $q=\e^{2\pi\ii/3}$ we take the normalisation factor $\kappa(z)$ to be identically equal to $1$. This choice for $\kappa$ causes $R$ to be invariant under negation of its arguments, so the crossing relation \eqref{eq:crossing} can be written without the negative sign,
\[ R(z,w)=R(qw,z).\]

When the above specialisations are taken, the constant in the expression for the Hamiltonian \eqref{eq:ham} disappears, and the rest simplifies to
\begin{align}
\label{eq:hamspec}H&=c_0(1-e_0)+c_L(1-e_L)+\sum_{j=1}^{L-1}(1-e_i),\\
\label{eq:ci}c_i&=\frac{3}{1+\zeta_i^2+\zeta_i^{-2}}.
\end{align}

For convenience, we keep the notation $q$, and make use of the fact that $q^3=1$. We also see that the relations of the 2BTL algebra can now be expressed as
\begin{align*}
 e_i^2&=e_i,\qquad \forall i,\\
 e_ie_{i\pm 1}e_i&=e_i,\qquad 1\leq i\leq L-1.
\end{align*}
This has a one-dimensional representation $\rho$ defined by
\be
\rho: e_i\mapsto 1,\qquad i=0,1,\ldots, L,
\label{eq:onedim}
\ee
which is a quotient of the link pattern representation, as it maps every link pattern to the identity (this is easily seen by viewing each link pattern as a word in the $e_i$, as described in \secref{linkp}). We choose the values $\omega_0=\omega_L=-2\gamma$ and $b=1$ in order to construct this representation. We note that $\rho(H)=0$, hence $0$ is an eigenvalue of $H$ in the link pattern representation. In fact, because the eigenvalues of $e_i$ are $0$ and $1$, for $c_0,c_L$ non-negative, $0$ is the lowest eigenvalue of $H$ and corresponds to the ground state of the O$(n=1)$ loop model.

Moreover, we can use a version of the Perron--Frobenius theorem to show that the ground state is unique. Consider the matrix $\mathcal H=\mathbb{I}L-H$, which has non-negative entries for $c_0,c_L$ non-negative, and which shares eigenvectors with $H$. The expansion of $\mathcal H^m$ includes every possible word in the generators $e_i$ of length $m$ or shorter, so for large enough $m$, acting with $\mathcal H^m$ on an arbitrary basis vector produces a vector with all entries non-zero. Thus $\mathcal H$ is irreducible, and the Perron--Frobenius theorem states that the eigenvector corresponding to eigenvalue $L$ (equal to the ground state of $H$) is unique. We will be interested in the ground state eigenvector $\ket{\Psi}$ as a function of the parameters $c_0$ and $c_L$,
\[
H\ket{\Psi(c_0,c_L)} =0.
\]

In the same way, the one-dimensional representation indicates that the ground state eigenvalue of the transfer matrix is equal to $1$, so the eigenvalue equation is
\be
\label{eq:evec1}T_L(w;z_1,\ldots,z_L)\ket{\Psi_L(z_1,\ldots,z_L)}=\ket{\Psi_L(z_1,\ldots,z_L)}.
\ee
In the homogeneous limit $z_i\rightarrow 1$, the transfer matrix $T_L$ becomes the probability matrix of the stochastic raise and peel model \cite{PearceRdGN02,Pyat04}, for which the steady state eigenvector \eqref{eq:evec1} is unique, again by the Perron--Frobenius theorem. The eigenvalue spectrum is continuous, so it can be argued that there exists an open set around $z_1=\ldots=z_L=1$ for which the eigenvector remains unique. We will thus assume that the eigenvector remains unique for generic values of $z_1,\ldots,z_L$.

The $q$KZ equation will allow us to obtain an explicit characterisation of $\ket{\Psi_L}$ for finite $L$. We will in particular be able to derive a closed form expression for the normalisation $Z_L$, which is the sum of all the components of $\ket{\Psi_L}$, as well as for a boundary-to-boundary correlation function defined in \chapref{Flow}. In order to do so we need a recursion relation for $\ket{\Psi_L}$, which we will discuss in \secref{estaterecur}.

It is worth noting that in the limit $e_L\mapsto 1$ the boundary Baxterised element $\check K_L$ maps to the identity, and this is the limit in which the two-boundary model maps to the one-boundary case. Similarly, in the limit $e_0,e_L\mapsto 1$, we obtain the trivial boundary case.

\newpage
\chapter{The ground state of the TL(1) loop model}
\label{chap:2BSoln}

The aim of this chapter is to calculate the elements of the ground state eigenvector of the TL($1$) loop model with two open boundaries, corresponding to the 2BTL algebra as discussed in the previous chapter. As will be shown, at $q^3=1$, the ground state eigenvalue equation \eqref{eq:evec1} for the inhomogeneous transfer matrix of the TL($1$) model is equivalent to a particular instance of the $q$-deformed Knizhnik--Zamolodchikov ($q$KZ) equation.

The $q$KZ equation is a set of finite difference equations, introduced by Frenkel and Reshetikhin \cite{FrenR92} in the context of the representation theory of quantum affine algebras. This equation depends on a number of parameters, traditionally $t$, $q$ (this is not the same as our $q$), and if applicable the boundary parameters $a,b,c$ and $d$. It also has two extra parameters, $k$ and $r$, called respectively the `rank' and `level' of the equation. It has been found that when these parameters satisfy
\be
\label{eq:restriction}
t^{k+1}q^{r-1}abcd=1,
\ee
the $q$KZ equation has polynomial solutions \cite{ShiU07,KasaP07,KasaT07} (see also \cite{FeiJMM02,FeiJMM03}).

Interesting recent developments \cite{KasaT07} relate polynomial solutions of the $q$KZ equation associated with $U_q(sl_k)$ to the polynomial representation of the double affine Hecke algebra \cite{Noumi95,Sahi99,Kasa08}. These solutions can be expressed in terms of Koornwinder or Macdonald polynomials with specialised parameters \cite{Baratta10,FeiJMM03}.

As mentioned, the ground state of the TL($1$) model is a solution of the $q$KZ equation with very specialised parameters. In particular, the parameters $a,b,c$ and $d$ are set to $1$, the parameter $t$ is equal to $q^2$ in our notation, and the parameter $q$ is related to a new parameter $s$. We take both the rank $k$ and the level $r$ to be $2$, and the restriction on the parameters \eqref{eq:restriction} becomes $q^6s^4=1$. Since the ground state eigenvalue equation is equivalent to the $q$KZ equation at $q^3=1$, this condition gives us the further restriction $s^4=1$. As we will see, the version of the $q$KZ equation used here has Laurent polynomials as solutions.

The ground states for TL($1$) models with a variety of boundary conditions have been studied in the past (see for example \cite{DF05,ZJ07,dGP07,MitraN04}), and each one is related to a version of the $q$KZ equation. In the models with zero or one open boundaries, there exists a highest weight vector, which is an element of the link pattern basis. The component of the ground state eigenvector corresponding to this basis element plays a special role in the solution, because it is possible to specify this component exactly, and from there calculate every other element of the vector. In the case of periodic boundaries, due to the cyclic symmetry there is not a unique highest weight vector, but an equivalence class of them. Again the corresponding components of the eigenvector can be fixed, and the $q$KZ equation provides the means to then calculate the other components.

However, in the case of two open boundaries, there is no such highest weight vector. This fact causes the calculation of the ground state eigenvector in the case of two open boundaries to be more challenging, and until recently little progress had been made.

In this chapter we deal with this case, and generalise the results for reflecting \cite{DF05} and mixed \cite{ZJ07} boundary conditions. It is very hard to find a general expression for all the components of the two boundary eigenvector, but a certain subset of them can be found, and for the others certain restrictions can be made. We also calculate the normalisation of the eigenvector, which is the sum of all its components. This can be found explicitly despite the lack of an exact formula for all the elements of the eigenvector. The main results in this chapter were also discovered independently and through slightly different means by Cantini \cite{Cantini09}, in the same year.

For small system sizes $L\leq 2$, we solve the $q$KZ equation explicitly for all components, up to an overall factor that we take to be $1$. For arbitrary size, we find an explicit expression for two special components of the eigenvector, using recursive properties of the TL($1$) transfer matrix. We then use the same recursive properties to find an expression for the overall ground state normalisation.

With the specialisation $s^4=1$, we find that the dependence on $s$ factors out of the final result. However we keep the notation for $s$ in the expressions for the components with a view to generalisation.

\section{The $q$-deformed Knizhnik--Zamolodchikov equation}
\label{sec:qKZ}
As stated above, the ground state eigenvector of the TL($1$) transfer matrix is a solution of the $q$KZ equation with a certain specialisation. This connection will provide the foundation for an explicit analysis of the ground state eigenvector for finite system size $L$. We will first describe the $q$KZ equation for open boundaries, and then prove the equivalence with the transfer matrix eigenvalue equation \eqref{eq:evec1}.

We consider a linear combination $\ket\Xi$ of states $\ket\alpha$ labelled by link patterns:
\[
\ket{\Xi(z_1,\ldots,z_{L})} = \sum_{\alpha}
\xi_\alpha(z_1,\ldots,z_{L}) \ket{\alpha}.
\]
Here the sum runs over the set of link patterns of size $L$, and the coefficient functions $\xi_\alpha$ are Laurent polynomials in the variables $z_1,\ldots,z_L$ with coefficients that are functions of the boundary parameters $\zeta_0$ and $\zeta_L$ as well as $q$ and $s$.
\begin{defn}
\label{def:qKZ}
The $q$-deformed Knizhnik--Zamolodchikov equation \cite{Smirn92,FrenR92} is a system of finite difference equations on the vector $\ket\Xi$. For open boundary conditions they can be written as \cite{DF05,ZJ07,Cantini09,Cher05},\symbolfootnote{We write the equations in a form used by Smirnov \cite{Smirn92} for the type $A$ case (see \apref{hecke} of this thesis for more details on the type classifications).}
\begin{align}
\check R_i(z_i/z_{i+1}) \ket\Xi &= \pi_i \ket\Xi,\qquad 1\leq i \leq L-1,\nn\\
\check K_0(1/z_1,\zeta_0) \ket\Xi  &= \pi_0 \ket\Xi,
\label{eq:qKZTL_TypeC2}\\
\check K_L(s z_L,s\zeta_L) \ket\Xi &= \pi_L\ket\Xi.\nn
\end{align}
\end{defn}
This definition uses the Baxterised elements $\check R_i$, $\check K_0$ and $\check K_L$ of the two-boundary Temperley--Lieb algebra, defined in \eqref{eq:R_z}. The operators $\check R_i(z_i/z_{i+1})$, $\check K_0(1/z_1,\zeta_0)$ and $\check K_L(sz_L,s\zeta_L)$  act on link patterns $\ket \alpha$, whereas the operators $\pi_i$ ($i=0,\dots , L$) act on the coefficient functions $\xi$ only;
\begin{align*}
\pi_i\ \xi(\ldots,z_i,z_{i+1},\ldots) &= \xi(\ldots,z_{i+1},z_{i},\ldots),\qquad 1\leq i\leq L-1,\\
\pi_0\ \xi(z_1,\ldots)&= \xi(1/z_1,\ldots),\\
\pi_L\ \xi(\ldots,z_L)&= \xi(\ldots,1/s^2z_L).
\end{align*}
For later convenience, we note that the $q$KZ equations can be rewritten
\be
\label{eq:qKZea}
e_i \ket\Xi = -a_i \ket\Xi,\qquad 0\leq i\leq L,
\ee
where $e_i$ are the generators of the 2BTL from \defref{2BTL}, and
\begin{align}
a_i &= (\pi_i+1) \frac{[z_i/qz_{i+1}]}{[z_i/z_{i+1}]},\qquad 1\leq i\leq L-1,\nn\\
\label{eq:adef}a_0 &= (\pi_0+1) \frac{k(1/z_1,\zeta_0)}{[q][z_1^2]},\\
a_L &= -(\pi_L+1) \frac{k(sz_L,s\zeta_L)}{[q][s^2z_L^2]},\nn
\end{align}
where $k$ was defined in \eqref{eq:k2}. The operators $a_i$ ($i=0,\ldots,L$) satisfy the relations of the Hecke algebra of affine type $C$ \cite{dGP07}, as well as those of the Hecke algebra of type $BC$. The Hecke algebra is described in more detail in \apref{hecke}.

For later convenience, we also define operators $s_i$, such that
\be
\label{eq:sdef}
\begin{split}
s_i &= (q+q^{-1})-a_i,\qquad 1\leq i\leq L-1,\\
s_k&=\frac{\sin{\omega_k}}{\sin{(\omega_k+\gamma)}}-a_k,\qquad k=0,L.
\end{split}
\ee
With our chosen specifications for $q$, $\omega_0$, and $\omega_L$ (see \secref{spec}), these definitions can be condensed to
\be
\label{eq:s}
s_i=-1-a_i, \quad \forall i.
\ee
The $s_i$ can be explicitly written as
\begin{align*}
s_i &= \frac{[qz_i/z_{i+1}]}{[z_i/z_{i+1}]}(1-\pi_i),\qquad 1\leq i\leq L-1,\\
s_0 &= \frac{k(z_1,\zeta_0)}{[q][z_1^2]}(\pi_0-1),\\
s_L &= \frac{k(1/sz_L,s\zeta_L)}{[q][s^2z_L^2]}(1-\pi_L).
\end{align*}

\subsection{Equivalence with the transfer matrix eigenvalue equation}
\label{sec:equiv}
At the special values of $q$ and $s$ previously mentioned, the elements of the eigenvector are polynomials in $z_1^\pm,\ldots,z_n^\pm$, and we use this fact to construct a definition for $\ket{\Psi_L}$ that is appropriately normalised.
\begin{defn}
\label{def:psi}
With the specialisations listed in \secref{spec}, the ground state eigenvector $\ket{\Psi_L(z_1,\ldots,z_L)}$ of the transfer matrix of the TL($1$) model of width $L$ has eigenvalue $1$, and is written in the link pattern basis as
\[ \ket{\Psi_L(z_1,\ldots,z_L)}=\sum_{\alpha\in \rm{LP}_L}\psi_\alpha(z_1,\ldots,z_L)\ket\alpha, \]
where the $\psi_\alpha$ are coprime polynomials.
\end{defn}
The basis orthogonal to the downward link patterns $\ket\alpha$ is given by the row vectors $\bra\beta$, with the usual inner product $\bra{\cdotp}\cdotp\rangle$ defined by
\[ \bra{\beta}\alpha\rangle=\delta_{\alpha,\beta}. \]
Written in this basis, the left eigenvector $\bra\Psi$ of the transfer matrix corresponding to eigenvalue $1$ is a row vector with every element equal to $1$. This is because the transfer matrix is a stochastic matrix, whose columns sum to $1$.
\begin{defn}
\label{def:Z}
The normalisation of the eigenvector is defined as
\[ Z_L=\bra{\Psi}\Psi\rangle, \]
or more explicitly,
\begin{align}
Z_L(z_1,\ldots,z_L)&=\sum_{\beta,\alpha}\psi_\alpha(z_1,\ldots,z_L)\ \bra{\beta}\alpha\rangle\nn\\
\label{eq:Zdef}&=\sum_{\alpha}\psi_\alpha(z_1,\ldots,z_L).
\end{align}
\end{defn}
For the next theorem we will also need the following lemma.
\begin{lm}
\label{lm:Hz}
In the almost homogeneous limit $z_1\rightarrow q$, $z_2=\ldots=z_L=\zeta_0=\zeta_L=1$, the Hamiltonian \eqref{eq:hamspec} can be normalised and expressed as
\[\widetilde H =e_1e_0-1.\]
The ground state eigenvector of this Hamiltonian, corresponding to eigenvalue $0$, is non-negative.
\end{lm}
\begin{proof}
At the point $z_2=\ldots=z_L=\zeta_0=\zeta_L=1$, we can calculate the Hamiltonian as was done in \secref{Ham}, obtaining
\[H=\left(\frac{3}{1+z_1^2+z_1^{-2}}-e_0\right)+\sum_{i=1}^{L-1}(1-e_i)+(1-e_L)+\frac{[z_1]}{[q/z_1]}e_1e_0+\frac{[1/z_1]}{[qz_1]}e_0e_1.\]
Normalising $H$ by $\frac{[q/z_1]}{[z_1]}$, and taking the limit as $z_1\rightarrow q$, this becomes
\begin{align*}
\widetilde H:=\lim_{z_1\rightarrow q}\frac{[q/z_1]}{[z_1]}\ H&=e_1e_0+\lim_{z_1\rightarrow q}\frac{3\ [q/z_1]}{(1+z_1^2+z_1^{-2})\ [z_1]}\\
&=e_1e_0-1.
\end{align*}

In matrix notation the non-diagonal entries of this Hamiltonian are non-negative, so according to the Perron--Frobenius theorem the ground state eigenvector must also be non-negative.
\end{proof}
\begin{thm}
\label{thm:PsiQKZ}
The specialised ground state eigenvector $\ket\Psi$ also satisfies the $q$KZ equation, \defref{qKZ}.
\end{thm}
The proof of this given in \cite{dGPS09} was incomplete. We give here the full proof.
\begin{proof}
We first act on both sides of the eigenvector equation \eqref{eq:evec1} with the Baxterised element $\check R(z_i/z_{i+1})$, and using the interlacing condition in \eqref{eq:interlace},
\begin{align*}
\check R_i(z_i/z_{i+1})\ket{\Psi_L(z_i,z_{i+1})}&=\check R_i(z_i/z_{i+1})T_L(w;z_i,z_{i+1})\ket{\Psi_L(z_i,z_{i+1})}\\
&=T_L(w;z_{i+1},z_i)\ \check R_i(z_i/z_{i+1})\ket{\Psi_L(z_i,z_{i+1})}.
\end{align*}
Since the eigenvector in \eqref{eq:evec1} is unique, this implies that
\be
\label{eq:prop1}
\check R_i(z_i/z_{i+1})\ket{\Psi_L(z_i,z_{i+1})}=\beta_i(z_1,\ldots,z_L) \ket{\Psi_L(z_{i+1},z_i)},
\ee
where $\beta_i$ is some rational function. By definition, none of the elements of $\ket\Psi$ have a denominator, so the denominator of $\beta_i(\ldots,z_i,z_{i+1},\ldots)$ must be the same as the denominator of $\check R_i(z_i/z_{i+1})$, i.e., $[qz_i/z_{i+1}]$. We can rearrange the above, using the unitarity relation \eqref{eq:RKcheckId} of $\check R$, and swap the parameters $z_i\leftrightarrow z_{i+1}$ to get
\be
\label{eq:Psibeta}
\check R_i(z_i/z_{i+1})\ket{\Psi_L(z_i,z_{i+1})}=\beta_i(\ldots,z_{i+1},z_i,\ldots,)^{-1} \ket{\Psi_L(z_{i+1},z_i)},
\ee
which leads directly to
\[
\beta_i(\ldots,z_i,z_{i+1},\ldots)=\beta_i(\ldots,z_{i+1},z_i,\ldots,)^{-1}.
\]
Since we already know the denominator of $\beta_i$, the above property gives us four choices for the numerator, namely
\be
\label{eq:beta}
\beta_i(z_i,z_{i+1})=\frac{\pm[qz_{i+1}/z_i]}{[qz_i/z_{i+1}]}\quad\text{or}\quad \pm 1.
\ee
In both cases the sign is fixed to $+1$ by setting $z_{i+1}=z_i$ in \eqref{eq:Psibeta}.

Assume $\beta_i$ to be the first choice in \eqref{eq:beta}. In the quotient $\rho:e_i\mapsto 1$ as defined in \eqref{eq:onedim}, every link pattern is projected onto the identity, and \eqref{eq:prop1} becomes
\[Z_L(z_i,z_{i+1})=\frac{[qz_{i+1}/z_i]}{[qz_i/z_{i+1}]}Z_L(z_{i+1},z_i).\]
Since $Z_L$ is a polynomial in the $z$'s, it must be of the form
\[Z_L=[qz_{i+1}/z_i]S_i(z_1,\ldots,z_L),\]
where $S_i$ is a polynomial symmetric in $z_i,z_{i+1}$. By considering the above argument for all other values of $i$ we find that $Z_L$ must be of the form
\[Z_L=\prod_{1\leq i<j\leq L}[qz_j/z_i]S(z_1,\ldots,z_L),\]
where $S$ here is a polynomial symmetric in all its arguments. This means that the eigenvector normalisation vanishes when $z_i=qz_j$, for any $j>i$. However, the specialisation in \lmref{Hz} has $z_1=qz_j$ for all $j>1$, and it is shown that $Z_L$ cannot be $0$ at that point. Our assumption must therefore be false and we must take $\beta_i=1$.

Thus, we have shown that
\[\check R_i(z_i/z_{i+1})\ket{\Psi_L(z_1,\ldots,z_L)}=\pi_i\ket{\Psi_L(z_1,\ldots,z_L)},\quad 1\leq i\leq L-1.\]
Similarly, we can show that
\begin{align*}
\check K_0(1/z_1,\zeta_0)\ket{\Psi_L(z_1,\ldots,z_L)} &= \pi_0\ket{\Psi_L(z_1,\ldots,z_L)},\\
\check K_L(sz_L,s\zeta_L)\ket{\Psi_L(z_1,\ldots,z_L)} &= \pi_L\ket{\Psi_L(z_1,\ldots,z_L)},
\end{align*}
where the proof of the last equation makes use of the fact that when $s^4=1$, $\check K_L(sa,sb)=\check K_L(a,b)$ and $R(s^2z,w)=R(z,w)$.
\end{proof}

\section{Recursions}
\label{sec:recur}
We will use the $q$KZ equation \eqref{eq:qKZTL_TypeC2} to calculate the elements of the ground state eigenvector of the TL(1) loop model for size $L$. However this equation does not contain enough information to fix the elements, so we will also use a recursion that is an inherent property of the loop model.

We begin by defining some maps between spaces of link patterns of different sizes. These maps will be very useful in constructing recursions satisfied by the transfer matrix and its eigenvector.
\begin{defn}
For $1\leq i\leq L+1$, let $\varphi_i$ be the map that takes a link pattern of size $L$, sends site $j$ to $j+2$ for $j\geq i$, and then inserts a link from site $i$ to $i+1$, thus creating a link pattern of size $L+2$. For example,
\[
\varphi_6:\quad \pic{-10pt}{30pt}{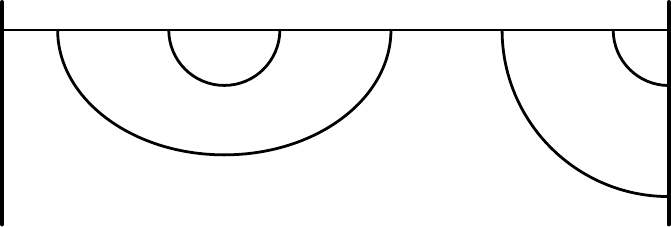}\quad \mapsto\quad \pic{-10pt}{30pt}{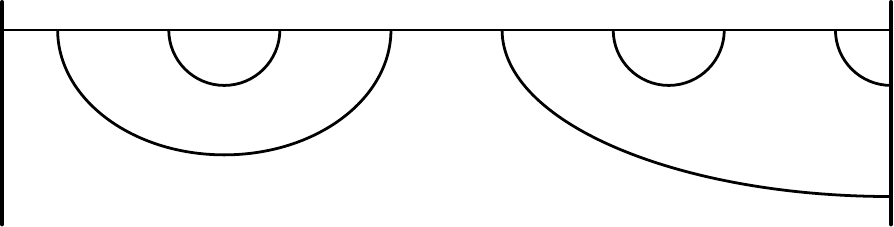}\quad.
\]

Let $\widetilde\varphi_0$ be the map that takes a link pattern of size $L$, sends site $j$ to site $j+1$ for all $j$, and inserts a `$)$' on the first site.

Let $\widetilde\varphi_{L+1}$ be the map that takes a link pattern of size $L$ and inserts a `$($' after the last site.
\end{defn}
We will also define similar maps on vectors in the link pattern space defined in \secref{linkp}. The action on the basis is a straightforward extension of the action on the link patterns, so we use the same notation for both maps.
\begin{defn}
\label{def:varphi}
Let $\varphi_i:$ LP$_L\rightarrow$ LP$_{L+2}$ be the map defined by
\[ \varphi_i\sum_{\alpha}c_{\alpha}\ket{\alpha}=\sum_{\alpha}c_{\alpha}\ket{\varphi_i\alpha}. \]
Similarly let $\widetilde\varphi_0:$ LP$_L\rightarrow$ LP$_{L+1}$ and $\widetilde\varphi_{L+1}:$ LP$_L\rightarrow$ LP$_{L+1}$ be defined by
\begin{align*}
\widetilde\varphi_0\sum_{\beta}c_{\beta}\ket{\beta}&=\sum_{\beta}c_{\beta}\ket{\widetilde\varphi_0\beta},\\
\widetilde\varphi_{L+1}\sum_{\beta}c_{\beta}\ket{\beta}&=\sum_{\beta}c_{\beta}\ket{\widetilde\varphi_{L+1}\beta}.
\end{align*}
\end{defn}
It is worth noting that the vector resulting from the action by $\varphi_i$ is of length $2^{L+2}$, but it has only $2^L$ non-zero entries, which are indexed by link patterns with a small link from $i$ to $i+1$. Similar statements can be made for $\widetilde\varphi_0$ and $\widetilde\varphi_L$.

\subsection{Transfer matrix recursion}
\label{sec:transmatrec}
\begin{prop}
\label{prop:Trecur}
The transfer matrix satisfies the following identity for general $q$:
\[
T_L(w;z_1,\ldots,z_{i+1}=qz_i,\ldots,z_L)\circ\varphi_i=\frac{[q/z_iw][q^2z_i/w]}{[q^2z_iw][qw/z_i]}\ \varphi_i\circ T_{L-2}(w;\ldots,z_{i-1},z_{i+2},\ldots).
\]
\end{prop}
The proof of this proposition is in \apref{bulkrecur}. A similar relation was proved in \cite{DF05} and for the case of periodic boundary conditions in \cite{DFZJ05}.
\begin{prop}
\label{prop:Tboundrecur}
Likewise, at the boundaries, and for $q=\e^{2\pi\ii/3}$, the transfer matrix satisfies
\begin{align}
\label{eq:0boundrecur}T_L(w;z_1=q\zeta_0,\ldots,z_L;\zeta_0,\zeta_L) \circ \widetilde\varphi_0 &= \widetilde\varphi_0 \circ T_{L-1}(w;z_2,\ldots,z_L;q\zeta_0,\zeta_L),\\
\label{eq:Lboundrecur}T_L(w;z_1,\ldots,z_L=\zeta_L/q;\zeta_0,\zeta_L) \circ \widetilde\varphi_L &= \widetilde\varphi_L \circ T_{L-1}(w;z_1,\ldots,z_{L-1};\zeta_0,\zeta_L/q).
\end{align}
\end{prop}
The first of these is proved in \apref{boundrecur}.

\subsection{Recursion of the eigenvector}
\label{sec:estaterecur}

In order to find a recursive definition for all components of $\ket{\Psi_L}$, we must refer to the recursive property of the transfer matrix described in \propref{Trecur}. We will suppress the arguments $z_1,\ldots,z_L$ of $T$ and $\ket{\Psi_L}$ except where detail is needed. The notation $\hat{z}_j$ will mean that $z_j$ is missing from the list $z_1,\ldots,z_L$. When we specify $q$ to be a third root of unity, the proportionality factor in \propref{Trecur} becomes $1$, so
\be
\label{eq:Trecur}
T_L(w;z_{i+1}=qz_i)\circ\varphi_i=\varphi_i\circ T_{L-2}(w;\hat z_i,\hat z_{i+1}).
\ee
Acting with both sides of this equation on the eigenvector $\ket{\Psi_{L-2}(\hat z_i,\hat z_{i+1})}$,
\[
T_L(w;z_{i+1}=qz_i)\ \Bigl(\varphi_i \ket{\Psi_{L-2}(\hat z_i,\hat z_{i+1})}\Bigr) = \varphi_i \ket{\Psi_{L-2}(\hat z_i,\hat z_{i+1})}\ ,
\]
which, by uniqueness of the eigenvector $\ket{\Psi_L}$, implies
\be
\label{eq:psiprop}\ket{\Psi_L(z_{i+1}=qz_i)}=p_i(z_i;z_1,\ldots,\hat z_i,\hat z_{i+1},\ldots,z_L)\ \varphi_i \ket{\Psi_{L-2}(\hat z_i,\hat z_{i+1})},
\ee
where $p_i$ is a proportionality factor. This proportionality implies that any component corresponding to a link pattern without a small link connecting $i$ and $i+1$\symbolfootnote{We use the term `small' or `little' link to mean a link connecting neighbouring sites.} vanishes when $z_{i+1}=qz_i$. In \secref{genLsoln} this property of the eigenvector will be derived in another way. Relation \eqref{eq:psiprop} was already proved for subcases of the most general open boundary conditions in \cite{DF05,ZJ07}, and for periodic boundary conditions in \cite{DFZJ05}.

Likewise, from the boundary recursions \eqref{eq:0boundrecur} and \eqref{eq:Lboundrecur} of the transfer matrix we deduce that
\begin{align*}
\ket{\Psi_L(z_1=q\zeta_0;\zeta_0)} &=r_0(z_2,\ldots,z_L;\zeta_0)\ \widetilde\varphi_0\ket{\Psi_{L-1}(\hat{z}_1;q\zeta_0)},\\[1mm]
\ket{\Psi_L(z_L=\zeta_L/q;\zeta_L)} &=r_L(z_1,\ldots,z_{L-1};\zeta_L)\ \widetilde\varphi_L\ket{\Psi_{L-1}(\hat{z}_L;\zeta_L/q)},
\end{align*}
where $r_0$ and $r_L$ are proportionality factors analogous to $p_i$.

The above recursions for the eigenvector imply the following recursions for the components of the eigenvector.
\begin{lm}
\label{lm:comprecur}
\begin{align*}
\psi^{L}_{(\varphi_i\circ\alpha)}(z_{i+1}=qz_i)&=p_i(z_i;z_1,\ldots,\hat z_i,\hat z_{i+1},\ldots,z_L)\ \psi^{L-2}_{\alpha}(\hat z_i,\hat z_{i+1}),\\
\psi^L_{(\widetilde\varphi_0\circ\beta)}(z_1=q\zeta_0)&=r_0(z_2,\ldots,z_L;\zeta_0)\ \psi^{L-1}_\beta(\hat z_1;q\zeta_0),\\
\psi^L_{(\widetilde\varphi_L\circ\beta)}(z_L=q/\zeta_L)&=r_L(z_1,\ldots,z_{L-1};\zeta_L)\ \psi^{L-1}_\beta(\hat z_L;\zeta_L/q),
\end{align*}
where $\alpha$ is any link pattern of length $L-2$ and $\beta$ is any link pattern of length $L-1$.
\end{lm}

Clearly the normalisation $Z$ given in \eqref{eq:Zdef} also satisfies these recursions. In \apref{propsym}, we show that $Z$ is symmetric in its arguments, and use this to prove that $p_i$ is symmetric in all its variables except $z_i$, and that $r_0$ and $r_L$ are symmetric in all the $z_i$. It is also shown that the function $p_i$ takes the same form for each $i$. We henceforth drop the index $i$ from $p$.

\section{Solutions to the $q$KZ equation for small $L$}
\label{sec:smallsize}
We now want to solve the $q$KZ equation to find the components of the ground state eigenvector. For this purpose we use the form of the equation given in \eqref{eq:qKZea}. We will be looking for the lowest degree solution for which all the $\psi_\alpha$ are not identically zero. Any higher degree solution must be a scalar multiple of the lowest degree solution, because of the uniqueness of the TL($1$) ground state eigenvector.

For $L=0$, the solution is trivial, but it will be needed for calculations of the recursions as described in \secref{estaterecur}. Obviously the eigenvector is of length one, i.e., a scalar. The definition of the eigenvector has the components being coprime, and so by analogy we take this scalar to be $1$.


\subsection{Example: $L=1$}
\label{sec:L=1}
When $L=1$, there are only two link patterns, denoted by `$)$' and `$($'. The ground state eigenvector of the TL($1$) loop model in the basis of link patterns is then
\[ \ket\Psi=\psi_)\ \big|)\big\rangle+\psi_(\ \big|(\big\rangle. \]
The $e_i$ act only on link patterns, leaving the polynomials unchanged. In particular, $e_0$ acts on both link patterns to produce `$)$', and $e_1$ acts on both to produce `$($'. Remembering that the $a_i$ act only on polynomials, the $q$KZ equations can be rewritten as
\begin{align*}
\left(\psi_(+\psi_)\right)\big|)\big\rangle &= -a_0\psi_(\ \big|(\big\rangle-a_0\psi_)\ \big|)\big\rangle,\\
\left(\psi_(+\psi_)\right)\big|(\big\rangle &= -a_1\psi_(\ \big|(\big\rangle-a_1\psi_)\ \big|)\big\rangle.
\end{align*}
Recalling that $s_i=-1-a_i$, $\forall i$, we can match up the coefficients of each link pattern to get a system of four equations,
\begin{align*}
a_0\psi_(=0, &\quad s_0\psi_)=\psi_(,\qquad a_1\psi_)=0,\quad s_1\psi_(=\psi_).
\end{align*}
These equations, with the definitions of $a_i$ in \eqref{eq:adef}, give us all the information we need to find the minimal degree solution. From the first equation, we know that
\[k(1/z_1,\zeta_0)\ \psi_(\]
is invariant under $z_1\rightarrow 1/z_1$, and we can deduce from this invariance that $\psi_($ must be of the form
\[ \psi_(=k(z_1,\zeta_0)f_1(z_1), \]
where $f_1$ is a polynomial invariant under $z_1\rightarrow 1/z_1$. Similarly, from the third equation we find that $\psi_)$ must be given by
\[ \psi_)=k(1/sz_1,s\zeta_L)\tilde f_1(z_1), \]
where $\tilde f_1$ is invariant under $sz_1\rightarrow 1/sz_1$.

We find that the remaining two equations are satisfied if $f_1$ and $\tilde f_1$ are both constants with $\tilde f_1=s^2f_1$, and this gives us the lowest degree solution. We want the components to be coprime, so we choose $f_1$ to be $-1$,\symbolfootnote{We choose $-1$ instead of $1$ with the benefit of hindsight --- at this stage it makes no difference to the solution, but this choice results in a neater expression for the final solution for general $L$.} and the solution for $L=1$ is
\be \begin{split}
\label{eq:L1soln}
\psi_(&=-k(z_1,\zeta_0),\\
\psi_)&=-s^2k(1/sz_1,s\zeta_L).
\end{split} \ee

\subsubsection{Recursion to $L=0$}
According to the recursions listed in \lmref{comprecur}, specialising the above solutions at $z_1=\zeta_L/q$ and $z_1=q\zeta_0$ respectively will produce the boundary proportionality factors $r_L$ and $r_0$ for $L=1$ (recall that the solution for $L=0$ is simply $1$). This gives
\be
\label{eq:rrecur1}
\begin{split}
r_L(\zeta_L)&=-s^2k(1/s\zeta_L,s\zeta_0),\\
r_0(\zeta_0)&=-k(\zeta_0,\zeta_L).
\end{split} \ee


\subsection{Example: $L=2$}
\label{sec:L=2}
For $L=2$, the eigenvector is of length four, and the link pattern basis consists of the elements $\big\{\ \big|((\big\rangle,\ \big|()\big\rangle,\ \big|)(\big\rangle,\ \big|))\big\rangle\ \big\}$. As before, in the $q$KZ equations the coefficients of each basis element can be collected together to form a system of equations. As an example, we take $i=0$ in \eqref{eq:qKZea}, and obtain
\begin{align*}
-\sum_{\alpha}a_0\psi_\alpha\ket\alpha&=\sum_{\alpha}\psi_\alpha e_0\ket\alpha\\
&=\left(\psi_{((}+\psi_{)(}\right)\big|)(\big\rangle+\left(\psi_{()}+\psi_{))}\right)\big|))\big\rangle.
\end{align*}
These equations can be written as
\be
\begin{split}
0&= a_0 \psi_{()} = a_0 \psi_{((},\\
\psi_{()} &= s_0 \psi_{))}, \qquad \psi_{((}= s_0\psi_{)(}\ .
\end{split}
\label{eq:froma0}
\ee
The rest of the system for $L=2$ is
\be
\begin{split}
0&= a_2 \psi_{()} = a_2 \psi _{))}, \\
\psi_{()}&= s_2 \psi_{((}, \qquad \psi_{))}= s_2\psi_{)(}\ ,
\end{split}
\label{eq:froma2}
\ee

\be
\begin{split}
0&= a_1 \psi_{))} = a_1 \psi _{)(} = a_1 \psi_{((}\ , \\
& s_1\psi_{()} = \psi_{))} + \psi_{((} + \psi_{)(}\ .
\end{split}
\label{eq:froma1}
\ee
As in the $L=1$ case, it is an easy consequence of the equation $a_0\psi_{()}=0$ that if $\psi_{()}\neq 0$, it should contain a factor $k(z_1,\zeta_0)$. Such vanishing conditions hold for all the components.
\begin{prop}
\label{prop:condL2}\hfill
\begin{itemize}
\item[i.] $\psi_{()}$ and $\psi_{((}$ vanish or contain a factor $k(z_1,\zeta_0)$, the remainder being invariant under $z_1 \leftrightarrow 1/z_1$.
\item[ii.] $\psi_{()}$ and $\psi_{))}$ vanish or contain a factor $k(1/sz_2,s\zeta_L)$, the remainder being invariant under $sz_2\leftrightarrow 1/sz_2$.
\item[iii.] $\psi_{))}$, $\psi_{)(}$ and $\psi_{((}$ vanish or contain a factor $[qz_1/z_2]$, the remainder being a symmetric function in $z_1$ and $z_2$.
\end{itemize}
\end{prop}

\subsubsection{Solution}
With the known factors and symmetries from \propref{condL2} above, we thus look for a solution of the form
\begin{align*}
\psi_{((} &= \prod_{i=1}^2 k(z_i,\zeta_0)\times [qz_1/z_2][q/z_1z_2]\times f_2(z_1,z_2), \\
\psi_{))} &= \prod_{i=1}^2 k(1/sz_i,s\zeta_L)\times [qz_1/z_2][qs^2z_1z_2]\times \tilde f_2(z_1,z_2),
\end{align*}
where $f_2(z_1,z_2)$ is a symmetric function invariant under $z_i\leftrightarrow 1/z_i$, and $\tilde f_2(z_1,z_2)$ is symmetric and invariant under $sz_i\leftrightarrow 1/sz_i$. Note that with $k$ as defined in \eqref{eq:k2} we could write this as
\begin{align*}
\psi_{((} &= \prod_{i=1}^2 k(z_i,\zeta_0)\times k(z_2,z_1)\times f_2(z_1,z_2), \\
\psi_{))} &= \prod_{i=1}^2 k(1/sz_i,s\zeta_L)\times k(1/sz_1,sz_2)\times \tilde f_2(z_1,z_2).
\end{align*}
The other two components may be determined from
\be
\label{eq:L2psi2}
\psi_{()} = s_0 \psi_{))},\qquad \psi_{)(} = s_1\psi_{()} - \psi_{))} - \psi_{((}.
\ee

We pause here to define an important polynomial, which will appear throughout the rest of this thesis. The symplectic character $\chi_\lambda$ of degree $\lambda=(\lambda_1,\ldots,\lambda_L)$ is defined by
\be
\chi_\lambda(z_1,\ldots,z_L) = \frac{\dt{z_i^{\lambda_j+L-j+1}-z_i^{-\lambda_j-L+j-1}}}{\dt{z_i^{L-j+1}-z_i^{-L+j-1}}}.
\label{eq:sympchar}
\ee
For convenience, we use the notation
\be
\tau_\lambda(z_1,\ldots,z_L) = \chi_\lambda(z_1^2,\ldots,z_L^2).
\label{eq:tau}
\ee
The symplectic character is symmetric, and for the degree $\lambda^{(L)}$ with $\lambda_j^{(L)}=\left\lfloor\frac{L-j}{2}\right\rfloor$ it satisfies the recursion
\be
\label{eq:chirecur}
\tau_{\lambda^{(L)}}(z_1, \ldots, z_L)|_{z_j=qz_i}=(-1)^L\prod_{k\neq i,j}k(z_i,z_k)\tau_{\lambda^{(L-2)}}(\hat z_i,\hat z_j).
\ee
The classical character $\chi_\lambda$, or equivalently the Schur polynomial of the type $C$ root system, with degree $\lambda=\lambda^{(L)}$ appears repeatedly in related studies on loop models \cite{DF05,ZJ07} and symmetry classes of alternating sign matrices \cite{Okada06}.

When $s^4=1$, $q=\e^{2\pi\ii/3}$ we find that the solution to equations \eqref{eq:froma0}--\eqref{eq:froma1} can be given explicitly by
\begin{align*}
f_2(z_1,z_2) &= -\tau_{(1,0,0)}(z_1, z_2, \zeta_L),\\
\tilde f_2(z_1,z_2) &= -\tau_{(1,0,0)}(s\zeta_0,sz_1,sz_2).
\end{align*}

\subsubsection{Recursion to $L=0$}
From Condition (iii) of \propref{condL2}, when we set $z_2=qz_1$, all components vanish except for $\psi_{()}$. From \eqref{eq:L2psi2},
\begin{align*}
\psi_{()}|_{z_2=qz_1}&=(-1-a_0)\psi_{))}|_{z_2=qz_1}\\
&=\pi_0\frac{k(1/z_1,\zeta_0)}{[q][z_1^2]}k(1/sz_1,s\zeta_L)k(1/sz_2,s\zeta_L)k(1/sz_1,sz_2)\\
&\quad\times\tau_{(1,0,0)}(sz_1,sz_2,s\zeta_0)|_{z_2=qz_1}\\
&=k(z_1,\zeta_0)^2k(z_1,\zeta_L)^2,
\end{align*}
where we have used the properties of $k$
\be
\label{eq:k}
k(s^{2}a,b)=s^{2}k(sa,sb)=k(a,1/b)=k(1/qa,b)=k(a,b).
\ee
Since the solution for $L=0$ is simply $1$, we can easily see that the proportionality factor $p$ in \lmref{comprecur} for $L=2$ is
\be
\label{eq:precur2}
p(z_1)=k(z_1,\zeta_0)^2k(z_1,\zeta_L)^2.
\ee

\subsubsection{Recursion to $L=1$}
Similarly, when we set $z_1=q\zeta_0$, the components $\psi_{()}$ and $\psi_{((}$ vanish, and $\psi_{))}$ becomes
\[\psi_{))}|_{z_1=q\zeta_0}=s^2\ k(1/sz_2,s\zeta_L)k(\zeta_0,\zeta_L)k(\zeta_0,z_2)^2.\]
Using our solution for $\psi_)(z_2)$ from \eqref{eq:L1soln}, we deduce that
\be
\label{eq:r0recur2}
r_0(z_2;\zeta_0)=-k(\zeta_0,\zeta_L)k(\zeta_0,z_2)^2.
\ee
In the same way, setting $z_2=\zeta_L/q$ in $\psi_{((}$ gives us
\be
\label{eq:rLrecur2}
r_L(z_1;\zeta_L)=-s^2k(1/s\zeta_L,s\zeta_0)k(1/s\zeta_L,sz_1)^2.
\ee

\subsection{Example: $L=3$}
It is computationally very intensive to compute explicitly the full solution for $L=3$. However, we can compute a solution for the subset of equations where $\psi_{)((}$ and $\psi_{))(}$ are not individually determined, but only their sum is (see \apref{L=3} for more details). We find
\begin{align*}
\psi_{(((}&=\prod_{0\leq i<j\leq 3}k(z_j,z_i)\ \tau_{(1,1,0,0)}(z_1,z_2,z_3,\zeta_L) g_3(z_1,z_2,z_3),\\
\psi_{)))}&=\ s^2 \prod_{1\leq i<j\leq 4}k(1/sz_i,sz_j)\ \tau_{(1,1,0,0)}(s\zeta_0,sz_1,sz_2,sz_3) g_3(sz_1,sz_2,sz_3),
\end{align*}
with $\tau$ as before, and where $g_3$ is symmetric and invariant under $z_i\leftrightarrow 1/z_i$. We have also introduced the notation $z_0=\zeta_0$ and $z_{L+1}=\zeta_L$. Imposing the boundary recursions in \lmref{comprecur} requires that the lowest degree solution is
\[
g_3(z_1,z_2,z_3) = \tau_{(1,0,0)}(z_1,z_2,z_3).
\]
\subsubsection{Recursion to $L=2$}
The above calculation also gives us
\be \begin{split}
\label{eq:rrecur3}
r_0(z_2,z_3;\zeta_0)&=-k(\zeta_0,\zeta_L)k(\zeta_0,z_2)^2k(\zeta_0,z_3)^2,\\
r_L(z_1,z_2;\zeta_L)&=-s^2k(1/s\zeta_L,s\zeta_0)k(1/s\zeta_L,sz_1)^2k(1/s\zeta_L,sz_2)^2.
\end{split} \ee

\subsubsection{Recursion to $L=1$}
Computing $\psi_{(()} = s_3 \psi_{(((}$, and setting $z_3=qz_2$, we find the recursion between size $L=3$ and size $L=1$:
\begin{align}
p(z_2;z_1)&=k(z_2,\zeta_0)^2k(z_2,\zeta_L)^2k(z_2,z_1)^4,\nn\\
\label{eq:precur3}\Rightarrow\qquad p(z_i;z_j)&=k(z_i,\zeta_0)^2k(z_i,\zeta_L)^2k(z_i,z_j)^4,\qquad j\neq i,i+1.
\end{align}

\section{Solution for general $L$}
\label{sec:genLsoln}
As in the case of $L=2$ in \secref{L=2}, for general $L$ we may derive factors for certain components. For each $i$ from $1$ to $L-1$, every link pattern in the LHS of the $q$KZ equation \eqref{eq:qKZea} will have a small link from $i$ to $i+1$ once $e_i$ has acted. The $q$KZ equation then says that $a_i\psi_\alpha=0$ if $\alpha$ does not have a small link from $i$ to $i+1$. This leads to the following conditions on $\psi_\alpha$.
\begin{prop}
\label{prop:condLgen}\hfill
\begin{itemize}
\label{eq:psiconditions}
\item[i.] If $\alpha$ does not have a small link from the left boundary to $1$, $\psi_\alpha$ vanishes or contains a factor $k(z_1,\zeta_0)$, the remainder being invariant under $z_1 \leftrightarrow 1/z_1$.
\item[ii.] If $\alpha$ does not have a small link from $L$ to the right boundary, $\psi_\alpha$ vanishes or contains a factor $k(1/sz_L,s\zeta_L)$, the remainder being invariant under $sz_L\leftrightarrow 1/sz_L$.
\item[iii.] If $\alpha$ does not have a small link from $i$ to $i+1$, $\psi_\alpha$ vanishes or contains a factor $[qz_i/z_{i+1}]$, the remainder being a symmetric function in $z_i$ and $z_{i+1}$.
\end{itemize}
\end{prop}
Using the above conditions, for general $L$ the component $\psi_{(\cdots(}$ is given by
\be
\label{eq:psidef1}\psi_{(\cdots(} = \prod_{0\leq i<j \leq L} k(z_j,z_i)\ f_L(z_1,\ldots,z_L),
\ee
where $f_L$ is symmetric and invariant under $z_i\rightarrow 1/z_i$. The majority of the factors in this expression are imposed by the symmetry conditions.

Likewise, the component $\psi_{)\cdots)}$ is expressed as
\begin{align}
\psi_{)\cdots)}
\label{eq:psidef2}&= \prod_{1\leq i<j \leq L+1} k(1/sz_i,sz_j)\ \tilde f_L(sz_1,\ldots,sz_L),
\end{align}
where $\tilde f_L$ is symmetric and invariant under $sz_i\rightarrow 1/sz_i$. Other components may be derived from the extremal components by acting with products of Baxterised versions of the operators $s_i$, as described in \apref{hecke} \cite{dGP07}. However, in the case under consideration it is not possible to derive every component of $\ket\Psi$ in this way. In \apref{L=3} we explain the reasons for this in detail for the case $L=3$.

To get more information about the polynomials $f_L$ and $\tilde f_L$, as well as about the other components, we use the recursive properties of the eigenvector, as we have previously for small $L$. As before, the recursions come from \lmref{comprecur}, which relates components of the eigenvector for $L$ to components of the eigenvector for $L-1$ and $L-2$.

\subsection{Recursions}
\label{sec:rec}

We have found the proportionality factors for small system sizes, given in \eqref{eq:rrecur1}, \eqref{eq:precur2}, \eqref{eq:r0recur2}, \eqref{eq:rLrecur2}, \eqref{eq:rrecur3} and \eqref{eq:precur3}. We make the assumption that these factors continue their pattern for larger system sizes, as stated in the following conjecture.
\begin{conj}
For general $L$ the proportionality factors are
\begin{align*}
p(z_i;z_1,\ldots,\hat z_i,\hat z_{i+1},\ldots,z_L)&=k(z_i,\zeta_0)^2k(z_i,\zeta_L)^2\prod_{j\neq i,i+1}k(z_i,z_j)^4,\\[-1mm]
r_0(z_2,\ldots,z_L;\zeta_0) &= -k(\zeta_0,\zeta_L) \prod_{i=2}^{L} k(\zeta_0,z_i)^2,\\[-1mm]
r_L(z_1,\ldots,z_{L-1};\zeta_L) &= -s^2 k(1/s\zeta_L,s\zeta_0)\prod_{i=1}^{L-1}k(1/s\zeta_L,sz_i)^2.
\end{align*}
\end{conj}

Using these factors, we can come up with associated recursions for the symmetric functions $f_L$ and $\tilde f_L$. For example, we can calculate the size $L$ eigenvector component $\psi^L_{(\cdots ()}$ using the $q$KZ equation and then, according to \lmref{comprecur}, setting $z_L=qz_{L-1}$ will give us $p(z_{L-1};z_1,\ldots,z_{L-2})$ multiplied by the size $L-2$ component $\psi^{L-2}_{(\cdots (}$. Recalling the definition \eqref{eq:s} of $s_i$, we have from the $q$KZ equation that
\[
\psi^L_{(\cdots()}=s_L \psi^L_{(\cdots(}\ ,
\]
and since $\left.\psi^L_{(\cdots(}\right|_{z_L=qz_{L-1}}=0$, it follows that
\begin{align*}
\left.\psi^L_{(\cdots()}\right|_{z_L=qz_{L-1}} &=\left.\left(\pi_L\frac{k(sz_L,s\zeta_L)}{[q][s^2z_L^2]}\prod_{0\leq i<j\leq L}k(z_j,z_i)f_L(z_1,\ldots,z_L)\right)\right|_{z_L=qz_{L-1}}\\
&=k(z_{L-1},\zeta_L)\prod_{i=0}^{L-2}k(z_{L-1},z_i)^2\prod_{0\leq i<j\leq L-2}k(z_j,z_i)f_L(z_1,\ldots,z_{L-1},s^2qz_{L-1})\\
&=k(z_{L-1},\zeta_0)^2k(z_{L-1},\zeta_L)\prod_{i=1}^{L-2}k(z_{L-1},z_i)^2\frac{f_L(z_1,\ldots,z_{L-1},s^2qz_{L-1})}{f_{L-2}(z_1,\ldots,z_{L-2})}\psi^{L-2}_{(\cdots(}.
\end{align*}
Here we have used the properties of $k$ given in \eqref{eq:k}. From above, the proportionality factor in this relation is given by $p(z_{L-1};z_1,\ldots,z_{L-2})$, so we arrive at a recursion for $f_L$,
\[
f_L(z_1,\ldots,z_{L-1},s^2qz_{L-1})=k(z_{L-1},\zeta_L)\prod_{j=1}^{L-2} k(z_{L-1},z_j)^2f_{L-2}(z_1,\ldots,z_{L-2}).
\]
Similarly, we can use the recursion from $\psi^L_{()\cdots)}$ to $\psi^{L-2}_{)\cdots)}$ to find a recursion for $\tilde f_L$. Due to the symmetry properties of both these functions, the recursions can be generalised to arbitrary $i$,
\be
\label
{eq:recurSym}\begin{split}
f_L(z_1,\ldots,z_i,s^2qz_i,\ldots,z_L)&=k(z_i,\zeta_L)\prod_{j\neq i,i+1} k(z_i,z_j)^2f_{L-2}(z_1,\ldots,\hat z_i,\hat z_{i+1},\ldots,z_L),\\
\tilde f_L(z_1,\ldots,z_i,s^2qz_i,\ldots,z_L)&=k(z_i,\zeta_0)\prod_{j\neq i,i+1} k(z_i,z_j)^2\tilde f_{L-2}(z_1,\ldots,\hat z_i,\hat z_{i+1},\ldots,z_L).\\
\end{split}
\ee

The boundary recursions from \lmref{comprecur} can be immediately applied to the extremal components \eqref{eq:psidef1} and \eqref{eq:psidef2}, and we find that $f_L$ and $\tilde f_L$ in addition satisfy
\be
\label{eq:recurSymbound}
\begin{split}
f_{L}(z_1,\ldots,z_{L-1},\zeta_L/q;\zeta_0,\zeta_L) &=-\prod_{j=1}^{L-1}k(1/\zeta_L,z_j)\ f_{L-1}(z_1,\ldots,z_{L-1};\zeta_0,\zeta_L/q),\\
\tilde f_{L}(q\zeta_0,z_2,\ldots,z_L;\zeta_0,\zeta_L) &=-s^2\prod_{j=2}^L k(\zeta_0,z_j)\ \tilde f_{L-1}(z_2,\ldots,z_L;q\zeta_0,\zeta_L),
\end{split}
\ee
where we have explicitly indicated the dependencies on $\zeta_0$ and $\zeta_L$.

\subsection{Degree}
Polynomial solutions of the $q$KZ can be labelled by their top degree $\mu$, where $\mu$ is a partition, $\mu_1\geq \mu_2 \geq \ldots \geq \mu_L\geq 0$. These solutions are of the form
\[
\sum_{\nu \in \widetilde W\cdot\mu} c_{\nu} z^{2\nu},
\]
where the notation $z^\nu$ stands for the product $z_1^{\nu_1}\ldots z_L^{\nu_L}$, the coefficients $c_\nu$ are polynomials in $\zeta_0$ and $\zeta_L$, and $\widetilde W\cdot \mu$ denotes the orbit of $\mu$ under the action of the group $\widetilde W$, defined as follows. The generators of $\widetilde W$ act on an $L$-tuple $\mu$ by the transformations
\[
\begin{array}{rlll}
t_i: & \quad (\ldots,\mu_i,\mu_{i+1},\ldots)\quad & \mapsto\quad (\ldots,\mu_{i+1},\mu_i,\ldots), \quad  & 1\leq i\leq n-1,\\
t_0: & \quad (\mu_1,\mu_2,\ldots) & \mapsto\quad (1-\mu_1,\mu_2,\ldots), & \text{if }\; \mu_1>0,\\
t_L: & \quad (\mu_1,\mu_2,\ldots) & \mapsto\quad (\ldots,\mu_{L-1},-\mu_L), &
\end{array}
\]
where $t_0$ acts as the identity if $\mu_1<0$. In effect, $W\cdot\mu$ is the set of all $L$-tuples that can be obtained from $\mu$ by any combination of the above actions.

We can use the recursions \eqref{eq:recurSym} and \eqref{eq:recurSymbound} to find the minimal degree of $f_L$ and $\tilde f_L$ for arbitrary size. The argument here is for $f_L$, but it is easily seen that it holds for $\tilde f_L$ as well, and therefore that they have the same degree. In \eqref{eq:recurSym}, consider $i=1$ and denote the top degree of $f_L$ by $\nu^{(L)}=(\nu_1^{(L)},\ldots,\nu_L^{(L)},0,0,\ldots)$. The (top) degree of $k(z_1,z_j)$ is $(1,0)$ in the variables $z_1^2$ and $z_j^2$, so the degree in $z_1^2$ on the right hand side of \eqref{eq:recurSym} is $2L-3$. Since the degree in $z_1^2$ on the LHS must be the same, the added degrees in $z_1^2$ and $z_2^2$ of $f_L(z_1,\ldots,z_L)$ must be greater than or equal to $2L-3$.

In addition, by comparing degrees of any $z_j^2$ in \eqref{eq:recurSymbound}, it immediately follows that $\nu_{j}^{(L)}$ is at least equal to $\nu_{j}^{(L-1)}+1$. We thus find that the following inequalities must hold,
\begin{align*}
\nu_1^{(L)} + \nu_2^{(L)} \geq 2L-3,
\\
\nu_j^{(L)} \geq \nu_{j}^{(L-1)}+1.
\end{align*}
For a minimal degree solution these inequalities become equalities, and using the solutions we explicitly constructed for the small system sizes $L=1,2,3$ in \secref{smallsize}, we find that
\[
\nu_j^{(L)} = L-j\qquad (j=1,\ldots,L).
\]
We will write $\nu^{(L)}=\lambda^{(L)}+\lambda^{(L+1)}$, where $\lambda^{(L)}$ is the partition of $|\lambda^{(L)}|= \left\lceil \frac{L}{2}\left( \frac{L}{2}-1\right)\right\rceil$ with
\be
\label{eq:lambda}
\lambda_j^{(L)} = \left\lfloor \frac{L-j}{2}\right\rfloor\qquad j=1,\ldots,L,
\ee
i.e.,
\[
\lambda^{(2n)}=(n-1,n-1,\ldots,1,1,0,0),\qquad \lambda^{(2n+1)}=(n,n-1,n-1\ldots,1,1,0,0).
\]

From the degree of $k(z_j,z_i)$ it immediately follows that the product of factors in the expressions for the extremal components, $\psi_{(\cdots(}$ and $\psi_{)\cdots)}$, amount to a degree of $\lambda^{(L+1)}+\lambda^{(L+2)}$. Solutions of the $q$KZ equation of minimal degree, which are relevant for the TL($1$) loop model with open boundaries, therefore have degree $\mu^{(L)}$, with
\[
\mu^{(L)} = \lambda^{(L)}+2\lambda^{(L+1)}+\lambda^{(L+2)},
\]
so that
\be
\mu_j^{(L)} = 2L+1-2j.
\label{eq:degree}
\ee
The total degree of these solutions is equal to $|\mu^{(L)}|=L^2$ and the degree in each variable $z_i^2$ is equal to $\mu_1 = 2L-1$.

\subsection{Eigenvector}
By using recursion and degree properties of the general solution, we can find expressions for $\tilde f_L$ and $f_L$. We emphasise again that we have taken $s^4=1$ and $q=\e^{2\pi\ii/3}$.

For $L=2$ and $L=3$, the solution contains a symmetric function that involves the symplectic character defined in \eqref{eq:sympchar}. The solution for general $L$ can also be expressed in terms of this symplectic character. It turns out that the following two functions satisfy the necessary recursions \eqref{eq:recurSym} and \eqref{eq:recurSymbound}, agree with the small size solutions, and have the correct degree $\nu^{(L)}$,
\be
\begin{split}
f_L(z_1,\ldots,z_L)&=(-1)^{\frac{L(L+1)}{2}}\tau_{\lambda^{(L+1)}}(z_1,\ldots,z_L,\zeta_L)\tau_{\lambda^{(L)}}(z_1,\ldots,z_L),\\
\tilde f_L(sz_1,\ldots,sz_L) &=(-1)^{\frac{L(L+1)}{2}}(s^2)^L\tau_{\lambda^{(L+1)}}(s\zeta_0,sz_1,\ldots,sz_L)\tau_{\lambda^{(L)}}(sz_1,\ldots,sz_L).
\end{split}
\label{eq:Symdef}
\ee
It is worthwhile noting that \eqref{eq:recurSym} and \eqref{eq:recurSymbound} are satisfied because of the recursion for the symplectic character in \eqref{eq:chirecur}, and the specification $s^4=1$.

\subsection{Eigenvector normalisation}
\label{sec:normalisation}

We have derived in \apref{propsym}, see \eqref{eq:Zrecur} and \eqref{eq:Zrecurbound}, the recursions for the normalisation $Z_L$ as defined in \eqref{eq:Zdef}. Using the recursion \eqref{eq:chirecur} for the symplectic character defined in \eqref{eq:sympchar}, we note that
\begin{align*}
\tau_{L+2}&(\zeta_0,z_1,\ldots,z_L,\zeta_L)\ \tau_{L+1}(\zeta_0,z_1,\ldots,z_L)\ \tau_{L+1}(z_1,\ldots,z_L,\zeta_L)\ \tau_L(z_1,\ldots,z_L)|_{z_{i+1}=qz_i}\\
&=k(z_i,\zeta_0)^2\ k(z_i,\zeta_L)^2\prod_{j\neq i,i+1} k(z_i,z_j)^4\ \tau_L(\zeta_0,\ldots,\hat z_i,\hat z_{i+1},\ldots,\zeta_L)\\
&\quad\times \tau_{L-1}(\zeta_0,\ldots,\hat z_i,\hat z_{i+1},\ldots)\ \tau_{L-1}(\ldots,\hat z_i,\hat z_{i+1},\ldots,\zeta_L)\ \tau_{L-2}(\ldots,\hat z_i,\hat z_{i+1},\ldots).
\end{align*}
This product of four symplectic characters therefore satisfies the same recursions as $Z_L$ (the boundary recursions are also easy to show). In addition, for $L=1$ and $L=2$ this product is equal to $Z_1$ and $Z_2$, respectively. Since the recursions \eqref{eq:Zrecur} and \eqref{eq:Zrecurbound} specify enough points to uniquely determine $Z_L$ of degree $\mu^{(L)}$, see \eqref{eq:degree}, up to a factor independent of $L$, we conclude that
\begin{align}
\label{eq:Z}
Z_L(z_1,\ldots,z_L)&=\tau_{L+2}(\zeta_0,z_1,\ldots,z_L,\zeta_L)\ \tau_{L+1}(\zeta_0,z_1,\ldots,z_L)\nn\\
&\quad\times \tau_{L+1}(z_1,\ldots,z_L,\zeta_L)\ \tau_L(z_1,\ldots,z_L).
\end{align}
In particular, the normalisation of the ground state of the Hamiltonian \eqref{eq:hamspec} is obtained by setting $z_i=1$, and is given by
\be
Z_L = \tilde{Z}_2(c_0,c_L)\ \tilde{Z}_{1}(c_0)\ \tilde{Z}_{1}(c_L)\ \tilde Z_0,
\label{eq:homonorm}
\ee
where
\begin{align*}
\tilde{Z}_{0} &= \tau_L(1,\ldots,1),\\
\tilde{Z}_{1}(c_i) &= \tau_{L+1}(\zeta_i,1,\ldots,1),\\
\tilde{Z}_{2}(c_0,c_L) &= \tau_{L+2}(\zeta_0,1,\ldots,1,\zeta_L),
\end{align*}
and $c_i$ is defined as
\[
c_i = \frac{3}{1+\zeta_i^2+\zeta_i^{-2}}\ ,
\]
as in \eqref{eq:ci}.

\newpage
\chapter{The boundary-to-boundary correlation function}
\label{chap:Flow}

In an attempt to rigorously prove conformal invariance of the scaling limit of critical two-dimensional lattice models, an interesting new class of parafermionic observables for Potts and loop models was recently introduced, see for example \cite{Smir06,Smir07,RivaC06,IkhlefC09}. These observables are expressed in terms of the loop representation of the Fortuin--Kasteleyn cluster expansion of the Potts model \cite{BaxKW76,Bax82}, and can be shown to be discretely holomorphic for certain parameter values. The operators corresponding to these observables carry a spin conjugate to the winding angles of the loops. In this chapter we will compute an exact closed form expression for the expectation of one such type of observable, with spin one, for the two-boundary TL($1$) loop model as described in \chapref{On}. The remarkable aspect of this result is the feasibility of an exact calculation of an expectation for a finite size model that is not free fermionic.

\begin{figure}[ht]
\centerline{\includegraphics[height=120pt]{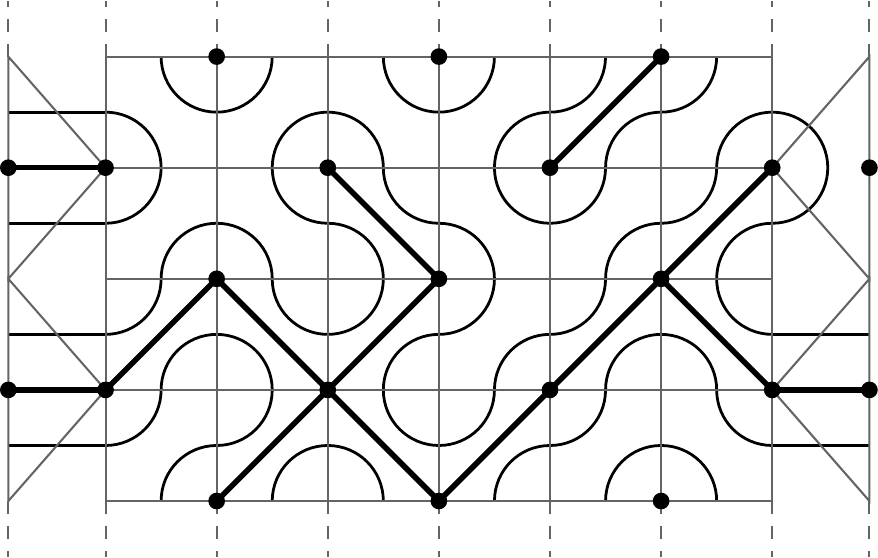}}
\caption{A percolating cluster with its associated hulls drawn as loops.}
\label{fig:wind}
\end{figure}

We consider the TL($1$) loop model as an alternative description of the two-dimensional critical bond percolation model \cite{PearceRZ06}. In this setting the percolation model is defined on a finite width, vertically infinite square lattice, as in \figref{wind}, where bonds are placed diagonally across the faces with a certain probability. Percolation occurs when a cluster of bonds stretches from the left boundary to the right. As previously mentioned, the loops in the TL($1$) model correspond to hulls drawn around the bond clusters, so a loop stretching from the left to the right implies a percolating cluster.

The expectation value calculated in this chapter can be interpreted as the density of percolation clusters passing in between two chosen vertices of the lattice, $x_1$ and $x_2$ \cite{MitraN04,Cardy00,DubailJS09b,Smir09}. The analogous calculation for the density of clusters closed around a cylinder has also been performed \cite{NienJdF10}. We will sometimes refer to this density as the boundary-to-boundary correlation function for the two-boundary TL($1$) model. Our calculation is valid for odd and even $L$, however for odd $L$ the boundary conditions of the TL($1$) model cause the interpretation in terms of the percolation model to break down.

This expectation value also corresponds to the spin current in a generalised Chalker--Coddington network model \cite{ChalkC88,Metz99,GruzbergLR99} for the quantum spin Hall effect. The Chalker--Coddington network model is based on the semi-classical picture of electrons in two dimensions moving under the influence of a strong perpendicular magnetic field in a long-ranged disorder potential. The network consists of a square lattice whose edges are unidirectional channels and whose vertices are scattering centres. A potential correlation length much larger than the magnetic length leads to the formation of clusters where the wave function amplitude changes only slightly. These clusters correspond to the clusters of bonds described above. The classical limit of the generalised Chalker--Coddington model \cite{Metz99} on the square lattice is therefore described by the solvable TL($1$) lattice model. In a further generalisation, Gruzberg et al.~\cite{GruzbergLR99} used a pseudo spin description of the particle and hole states.  When the $SU(2)$ spin symmetry is observed, the full quantum mechanical spin current corresponds to the observable we study here.

In this chapter we will exactly calculate the correlation function for finite system size $L$. We will first give an explicit description of this quantity, before defining it as an expectation value in \secref{flowdef}. We then discuss the symmetries and recursion relations satisfied by the function, and we use these properties along with a degree argument to prove an explicit form for the correlation function, given in \secref{flowsoln}. Our proof is complete except for two conjectures, both of which we have observed to be true for small $L$, and which we expect to hold in general.

\section{Definition of the correlation function}

Given a configuration $C$ of bonds on the lattice, and any two marked vertices (at positions $x_1$ and $x_2$), we consider all the loops (hulls) that pass between these two vertices. We give a weight of $0$ to those that close on themselves or have both ends connected to the same boundary, and a non-zero weight to loops that connect the left boundary to the right. The weight given is either $1$ or $-1$ depending on the winding of the loop, so that a percolating cluster with a hull that crosses back and forth between the marked vertices is only counted once. The sum of the loop weights gives the number of percolating clusters passing in between $x_1$ and $x_2$ in the configuration, multiplied by the sign relating to the winding. The average of this value over all configurations, weighted by the probability of obtaining each configuration, then gives the average density of percolating clusters between the marked points, which we refer to as the correlation function $F^{\{x_1,x_2\}}$.

\begin{figure}[ht]
\centerline{\includegraphics[height=120pt]{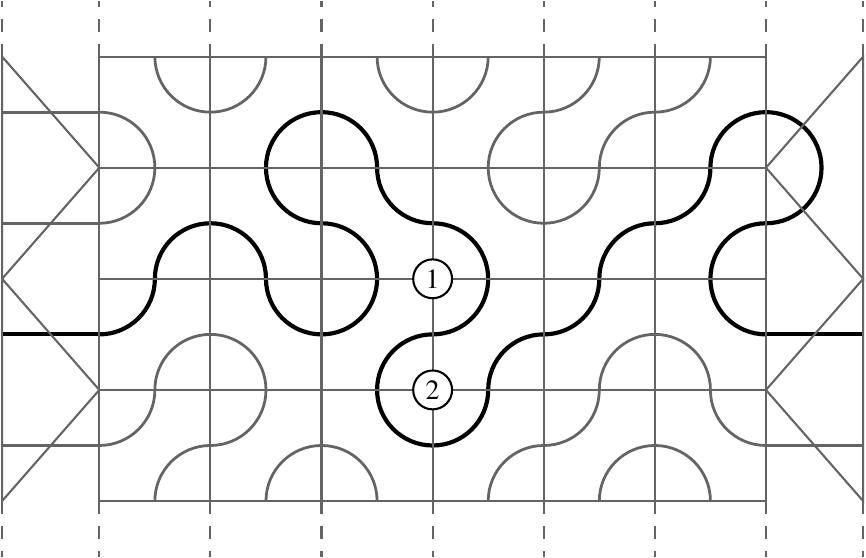}}
\caption[Path contributing to $Y_L^{(4)}$]{Path between markers 1 and 2 (at positions $x_1$ and $x_2$ respectively) with phase $-1$, arising from \figref{wind}, contributing to $Y_L^{(4)}$.}
\label{fig:Vflow_O(1)}
\end{figure}

This correlation function is given explicitly by
\be
F^{\{x_1,x_2\}} = \sum_{C\in\Gamma} P(C) N_C^{\{x_1,x_2\}} \sign^{\{x_1,x_2\}}_C.
\label{eq:flow_gendef}
\ee
Here $\Gamma$ is the set of configurations, and the probability $P(C)$ of a configuration $C$ is the product of weights of all the individual face configurations as defined by the $R$ and $K$-matrices in \secref{bax}.\symbolfootnote{Throughout this chapter we take $q$ to be the third root of unity $\e^{2\ii\pi/3}$  and $\omega_0=\omega_L=-4\pi/3$ as introduced in \secref{spec}.} $N_C^{\{x_1,x_2\}}$ is the number of paths passing in between the points $x_1$ and $x_2$ and running from the left to the right boundary, and $\sign_C^{\{x_1,x_2\}}$ is $+1$ if $x_1$ lies in the region above the paths, and $-1$ if it lies below. The sign of the path in \figref{Vflow_O(1)} thus is $-1$.

\subsection{The observables}
\label{sec:flowdef}
\begin{defn}
\label{def:Fhat}
We define the operator $\widehat{F}_L^{\{x_1,x_2\}}$ by marking two vertices inside the transfer matrix $T_L$, from \defref{transmat}. The locations of the markers are given by $x_1$ and $x_2$, each of which is an ordered pair $(k,y)$. The first coordinate, $k$, can take any value in $\{1,\ldots,L+1\}$, corresponding to the possible positions along the width of lattice. The second coordinate $y$ can take one of three values: {\rm b} (bottom), {\rm m} (middle), or {\rm t} (top). The markers give weights to loops passing between them as described in the previous section.
\end{defn}
As an example of the observable we have
\[
\widehat{F}_L^{\{(3,{\rm m}),(3,{\rm b})\}}=\raisebox{-40pt}{\includegraphics[height=80pt]{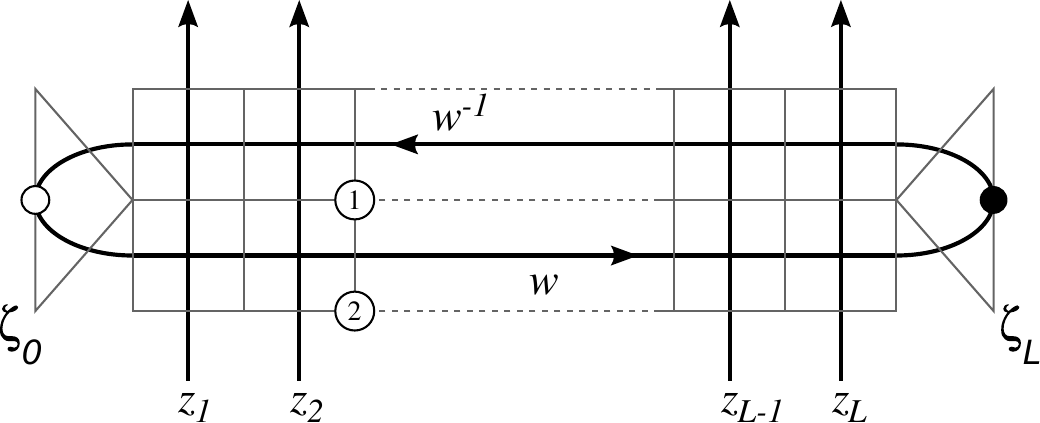}}\;.
\]
The boundary-to-boundary correlation function is given by the expectation value $F_L^{\{x_1,x_2\}}$ of this observable. We will now derive an expression for this expectation value.

The probability distribution of downward link patterns at the bottom of the transfer matrix is given by the ground state eigenvector $\ket{\Psi(\zeta_0,\zeta_L;z_1,\ldots,z_L)}$, calculated in the previous chapter, which is an element of the space LP$_L$ as defined in \secref{linkp}. To describe the probabilities for the upward link patterns at the top of the transfer matrix, we must first define a new space, $\widetilde{\rm LP}_L$, which is spanned by the upward link patterns of length $L$. The elements of this space are denoted $\langle{\widetilde\alpha}|$, not to be confused with the notation used for the left eigenvector of the transfer matrix defined in \secref{equiv}. For upward link patterns, we use the same shorthand notation in terms of opening and closing brackets as was defined in \secref{linkp} for downward link patterns.
\begin{defn}
\label{def:ip}
Between $\widetilde{\rm LP}_L$ and LP$_L$ we define the LP-inner product as follows. Let $\ket\alpha\in{\rm LP}_L$ be a downward link pattern and $\langle\widetilde\alpha|\in\widetilde{\rm LP}_L$ be an upward link pattern. Then
\[\langle\widetilde\alpha\ket\alpha_{\rm LP}=1,\quad \forall\ \alpha,\widetilde\alpha.\]
\end{defn}
This inner product can be viewed as the result obtained when an upward and a downward link pattern are attached to each other. All resulting closed loops and loops connected to the boundaries disappear, giving a weight of $1$. This is best understood pictorially; for example, choosing $\widetilde\alpha=$`$)()())(($' and $\alpha=$`$(())(()($', we have
\[\pic{-32pt}{70pt}{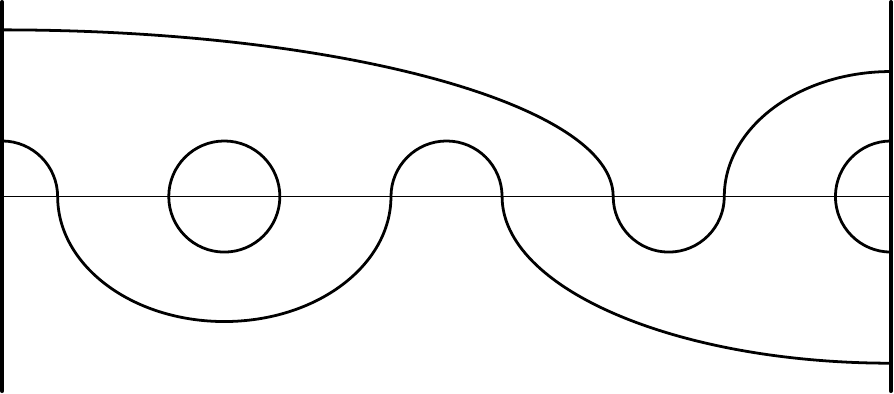}\quad=1.\]

The probabilities for the upward link patterns at the top of the transfer matrix are prescribed by $\langle{\widetilde \Psi}|$, which is the groundstate eigenvector of the adjoint transfer matrix,
\[ \langle\widetilde\Psi|T_L:=T_L^\dag\left(\langle\widetilde\Psi|\right)=\langle\widetilde\Psi|. \]
\begin{lm}
\label{lm:Ttranspose}
The adjoint transfer matrix is given by
\[
T_L^\dag(w;\zeta_0,\zeta_L;z_1,\ldots,z_L) = T_L(qw^{-1};\zeta_L,\zeta_0;z_L,\ldots,z_1).
\]
\end{lm}
\begin{proof}
This is seen by considering the result of rotating the transfer matrix by $180^\circ$. Each vertical pair of $R$-matrices can be transformed by the crossing relation \eqref{eq:crossing} into
\[
\pic{-29pt}{70pt}{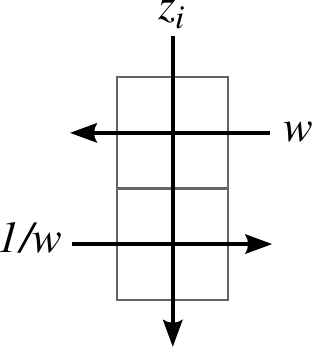}\quad=\quad\pic{-36pt}{70pt}{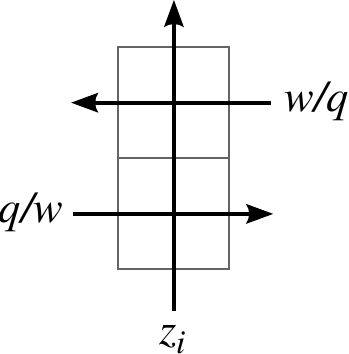}\quad.
\]
The right boundary $K$-matrix $K_L(w)$, when rotated by $180^\circ$, is identical to $K_0(w/q)$ by \eqref{eq:KK}. Similarly, $K_0(1/w)$ rotated by $180^\circ$ is identical to $K_L(q/w)$. Thus, the adjoint transfer matrix can be obtained by a swapping of parameters; $\zeta_0\leftrightarrow\zeta_L$, $z_1\leftrightarrow z_L$, $z_2\leftrightarrow z_{L-1},\ldots$, and replacing $w\rightarrow q/w$.
\end{proof}
\begin{lm}
\label{lm:dual}
The elements of the vector $\langle{\widetilde \Psi}|$ are related to those of $\ket\Psi$ by the same swapping of parameters as in \lmref{Ttranspose}, as well as a horizontal reflection of the link patterns, e.g., $\alpha=$`$)(()()$' goes to $\widetilde\alpha=$`$()())($' for $L=6$. Thus,
\[ \widetilde\psi_{\widetilde\alpha}(\zeta_0,\zeta_L;z_1,\ldots,z_L)=\psi_\alpha(\zeta_L,\zeta_0;z_L,\ldots,z_1). \]
\end{lm}

The expectation value $F_L^{\{x_1,x_2\}}$, as defined in \eqref{eq:flow_gendef}, is thus given for the infinite lattice by
\be
\label{eq:Expectation}
F_L^{\{x_1,x_2\}}=\frac{\langle{\widetilde \Psi}|\ \widehat{F}_L^{\{x_1,x_2\}}\ \ket{\Psi}_{\rm LP}}{\langle{\widetilde \Psi}|\Psi\rangle_{\rm LP}},
\ee
with $\widehat{F}_L^{\{x_1,x_2\}}$ as defined in \defref{Fhat}. As $\bra{\widetilde\alpha}\alpha\rangle_{\rm LP}=1$, $\forall \alpha,\widetilde\alpha$, we have that
\[
\langle{\widetilde \Psi}|\Psi\rangle_{\rm LP} = \sum_{\widetilde\alpha,\alpha}\widetilde\psi_{\widetilde\alpha}\ \psi_{\alpha} = \widetilde Z_L\ Z_L,
\]
where $Z_L=\sum_\alpha \psi_\alpha$ is the normalisation of the eigenvector, given by \eqref{eq:Z}. Since $Z_L$ is symmetric in all the $z_i$, as well as between $\zeta_0$ and $\zeta_L$, we have $\widetilde Z_L=Z_L$, so the denominator of \eqref{eq:Expectation} becomes $Z_L^2$.

\begin{lm}
The function $F_L$ is antisymmetric, $F_L^{\{x_1,x_2\}}=-F_L^{\{x_2,x_1\}}$ and additive, $F_L^{\{x_1,x_2\}} + F_L^{\{x_2,x_3)} = F_L^{\{x_1,x_3\}}$.
\end{lm}
\begin{proof}
These properties are easily seen from the definition.
\end{proof}
Because of the additivity property of $F_L$, it suffices to concentrate on the case where the two markers are placed on adjacent sites, so that $N_C^{\{x_1,x_2\}}$ can only be $0$ or $1$.  The markers can be separated by a horizontal lattice edge, or a vertical one as in \figref{Vflow_O(1)}.

There are a number of possible positions for the pair of markers in the transfer matrix. If the markers are vertically separated, they can be situated on the top or bottom row, and at any of the $L+1$ positions along the width of the lattice. Recalling \defref{Fhat}, for the bottom row we define
\[
\widehat Y_L^{(k)}=\widehat F_L^{\{(k,\text m),(k,\text b)\}},
\]
and from the expectation value $Y_L^{(k)}$, we can calculate $F_L^{\{(k,\text t),(k,\text m)\}}$ by
\be
\label{eq:YF}
F_L^{\{(k,\text t),(k,\text m)\}}(z_1,\ldots,z_L)=Y_L^{(k)}(1/z_1,\ldots,1/z_L).
\ee
This fact comes directly from the definition of the transfer matrix. In addition, we will show (see \secref{symmY}) that $Y_L$ is invariant under horizontal translation of the markers, so we sometimes drop the label $k$.

Horizontally separated markers can straddle any of the $L$ horizontal positions, and can be in one of three vertical positions. With the markers at the very bottom we define
\[
\widehat X_L^{(k)}=\widehat F_L^{\{(k,\text b),(k+1,\text b)\}}.
\]
It will be shown in \secref{symmX} that the expectation value $X_L^{(k)}$ is invariant under vertical translation of the markers, i.e., $F_L^{\{(k,\text t),(k+1,\text t)\}}=F_L^{\{(k,\text m),(k+1,\text m)\}}=X_L^{(k)}$. Furthermore, we can write the operator $\widehat X_L^{(k)}$ as a product, $\widehat{X}^{(k)}_L = T_L\ \kappa^{(k)}_X$, where $\kappa_X$ is the operator consisting of the two markers. Hence,
\[
X^{(k)}_L =\frac1{Z_L^2}\langle{\widetilde \Psi}| \widehat{X}^{(k)}_L\ket\Psi_{\rm LP}
=\frac1{Z_L^2}\langle{\widetilde \Psi}|T_L\ \kappa^{(k)}_X \ket\Psi_{\rm LP}
=\frac1{Z_L^2}\langle{\widetilde \Psi}|\kappa^{(k)}_X\ket\Psi_{\rm LP}.
\]
This also implies that the expectation value $X^{(k)}_L$ does not depend on the spectral parameter $w$.

We will at times suppress some of the arguments of polynomials. Unless otherwise stated, $Y_L=Y_L(w;\zeta_0,\zeta_L;z_1,\ldots,z_L)$ with a special dependence on $w$, and $X_L^{(k)}=X_L^{(k)}(\zeta_0,\zeta_L;z_1,\ldots,z_L)$ with a special dependence on $z_k$. Where relevant, variables will be given explicitly, and the notation $\hat z_i$ in a list of arguments will mean that $z_i$ is missing from the list (a convention used in the previous chapter).

\subsection{Relations between the observables $\widehat Y_L$ and $\widehat X_L$}
Relations between $\widehat{X}_L$ and $\widehat{Y}_L$ are obtained by looking at the way the markers can move within configurations with fixed local orientations of paths.
\begin{prop}
\label{prop:recur1}
$\widehat Y_L$ is related to $\widehat X_L^{(k)}$ by
\[\widehat Y_L(w=z_i;z_1,\ldots,z_L)=\widehat X^{(i)}_L(z_1,\ldots,z_L),\]
and $\widehat X_L^{(k)}$ is related to $\widehat Y_L$ by
\begin{align*}
\widehat{X}^{(i+1)}_L(\ldots,z_i=wq^{-1},z_{i+1}=w,\ldots)\circ\varphi_i &=\varphi_{i}\circ\widehat{Y}_{L-2}(w;\ldots,\hat{z}_{i},\hat{z}_{i+1},\ldots),\\
\widehat{X}^{(i)}_L(\ldots,z_i=wq^{-1},z_{i+1}=w,\ldots)\circ\varphi_i &=-\varphi_{i}\circ\widehat{Y}_{L-2}(w;\ldots,\hat{z}_{i},\hat{z}_{i+1},\ldots),
\end{align*}
where $\varphi_i$ is the map from \defref{varphi}.
\end{prop}
\begin{proof}
These relations can be proved pictorially. We first prove the relation that expresses $\widehat{Y}_{L}$ in terms of $\widehat{X}_L$. At the specialisation $w=z_i$, the lower face at position $i$ only has one possible configuration,
\[ \widehat{Y}_L|_{w=z_i}=\raisebox{-45pt}{\includegraphics[height=100pt]{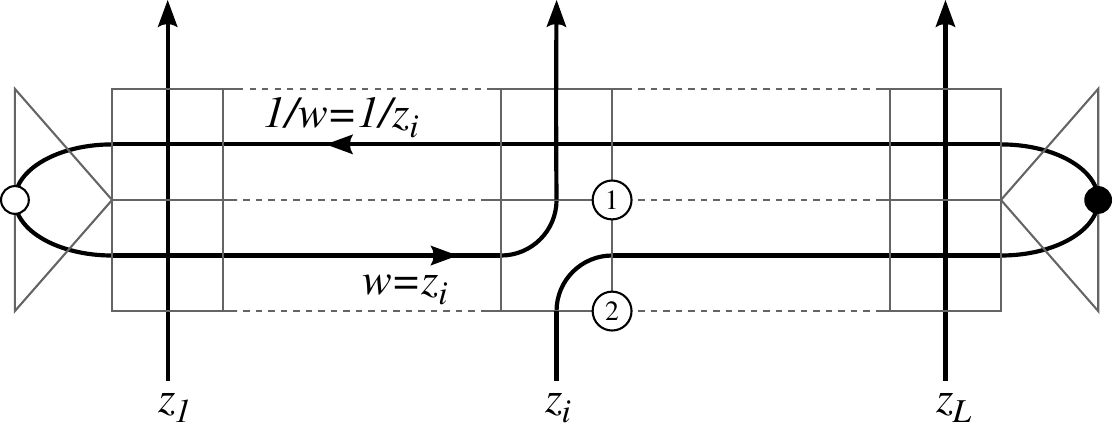}}\quad.\]
The marker $x_1$ can be moved along the loop on this face, producing
\[ \raisebox{-45pt}{\includegraphics[height=100pt]{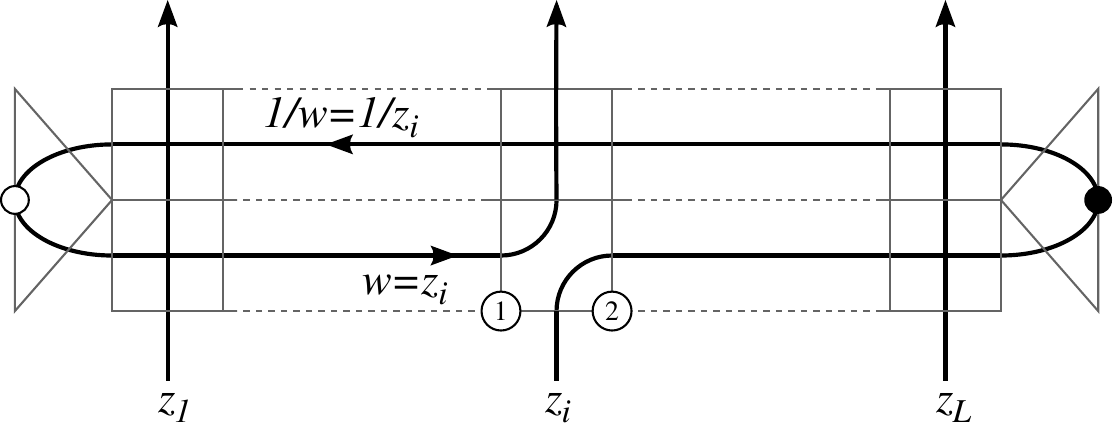}}\quad,\]
which is just $\widehat{X}^{(i)}_L$.

The relations that express $\widehat{X}_L$ in terms of $\widehat{Y}_{L-2}$ are similarly proved. For the first of these, the specialisation $z_{i+1}=qz_i=w$ imposes one possible configuration on the four central faces, which acts on the small loop introduced by $\varphi_i$,
\[ \widehat{X}^{(i+1)}_L\Big|_{\begin{subarray}{l} z_{i+1}=w\\z_i = wq^{-1} \end{subarray}}\circ\varphi_i =\raisebox{-45pt}{\includegraphics[height=100pt]{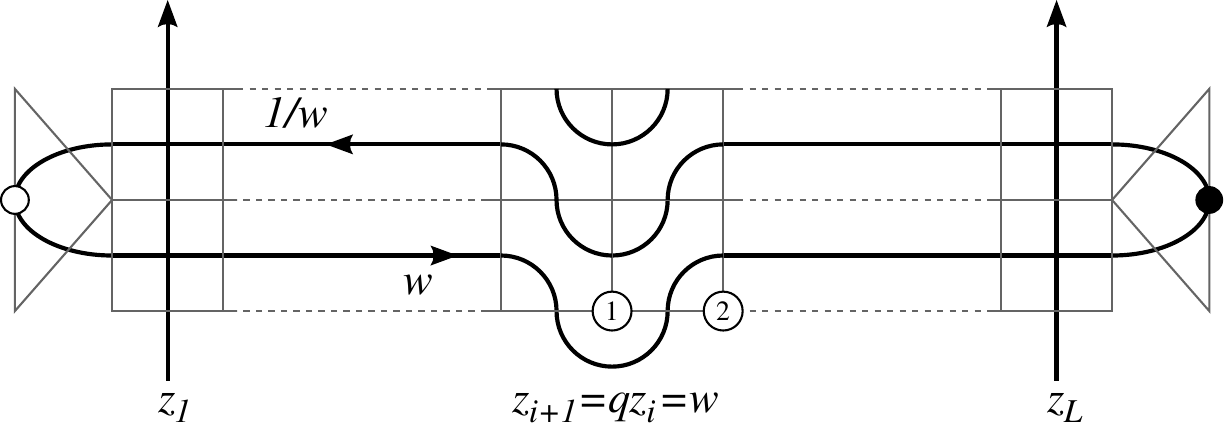}}\quad.\]
Again the marker $x_1$ can be moved along the loop, and as in \propref{Trecur} (remembering that the chosen specialisation for $q$ makes the proportionality factor $1$), this leaves
\[ \raisebox{-45pt}{\includegraphics[height=100pt]{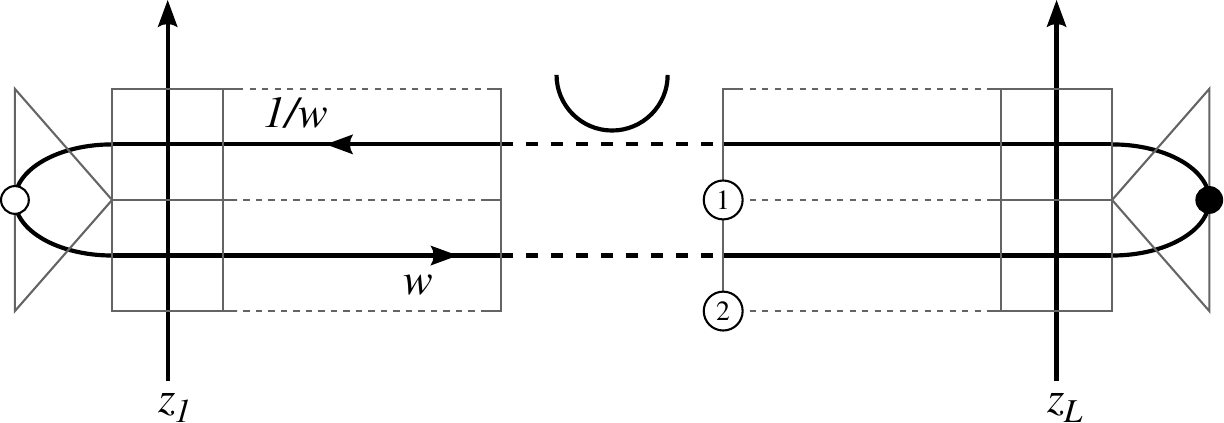}}\quad, \]
which is exactly $\varphi_{i}$ acting on $\widehat{Y}_{L-2}(w;\ldots,\hat{z}_{i},\hat{z}_{i+1},\ldots)$. The proof of the second relation is analagous.
\end{proof}
\section{Symmetries}
\label{sec:symm}
The expectation values $Y_L$ and $X_L^{(k)}$ have a number of symmetries, most arising from properties of the transfer matrix. In this section we will describe these symmetries.
\begin{lm}
\label{lm:Xinterlace}
It is easily seen that the interlacing conditions listed in \eqref{eq:interlace} still hold to the left and right of the markers in the operators $\widehat X_L^{(k)}$ or $\widehat Y_L^{(k)}$, for example
\[
\check R(z_i/z_{i+1})\widehat{X}^{(k)}_L(\ldots,z_i,z_{i+1},\ldots) = \widehat{X}^{(k)}_L(\ldots,z_{i+1},z_i,\ldots)\check R(z_i/z_{i+1}),
\]
for all values $k\neq i,i+1$.
\end{lm}
Recall that the ground state eigenvector $\ket\Psi$ satisfies the $q$KZ equation for open boundaries \eqref{eq:qKZTL_TypeC2}, explicitly
\begin{align}
\check R(z_i/z_{i+1})\ket{\Psi(\ldots,z_i,z_{i+1},\ldots)}&=\ket{\Psi(\ldots,z_{i+1},z_i,\ldots)},\nn\\
\check K_0(z_1^{-1},\zeta_0)\ket{\Psi(z_1,\ldots)}&=\ket{\Psi(z_1^{-1},\ldots)},
\label{eq:qKZ}\\
\check K_L(z_L,\zeta_L)\ket{\Psi(\ldots,z_L)}&=\ket{\Psi(\ldots,z_L^{-1})}.\nn
\end{align}
Similarly, the vector $\langle{\widetilde \Psi}|$ satisfies
\begin{align}
\langle{\widetilde \Psi(\ldots,z_i,z_{i+1},\ldots)}| \check R(z_{i+1}/z_i)&=\langle{\widetilde \Psi(\ldots,z_{i+1},z_i,\ldots)}|,\nn\\
\langle{\widetilde \Psi(z_1,\ldots)}|\check K_0(z_1,\zeta_0)&=\langle{\widetilde \Psi(z_1^{-1},\ldots)}|,
\label{eq:qKZ*}\\
\langle{\widetilde \Psi(\ldots,z_L)}|\check K_L(z_L^{-1},\zeta_L)&=\langle{\widetilde \Psi(\ldots,z_L^{-1})}|.\nn
\end{align}
These properties will be used to prove the two propositions below.
\begin{prop}
\label{prop:Xsym}
$X_L^{(k)}$ has the following symmetries:
\begin{align*}
X_L^{(k)}(\ldots,z_i,z_{i+1},\ldots)&= X_L^{(k)}(\ldots,z_{i+1},z_{i},\ldots),\qquad\text{for}\;k\neq i,i+1,\\
X_L^{(k)}(z_1,\ldots)&= X_L^{(k)}(z_1^{-1},\ldots),\qquad\text{for}\;k\neq 1,\\
X_L^{(k)}(\ldots,z_L)&=X_L^{(k)}(\ldots,z_L^{-1}),\qquad\text{for}\;k\neq L.
\end{align*}
Therefore, $X^{(k)}_L\in\mathbb{C}[z_1^{\pm},\ldots,z_{k-1}^{\pm}]^{\mathcal W_{\rm B}}$, i.e., regarding the variables $z_k,\ldots,z_L$ as complex numbers, $X^{(k)}_L$ is invariant under the action of the Weyl group of type $B_{k-1}$ and thus is symmetric in $\cup_{j=1}^{k-1}\{z_j\}$ as well as invariant under $z_j \leftrightarrow 1/z_j$, $\forall 1\leq j\leq k-1$. Similarly, we also have that $X^{(k)}_L\in\mathbb{C}[z_{k+1}^{\pm},\ldots,z_{L}^{\pm}]^{\mathcal W_{\rm B}}$.
\end{prop}
\begin{proof}
For $k\neq i,i+1$, we use the unitarity relation \eqref{eq:RKcheckId} of the $R$-matrix to insert an identity into the numerator of $X_L^{(k)}$, and by the interlacing condition in \lmref{Xinterlace} and the $q$KZ equations \eqref{eq:qKZ} and \eqref{eq:qKZ*},
\begin{align*}
\langle{\widetilde \Psi}| \widehat{X}^{(k)}_L\ket\Psi_{\rm LP} &=\langle{\widetilde \Psi}| \check R(z_{i+1}/z_i)\check R(z_i/z_{i+1})\widehat{X}^{(k)}_L\ket\Psi_{\rm LP}\\
&=\langle{\widetilde \Psi}|\check R(z_{i+1}/z_i)\widehat{X}^{(k)}_L(z_{i+1},z_i)\check R(z_i/z_{i+1})\ket\Psi_{\rm LP}\\
&=\langle{\widetilde \Psi(z_{i+1},z_i)}|\widehat{X}^{(k)}_L(z_{i+1},z_i)\ket{\Psi(z_{i+1},z_i)}_{\rm LP}.
\end{align*}
Since the denominator of $X_L^{(k)}$ is symmetric, we have shown that $X_L^{(k)}$ is also symmetric under $z_i\leftrightarrow z_{i+1}$.

Similarly, for $k\neq 1$ and $k\neq L$ respectively,
\begin{align*}
\langle{\widetilde \Psi}| \widehat{X}^{(k)}_L\ket\Psi_{\rm LP}  &= \langle{\widetilde \Psi}| \check K_0(z_1,\zeta_0)\check K_0(z_1^{-1},\zeta_0)\widehat{X}^{(k)}_L\ket\Psi_{\rm LP} \\
&= \langle{\widetilde \Psi(z_1^{-1},\ldots,)}|\widehat{X}^{(k)}_L(z_1^{-1},\ldots) \check K_0(z_1^{-1},\zeta_0)\ket\Psi_{\rm LP} \\
&= \langle{\widetilde \Psi(z_1^{-1},\ldots,)}|\widehat{X}^{(k)}_L(z_1^{-1},\ldots)\ket{\Psi(z_1^{-1},\ldots,)}_{\rm LP},\\
\langle{\widetilde \Psi}| \widehat{X}^{(k)}_L\ket\Psi_{\rm LP} &= \langle{\widetilde \Psi(\ldots,z_L^{-1})}|\widehat{X}^{(k)}_L(\ldots,,z_L^{-1})\ket{\Psi(\ldots,z_L^{-1})}_{\rm LP}.
\end{align*}
Again, the denominator of $X_L^{(k)}$ is invariant under $z_1\rightarrow 1/z_1$ and $z_L\rightarrow 1/z_L$, so the result is proved.
\end{proof}

\begin{prop}
\label{prop:Ysym}
$Y_L^{(k)}$ has the following symmetries:
\begin{align*}
Y_L^{(k)}(\ldots,z_i,z_{i+1},\ldots)&= Y_L^{(k)}(\ldots,z_{i+1},z_{i},\ldots),\qquad\text{for}\;k\neq i+1,\\
Y_L^{(k)}(z_1,\ldots)&= Y_L^{(k)}(z_1^{-1},\ldots),\qquad\text{for}\;k\neq 1,\\
Y_L^{(k)}(\ldots,z_L)&=Y_L^{(k)}(\ldots,z_L^{-1}),\qquad\text{for}\;k\neq L+1.
\end{align*}
Therefore we have that $Y^{(k)}_L\in\mathbb{C}[z_1^{\pm},\ldots,z_{k-1}^{\pm}]^{\mathcal W_{\rm B}} \cap \mathbb{C}[z_{k}^{\pm},\ldots,z_{L}^{\pm}]^{\mathcal W_{\rm B}}$.
\end{prop}
\begin{proof}
The proof is exactly analagous to that of \propref{Xsym}.
\end{proof}
In the next section we will deduce further symmetries of $Y_L^{(k)}$. Then in \secref{symmX} these symmetries of $Y_L$ will be used to find further symmetries of $X_L^{(k)}$.
\subsection{Symmetries of $Y_L$}
\label{sec:symmY}
\begin{prop}
\label{prop:Ysym2}
$Y_L^{(k)}$ is independent of $k$, and is symmetric in the combined set of inhomogeneities, $Y^{(k)}_L\in\mathbb{C}[z_1^{\pm},\ldots,z_{L}^{\pm}]^{\mathcal W_{\rm B}}$.
\end{prop}
\begin{proof}
Consider an infinite lattice made up of double-row transfer matrices. Let $P$ be the set of all paths $p$ that start from the left boundary and end at the right. Since $Y_L^{(k)}$ depends only on the spectral parameter of one double-row transfer matrix, we can assume that the internal spectral parameters of all transfer matrices are equal. Let us call this common value $w$. Therefore, a path is independent of its vertical position and we shall abuse notation to also denote by $p$ the equivalence class of all vertical translates of a particular path $p$. The weight $\omega_p$ of each path is the product of the Boltzmann weights associated with the local path orientations from each $R$ and $K$-matrix involved in fixing $p$.

The set $P$ can be split into two mutually exclusive subsets: $P_{\rm t}$, the set of paths that start in the top half of the $K$-matrix on the left, and $P_{\rm b}$, the set of paths that start in the bottom half. These sets have a $1$--$1$ correspondence: each path $p\in P_{\rm t}$ can be transformed into a path $\tilde p\in P_{\rm b}$ by a vertical flip. The weights of the two paths are related by $\omega_p(z_1,\ldots,z_L)=\omega_{\tilde p}(z_1^{-1},\ldots,z_L^{-1})$.

Consider now $Y_L^{(k)}, k\in\{1,\ldots,L+1\}$, and a path $p\in P_{\rm b}$. Because we must consider all of the possible vertical positions of $p$, it may have multiple contributions to $Y_L$. The number of contributions $m_p$ depends on the number of times the path crosses back and forth over the vertical line on which the markers of $\widehat Y_L$ are placed. It can be seen that when $k$ is odd, all paths in $P_{\rm b}$ contribute positively, because each path enters $\widehat Y_L^{(k)}$ from the left. They all contribute negatively, i.e., enter from the right, when $k$ is even. The opposite is true for the paths in $P_{\rm t}$.

For example, for $k=1$, the path $p_1\in P_{\rm b}$ in \figref{pathsa} has one vertical translate that also winds through the two markers, hence its multiplicity $m_{p_1}=2$. The sister path $\tilde p_1$, depicted in \figref{pathsb}, is obtained by flipping the complete configuration (but not the markers), and has $m_{\tilde p_1}=1$.
\begin{figure}[ht]
\centering{
\subfigure[path in $P_{\rm b}$]{
\pic{0pt}{70pt}{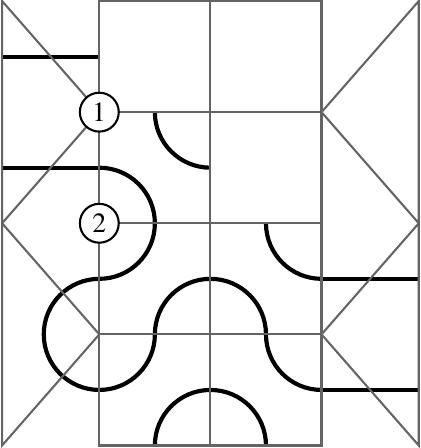}
\label{fig:pathsa}
}
\hspace{1cm}
\subfigure[path in $P_{\rm t}$]{
\pic{0pt}{70pt}{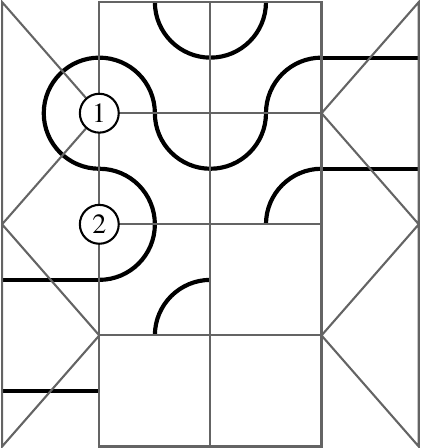}
\label{fig:pathsb}
}
}
\caption{Paths contributing to $Y^{(1)}_2$.}
\end{figure}

Since a path must cross position $k$ from left to right exactly once more than it crosses from right to left, there is always one more path contributing positively than negatively. As a consequence of this, for $p\in P_{\rm b}$ and its sister path $\tilde p\in P_{\rm t}$, it holds that $m_{\tilde p}=m_p-1$ when $k$ is odd and $m_{\tilde p}=m_p+1$ when $k$ is even. Consequently, $Y_L$ is given by
\begin{align*}
Y^{(k)}_L(z_1,\ldots,z_L)&=(-1)^{k} \left(\sum_{\tilde p\in P_t} m_{\tilde p}\ \omega_{\tilde p}(z_1,\ldots,z_L)-\sum_{p\in P_b}m_p\ \omega_p(z_1,\ldots,z_L) \right)\\
&=(-1)^k\sum_{p\in P_b} \Big((m_p+(-1)^k)\omega_p(z_1^{-1},\ldots,z_L^{-1})-m_p\ \omega_p(z_1,\ldots,z_L)\Big)\ .
\end{align*}
As we showed in the previous section, $Y^{(k)}_L$ is invariant under $z_i\rightarrow 1/z_i$, for $i<k$ as well as for $i\geq k$, so consider the construction
\begin{align*}
Y^{(k)}_L(z_1,\ldots,z_L) &= \frac12\Big(Y_L^{(k)}(z_1,\ldots,z_L)+Y_L^{(k)}(z_1^{-1},\ldots,z_L^{-1})\Big)\\
&=\frac12(-1)^{k}\sum_{p\in P_b}\Big((m_p+(-1)^k)\omega_p(z_1^{-1},\ldots,z_L^{-1})-m_p\ \omega_p(z_1,\ldots,z_L) \\
&\qquad+(m_p+(-1)^k)\omega_p(z_1,\ldots,z_L)-m_p\ \omega_p(z_1^{-1},\ldots,z_L^{-1}) \Big)\\
&= \frac12 \sum_{p\in P_b}\Big(\omega_p(z_1^{-1},\ldots,z_L^{-1})+\omega_p(z_1,\ldots,z_L)\Big).
\end{align*}
It thus follows that each translational equivalence class of paths $p$ in $P_{\rm b}$ has only one contribution to $Y_L$, which is equal to the average of the weights of $p$ and $\tilde p$. It is also clear from the above that $Y_L^{(k)}$ is in fact independent of $k$. Since $Y_L$ is independent of its horizontal position, there is no longer a restriction on the symmetries, so it is symmetric in $z_i$ and invariant under $z_j\rightarrow 1/z_j$, $\forall i,j$.
\end{proof}
\begin{prop}
\label{prop:Ysym3}
As a function of the spectral parameter $w$, $Y_L$ is symmetric under $w\rightarrow q/w$ and a swapping of boundary parameters,
\[Y_L(w;\zeta_0,\zeta_L;z_1\ldots,z_L) = Y_L(qw^{-1};\zeta_L,\zeta_0;z_1\ldots,z_L).\]
\end{prop}
\begin{proof}
From \lmref{Ttranspose} we know that
\[Y_L(w;\zeta_0,\zeta_L;z_1,\ldots,z_L)=F_L^{\{(k,{\rm t}),(k,{\rm m})\}}(qw^{-1};\zeta_L,\zeta_0;z_L,\ldots,z_1),\]
for some $k$, and by \eqref{eq:YF} this becomes
\[Y_L(w;\zeta_0,\zeta_L;z_1,\ldots,z_L)=Y_L(qw^{-1};\zeta_L,\zeta_0;1/z_L,\ldots,1/z_1).\]
Then use of the symmetries given in \propref{Ysym2} completes the proof.
\end{proof}

\subsection{Symmetries of $X^{(k)}_L$}
\label{sec:symmX}
\begin{prop}
\label{prop:Xsym2}
$X^{(k)}_L$ is an element of $\mathbb{C}[z_1^{\pm},\ldots,\hat{z}_k,\ldots,z_{L}^{\pm}]^{\mathcal W_{\rm B}}$, and has the property $X_L^{(k)}=X_L^{(j)}|_{z_k\leftrightarrow z_j}$.
\end{prop}
\begin{proof}
This comes from the symmetries of $Y_L$ in \propref{Ysym2}, and the second relationship in \propref{recur1}.
\end{proof}
\begin{prop}
\label{prop:Xsym3}
$X^{(k)}_L$ has the antisymmetry
\[X^{(k)}_L(\ldots,z_k,\ldots) = -X^{(k)}_L(\ldots,z_k^{-1},\ldots).\]
\end{prop}
\begin{proof}
If a path $p\in P_b$ has a contribution of $\omega_p(z_1,\ldots,z_L)$ to $X^{(k)}_L$, then its sister path has the contribution $-\omega_p(z_1^{-1},\ldots,z_L^{-1})$. As a result, $X^{(k)}_L$ is antisymmetric when all $z_i$ are sent to $1/z_i$. From the symmetries already established in \propref{Xsym2}, this antisymmetry must be attributed to $z_k\rightarrow 1/z_k$.
\end{proof}
Finally, note the additivity property around an elementary plaquette,
\[
\widehat X_L^{(k)}-\widehat Y_L^{(k+1)}-\widehat F_L^{\{(k,\text m),(k+1,\text m)\}}+\widehat Y_L^{(k)}=0.
\]
The invariance of $Y_L$ implies that the operator $\widehat F_L^{\{(k,\text m),(k+1,\text m)\}}$, with the markers in the middle of the transfer matrix, has the same expectation value as the operator $\widehat X_L^{(k)}$.
\section{Recursion relations}
\label{sec:recurXY}
The transfer matrix allows us to deduce recursion relations that the expectation values should satisfy, and the symmetries listed in the previous section can be used to generalise these relations. In this section, we will list relations that are satisfied by the expectation values $X_L^{(k)}$ and $Y_L$, in order to show that there are enough such relations, given their respective degrees, to determine these polynomials. Most of these relations are based on the recursions satisfied by the transfer matrix and the normalisation, as described in \secref{recur}.

\begin{prop}
\label{prop:Frecur}
If the markers in the observable $\widehat F_L$ are located sufficiently far away from the region where we specialise $z_i$, the following recursions hold,
\begin{align*}
F_L(z_{i+1}=qz_i)&=F_{L-2}(\hat z_i,\hat z_{i+1}),\\
F_L(\zeta_0;z_1=q\zeta_0)&=F_{L-1}(q\zeta_0;\hat z_1),\\
F_L(\zeta_L;z_L=q\zeta_L^{-1})&=F_{L-1}(q\zeta_L^{-1};\hat z_L).
\end{align*}
For $Y_L$, these properties always hold, as by \propref{Ysym2} we can move the markers horizontally away from the specialised position. For $X_L^{(k)}$, the first recursion holds if $k\neq i,i+1$, and the second and third hold if $k\neq 1$ and $k\neq L$ respectively.
\end{prop}
The proof of these recursions given in \cite{dGNP10} did not describe the recursion of the vector $\langle\widetilde\Psi|$. Here the full proof is given.
\begin{proof}
We will prove here the first recursion. The other two are similarly proved, but also rely on the fact that the components of the eigenvector are invariant under $\zeta_0\rightarrow \zeta_0^{-1}$ and $\zeta_L\rightarrow \zeta_L^{-1}$, see \eqref{eq:psidef1}, \eqref{eq:psidef2}, \eqref{eq:Symdef}, and the definition of the $q$KZ equation in \secref{qKZ}.

From \secref{estaterecur}, the eigenvector $\ket\Psi$ has a recursion at the point $z_{i+1}=q z_i$,
\[ \ket{\Psi_L(z_{i+1}=qz_i)}=p(z_i;z_1,\ldots,\hat z_i,\hat z_{i+1},\ldots,z_L)\ \varphi_i\ket{\Psi_{L-2}(\hat z_i,\hat z_{i+1})}.\]
From \defref{Fhat}, it is clear that the observable $\widehat F_L$ satisfies the same recursions as the transfer matrix in \eqref{eq:Trecur}, \eqref{eq:0boundrecur} and \eqref{eq:Lboundrecur},
\[ \widehat F_L(z_{i+1}=qz_i)\circ\varphi_i=\varphi_i\circ \widehat F_{L-2}(\hat z_i,\hat z_{i+1}),\]
provided that the markers are not too close to $i$, as stated in the proposition. It remains to show
\be
\label{eq:dualrecur}
\langle{\widetilde \Psi_L(z_{i+1}=qz_i)}|\ \varphi_i\propto \langle{\widetilde \Psi_{L-2}(\hat z_i,\hat z_{i+1})}|.
\ee

We define $\varphi_i^\dag$ to be a small upward link from position $i$ to $i+1$, acting on a downward link pattern. We will prove here the property
\[ \varphi_i^\dag\ket{\Psi_L(z_i=qz_{i+1})} \propto \ket{\Psi_{L-2}(\hat z_i,\hat z_{i+1})},\]
and then call on \lmref{dual} to prove \eqref{eq:dualrecur}.

The $q$KZ equation \eqref{eq:qKZea} can be written explicitly as
\[ e_i\ket\Psi=\sum_{\alpha}\psi_\alpha\ e_i\ket\alpha=\sum_{\alpha}\left(-a_i\psi_\alpha \right)\ket\alpha, \]
and from \secref{genLsoln}, we know that the only non-zero terms on the RHS are those where $\alpha=\varphi_i\beta$ for some size $L-2$ link pattern $\beta$. Thus,
\[ \sum_{\alpha}\psi_\alpha\ e_i\ket\alpha=\sum_{\beta}\left(-a_i\psi_{\varphi_i\beta} \right)\ket{\varphi_i\beta}, \]
where every link pattern on the RHS has a little link from $i$ to $i+1$.

Another way of thinking of $\varphi_i^\dag$ is as the bottom half of an $e_i$ operator, which acts on $\ket\alpha\in{\rm LP}_L$ to give $\ket{\alpha'}\in{\rm LP}_{L-2}$, where $\varphi_i\alpha'=e_i\alpha$. By acting on $\ket\Psi$ with $\varphi_i^\dag$ instead of $e_i$, the little link from $i$ to $i+1$ is not introduced, so in the place of the $q$KZ equation we have
\[ \sum_{\alpha}\psi_\alpha\ \varphi_i^t \ket\alpha=\sum_{\beta}\left(-a_i\psi_{\varphi_i\beta} \right)\ket{\beta}, \]
and we want to show that when specified to $z_i=qz_{i+1}$, the RHS of this equation is proportional to the ground state eigenvector $\ket{\Psi_{L-2}(\hat z_i,\hat z_{i+1})}$. Recalling the definitions of $a_i$ in \eqref{eq:adef}, we have
\[ -a_i\psi^L_{\varphi_i\beta}(z_i,z_{i+1})=\frac{[z_{i+1}/qz_i]}{[z_i/z_{i+1}]}\psi^L_{\varphi_i\beta}(z_{i+1},z_i)+\frac{[qz_{i+1}/z_i]}{[z_i/z_{i+1}]}\psi^L_{\varphi_i\beta}(z_i,z_{i+1}). \]
Specialising $z_i=qz_{i+1}$, the second term disappears. In the first, the coefficient becomes $1$ (recall $q^3=1$), and the first component recursion from \lmref{comprecur} can be applied  to get
\[ \left.-a_i\psi^L_{\varphi_i\beta}(z_i,z_{i+1})\right|_{z_i=qz_{i+1}}=p(z_{i+1};\ldots,\hat z_i,\hat z_{i+1},\ldots)\ \psi^{L-2}_\beta(\hat z_i,\hat z_{i+1}), \]
and thus,
\[ \varphi_i^t\ket{\Psi_L(q z_{i+1},z_{i+1})}=p(z_{i+1};\ldots,\hat z_i,\hat z_{i+1},\ldots)\ket{\Psi_{L-2}(\hat z_i,\hat z_{i+1})}. \]
Using \lmref{dual}, this result implies \eqref{eq:dualrecur}, where the proportionality factor is $p(z_{i})$.

Finally, the proportionality factors obtained in the numerator and denominator of $F_L$ cancel, resulting in the desired recursion.
\end{proof}
A description of the derivation of the bulk and left boundary recursions for the transfer matrix can be found in \apref{recursion}.

\subsection{Relations for $Y_L$}
\begin{prop}
\label{prop:Yrecur}
$Y_L$ has the following $4L-4$ bulk recursions,
\[Y_L\Big|_{z_i=q^{\pm1}z_j^{\pm1}}=Y_{L-2}(\ldots,\hat z_i,\ldots,\hat z_j,\ldots),\qquad j\neq i,\]
and for the boundaries there are $4$ recursions,
\begin{align*}
Y_L(\zeta_0;\ldots,z_i,\ldots)\Big|_{z_i=(q\zeta_0)^{\pm1}}&=Y_{L-1}((q\zeta_0)^{\pm1};\ldots,\hat z_i,\ldots),\\
Y_L(\zeta_L;\ldots,z_i,\ldots)\Big|_{z_i=(\zeta_L/q)^{\pm1}}&=Y_{L-1}((\zeta_L/q)^{\pm1};\ldots,\hat z_i,\ldots).
\end{align*}
\end{prop}
\begin{proof}
By the symmetries of $Y_L$ in \propref{Ysym2}, the recursions in \propref{Frecur} can be generalised; for instance, the bulk recursions are true with $z_i = q z_j$, not just $z_i = q z_{i-1}$.
\end{proof}
\begin{prop}
\label{prop:YtoX}
We also have the relations
\begin{align*}
Y_L(w;\zeta_0,\zeta_L;z_1,\ldots,z_L)\Big|_{w=z_i^{\pm1}} &= \pm X^{(i)}_L(\zeta_0,\zeta_L;z_1,\ldots,z_L),\\
Y_L(w;\zeta_0,\zeta_L;z_1,\ldots,z_L)\Big|_{w=q z_i^{\pm1}} &= \mp X^{(i)}_L(\zeta_L,\zeta_0;z_1,\ldots,z_L).
\end{align*}
\end{prop}
\begin{proof}
These come from \propref{recur1}, the antisymmetry of $X^{(k)}_L$ in \propref{Xsym3} and the symmetry of $Y_L$ in \propref{Ysym3}.
\end{proof}
Viewing the above relations as specialisations for $z_i$, and assuming that we know $X_L$, $Y_{L-1}$ and $Y_{L-2}$, we thus find from \propref{Yrecur} and \propref{YtoX} the value of $Y_L$ in $4L+4$ values of $z_i$.

\subsection{Relations for $X^{(k)}_L$}
\begin{prop}
\label{prop:Xrecur}
$X_L^{(k)}$ satisfies the $4L-8$ bulk recursions (we have taken $i>k$ without loss of generality),
\begin{align*}
X^{(k)}_L\Big|_{z_i=q^{\pm1}z_j^{\pm1}} &= X^{(k)}_{L-2}(\ldots,\hat z_i,\ldots,\hat z_j,\ldots),\qquad j>k,\\
X^{(k)}_L\Big|_{z_i=q^{\pm1}z_j^{\pm1}} &= X^{(k-1)}_{L-2}(\ldots,\hat z_j,\ldots,\hat z_i,\ldots),\qquad j<k,
\end{align*}
and $4$ recursions for the boundaries,
\begin{align*}
X^{(k)}_L(\zeta_0;\ldots,z_i,\ldots)\Big|_{z_i=(q\zeta_0)^{\pm1}}&=X^{(k)}_{L-1}((q\zeta_0)^{\pm1};\ldots,\hat z_i,\ldots),\\
X^{(k)}_L(\zeta_L;\ldots,z_i,\ldots)\Big|_{z_i=(\zeta_L/q)^{\pm1}}&=X^{(k)}_{L-1}((\zeta_L/q)^{\pm1};\ldots,\hat z_i,\ldots).
\end{align*}
\end{prop}
\begin{proof}
Again, the recursions in \propref{Frecur} can be generalised by the symmetries of $X_L^{(k)}$ in \propref{Xsym2}, provided $i,j\neq k$.
\end{proof}
\begin{prop}
\label{prop:XtoY}
$X_L^{(k)}$ also satisfies the relations
\begin{align*}
X^{(k)}_L\Big|_{z_i=(z_k/q)^{\pm1}} &=Y_{L-2}(z_k;\hat z_i,\hat z_k),\\
X^{(k)}_L\Big|_{z_i=(qz_k)^{\pm1}} &=-Y_{L-2}(z_k^{-1};\hat z_i,\hat z_k),
\end{align*}
provided $k\neq i$.
\end{prop}
\begin{proof}
These come from \propref{recur1}, and the symmetry of $Y_L$ under $w\rightarrow q/w$.
\end{proof}
Therefore, assuming we know $Y_{L-2}$, $X^{(k)}_{L-2}$ and $X^{(k)}_{L-1}$, we know the value of $X^{(k)}_L$, viewed as a polynomial in $z_i$, in $4L$ points.

\section{Degrees of $\widehat Y_L$ and $\widehat X_L$}
\label{sec:degree}
We recall definition \eqref{eq:Expectation} of the expectation value
\[ F_L^{\{x_1,x_2\}}=\frac{1}{Z_L^2} \langle{\widetilde \Psi}|\ \widehat{F}_L^{\{x_1,x_2\}}\ \ket{\Psi}_{\rm LP}. \]
\begin{conj}
\label{conj:factor}
The numerator of $F_L$ contains a factor $Z_L$, cancelling one of such factors in the denominator. Equivalently, $F_L$ is of the form $P/Z_L$, where $P$ is a polynomial.
\end{conj}
For small systems we have observed this to be true, but it is still an outstanding problem for general $L$.

\begin{prop}
\label{prop:degree}
Conditioned on \conjref{factor}, the numerator of $X^{(k)}_L$ must be of degree width $4L-2$ in each variable, and the numerator of $Y_L$ must be of degree width $4L+2$. Thus we have enough recursion relations, listed in the previous section, to fix the numerators of both expectation values.
\end{prop}
\begin{proof}
For both expectation values $X^{(k)}_L$ and $Y_L$, the degree width of the numerator must equal the degree width of the denominator. The degree width of $Z_L$ is $4L-2$ in each variable $z_i$. By the above conjecture, this means for $X^{(k)}_L$ that the numerator must also be of degree width $4L-2$ in each variable.

For $Y_L$, we must also take into account the action of the transfer matrix. The numerator and denominator of each term in the transfer matrix have degree width $4$ in each variable. We can factor the denominator out, so the degree width of both the numerator and the denominator of $Y_L$ are now $4L+2$.
\end{proof}

\section{Solution}
\label{sec:flowsoln}
Before we present the exact expressions for $X_L^{(k)}$ and $Y_L$, we must introduce some auxiliary functions.
\begin{defn}
\label{def:Sdef}
With a slight change in notation from \eqref{eq:tau}, let
\[ \tau_L(z_1,\ldots,z_L) = \chi_{\lambda^{(L)}}(z_1^2,\ldots,z_L^2),\qquad \lambda^{(L)}_j = \left\lfloor \frac{L-j}{2}\right\rfloor, \]
where the symplectic character $\chi_\lambda$ is defined in \eqref{eq:sympchar}.
\end{defn}
\begin{defn}
\label{def:u}
We also define
\[ u_L(\zeta_0,\zeta_L;z_1,\ldots,z_L)=c_L \log \left[ \frac{\tau_{L+1}(\zeta_0,z_1,\ldots,z_L)\ \tau_{L+1}(\zeta_L,z_1,\ldots,z_L)} {\tau_{L}(z_1,\ldots,z_L)\ \tau_{L+2}(\zeta_0,\zeta_L,z_1,\ldots,z_L)} \right], \]
where the constant $c_L$ is given by
\[
c_L=(-1)^L \ii \frac{\sqrt{3}}{2}.
\]
\end{defn}
It is interesting to note that the function $u_L$ has the form of a Toda lattice wave function; we refer the interested reader to \cite{JimboM83}.

\begin{thm}
\label{thm:flowresult}
Conditioned on \conjref{factor} and \conjref{wqwsymm} (below), the expectation values $X_L^{(k)}$ and $Y_L$ are given by
\begin{align}
X^{(k)}_L &= \ z_k \frac{\partial}{\partial z_k}\ u_L(\zeta_0,\zeta_L;z_1,\ldots,z_L),
\label{eq:Xval}\\
Y_L & = \left.\ w \frac{\partial}{\partial w}\ u_{L+2}(\zeta_0,\zeta_L;z_1,\ldots,z_L,vq^{-1},w)\right|_{v=w},
\label{eq:Yval}
\end{align}
where $v$ is a dummy variable.
\end{thm}
This will be proved in the next section.

\subsection{Proof of the main result}
It is easy to see that the degree of the numerators of \eqref{eq:Xval} and \eqref{eq:Yval} are $4L-2$ and $4L+2$ respectively, as required by \propref{degree}. In addition, the denominator of \eqref{eq:Xval} is $Z_L$, and the denominator of \eqref{eq:Yval} is $Z_L$ times the denominator of the transfer matrix. We have also calculated by brute force the expectation values for system sizes $L=1,2$ and $3$ and verified the solution for these cases. Thus, in order to prove that these expressions are the correct forms for $X^{(k)}_L$ and $Y_L$ in general, we must prove that they satisfy all the required recursions as set out in \secref{recurXY}.
\begin{lm}
\label{lm:recuru}
The function $u$ has a list of simple recursions, based on the recursion for $\tau$ in \eqref{eq:chirecur}:
\begin{align*}
u_L(\zeta_0,\zeta_L;\ldots,z_{L-2},z,qz) &= u_{L-2}(\zeta_0,\zeta_L;\ldots,z_{L-2}),\\
u_L(\zeta_0,\zeta_L;z_1=q\zeta_0,z_2,\ldots) &= -u_{L-1}(q\zeta_0,\zeta_L;z_2,\ldots)-c_L\log\left[-k(\zeta_0,\zeta_L )\right ],\\
u_L(\zeta_0,\zeta_L;\ldots,z_{L-1},z_L=q^{-1}\zeta_L) &= -u_{L-1}(\zeta_0,q^{-1}\zeta_L;\ldots,z_{L-1})-c_L\log\left[-k(q^{-1}\zeta_L,\zeta_0 )\right ].
\end{align*}
\end{lm}
\begin{proof}
The first of these is easily obtained by using the recursion \eqref{eq:chirecur} on each of the $\tau$-functions in the definition of $u_L$, and noting that the proportionality factors on the top and bottom cancel exactly. The second recursion we prove now, and the third works in a similar way:
\begin{align*}
u_L(z_1=q\zeta_0) &=c_L \log \left[ \frac{\tau_{L+1}(\zeta_0,q\zeta_0,z_2,\ldots,z_L)\ \tau_{L+1}(\zeta_L,q\zeta_0,z_2,\ldots,z_L)} {\tau_{L}(q\zeta_0,z_2,\ldots,z_L)\ \tau_{L+2}(\zeta_0,\zeta_L,q\zeta_0,z_2,\ldots,z_L)} \right]\\
&=c_L \log \left[ \frac{-1}{k(\zeta_0,\zeta_L)}\frac{\tau_{L-1}(z_2,\ldots,z_L)\ \tau_{L+1}(q\zeta_0,\zeta_L,z_2,\ldots,z_L)} {\tau_{L}(q\zeta_0,z_2,\ldots,z_L)\ \tau_{L}(\zeta_L,z_2,\ldots,z_L)} \right]\\
&=-c_L \log \left[ \frac{\tau_{L}(q\zeta_0,z_2,\ldots,z_L)\ \tau_{L}(\zeta_L,z_2,\ldots,z_L)} {\tau_{L-1}(z_2,\ldots,z_L)\ \tau_{L+1}(q\zeta_0,\zeta_L,z_2,\ldots,z_L)} \right] -c_L \log\left[-k(\zeta_0,\zeta_L)\right]\\
&=-u_{L-1}(q\zeta_0,\zeta_L;z_2,\ldots)-c_L \log\left[-k(\zeta_0,\zeta_L)\right].\qedhere
\end{align*}
\end{proof}
These recursions immediately imply the recursions for $Y_L$ and $X_L^{(k)}$ that do not involve the marked parameter; that is, the $4L$ recursions in \propref{Yrecur} for $Y_L$, as well as the $4L-4$ recursions in \propref{Xrecur} for $X_L$. The remaining recursions, from \propref{YtoX} and \propref{XtoY}, which respectively complete the degree arguments for $Y_L$ and $X_L^{(k)}$, are more complicated. These two recursions are addressed in the next section.

\subsubsection{Relations between $Y_L$ and $X_L^{(k)}$ in \propref{YtoX} and \propref{XtoY}}
\begin{conj}
\label{conj:wqwsymm}
The proposed expression for $Y_L$ in \eqref{eq:Yval} is invariant under the transformation $w\rightarrow q/w$.
\end{conj}
The proof of this in \cite{dGNP10} was found to be flawed, and a correct proof has not yet been found, so we merely assume this symmetry. It can be shown explicitly that $Y_L$ satisfies this symmetry for small $L$ (we have confirmed it for $L=2$ and $3$), and we expect this symmetry to hold for arbitrary $L$.
\begin{prop}
The expressions \eqref{eq:Xval} and \eqref{eq:Yval} satisfy the first relation in \propref{YtoX}, where $w=z_i^{\pm 1}$. Conditioned on \conjref{wqwsymm} they also satisfy the second relation, where $w=qz_i^{\pm 1}$.
\end{prop}
\begin{proof}
Taking expression \eqref{eq:Yval} for $Y_L$ at the point $w=z_i$, and using the first recursion in \lmref{recuru},
\begin{align*}
&\left.w \frac{\partial}{\partial w}\ u_{L+2}(z_1,\ldots,z_i,\ldots,z_L,qv^{-1},w)\right|_{\begin{subarray}{l}v=z_i\\w=z_i\end{subarray}}\\
&=\left.w \frac{\partial}{\partial w}\ u_{L}(z_1,\ldots,\hat z_i,\ldots,z_L,w)\right|_{w=z_i}\\
&=z_i\ \frac{\partial}{\partial z_i}\ u_{L}(z_1,\ldots,z_L),
\end{align*}
which is exactly $X_L^{(i)}$ from \eqref{eq:Xval}. The other relations in \propref{YtoX} follow by the antisymmetry of \eqref{eq:Xval} under $z_i\rightarrow 1/z_i$ and \conjref{wqwsymm}.
\end{proof}
\begin{prop}
The expressions \eqref{eq:Xval} and \eqref{eq:Yval} satisfy the relations listed in \propref{XtoY}.
\end{prop}
\begin{proof}
We will prove the cases where $z_i=z_kq^{-1}$ and $z_i=(qz_k)^{-1}$, and the other two cases follow from symmetry under $z_i\rightarrow z_i^{-1}$. We take the expression for $X_L^{(k)}$ \eqref{eq:Xval} at the point $z_i=z_kq^{-1}$,
\begin{align*}
&\left.z_k \frac{\partial}{\partial z_{k}}\ u_{L}(\ldots,z_i,\ldots,z_k,\ldots)\right|_{z_i=z_kq^{-1}}\\
&= \left.z_k \frac{\partial}{\partial z_k}\ u_{L}(z_1,\ldots,\hat z_i,\ldots,\hat z_k,\ldots,z_L,vq^{-1},z_k)\right|_{v=z_k},
\end{align*}
which is exactly the expression for $Y_L$ in \eqref{eq:Yval} at the point $w=z_k$. Similarly, at $z_i=(q z_k)^{-1}$, \eqref{eq:Xval} becomes
\begin{align*}
&\left.z_k \frac{\partial}{\partial z_{k}}\ u_{L}(\ldots,z_i,\ldots,z_k,\ldots)\right|_{z_i=(qz_k)^{-1}}\\
&= -\left.z_k^{-1} \frac{\partial}{\partial (z_k^{-1})}\ u_{L}(z_1,\ldots,\hat z_i,\ldots,\hat z_k,\ldots,z_L,vq^{-1},z_k^{-1})\right|_{v=z_k^{-1}},
\end{align*}
which is exactly \eqref{eq:Yval} at the point $w=z_k^{-1}$.
\end{proof}
We expect both \conjref{factor} and \conjref{wqwsymm} to hold true for arbitrary system size. Under these conditions \thmref{flowresult} is proved.

\newpage
\chapter{Separation of variables for the symplectic character}
\label{chap:Branching}

In this thesis we have found exact finite size expressions for two quantities related to the two-boundary TL($1$) loop model: the normalisation $Z$ of the ground state eigenvector, see \eqref{eq:Z}; and the boundary-to-boundary correlation function $F$, see \eqref{eq:Xval} and \eqref{eq:Yval}. Both of these quantities are expressed in terms of the polynomial character of the symplectic group, $\chi$, defined in \eqref{eq:sympchar}.

For each version of the TL($1$) loop model with left and right boundary conditions, physical quantities can be expressed in terms of this character. The function $\chi$ depends on the bulk inhomogeneity parameters as well as, if applicable, the parameters corresponding to the boundaries of the lattice. Analogous calculations in the periodic version of the TL($1$) loop model result in physical quantities that can be expressed in terms of the Schur function, a multivariate polynomial well-known in representation theory.

The Schur function, corresponding to the root system of type $A$ (in the classification of Dynkin diagrams), is symmetric in all its variables, and it is the polynomial character of the general linear group. The symplectic character is an analogue of the Schur function for the root system of type $C$; that is, not only is it a symmetric function, but it is also invariant under inversion of its arguments. The boundary loop models have this symmetry because of the effect of the boundaries, which do not exist in the periodic case.

In addition to appearing in solutions of statistical mechanical models, both the symplectic character and the Schur function play an important role in the representation theory of the classical groups \cite{FultonH91}. They are also used as generating functions for counting problems in enumerative combinatorics; for example, each one provides a generalised enumeration of alternating sign matrices with various boundary conditions, as well as plane partitions and rhombus tilings \cite{CiucuK09,Okada06,ElkiesKLP92}.

For all of these applications it is important to understand the asymptotic behaviour of the polynomials, as the number of variables goes to infinity. In the case of the two-boundary TL($1$) loop model, the calculation of the asymptotic limit of the symplectic character will give the thermodynamic limit of the physical quantities $Z$ and $F$. In particular we would like to analyse the effect of the two boundaries in the asymptotic limit, which involves setting the bulk parameters to $1$ and keeping the boundary parameters general.

Okounkov and Olshanski studied the asymptotic limit of type $A$ Jack polynomials and type $BC$ orthogonal polynomials with all but a finite subset of variables set to $1$ \cite{OkouO98,OkouO06}. For both of these problems, the authors considered polynomials with a degree growing linearly with the number of variables. In contrast, the symplectic character that we would like to analyse has a degree that is quadratic in the number of variables (see \eqref{eq:lambda}) and, as far as we are aware, its asymptotic behaviour is an open problem.

With the asymptotic limit described above as our motivation, in this chapter we study a method for the separation of variables (SoV) of symplectic characters, initiated by Sklyanin \cite{Skl88} and extended by Kuznetsov, Mangazeev and Sklyanin \cite{KuzSkl95,KuzSkl98,KuzMS03,KuzSkl06}. This SoV method makes use of the $Q$-operator introduced by Baxter \cite{Bax82} for the solution of the 8-vertex model. The aim of the so-called $Q$-operator method is to transform a chosen multi-variable polynomial into a product of single variable polynomials. Using this method, the problem of the asymptotics will be reduced to that of finding the asymptotics for each factorised part. In order for this approach to be useful the $Q$-operator method must also be reversible, which is one of the main technical hurdles in SoV.

Symmetric polynomials are eigenfunctions of certain multivariate differential operators, or Hamiltonians $\{H_i\}$, which form a quantum integrable system. The $Q$-operator for a quantum integrable system is a quantisation of the B\"acklund transformation for the corresponding classical integrable system. This connection was first found by Pasquier and Gaudin \cite{PasqG92}, who discovered a correspondence between the classical B\"acklund transformation and the $Q$-operator for the periodic Toda lattice. The operator $Q$ can be realised as an integral operator with a simple kernel. In the quasi-classical limit, this kernel turns into the generating function of the canonical transform \cite{KuzSkl98,PasqG92}.

A number of examples are given in \cite{KuzSkl06} for the application of the $Q$-operator method to systems of symmetric polynomials associated with the root system of type $A$. In particular, \cite{KuzSkl06} discusses SoV for Schur functions, being a special case of Jack polynomials. As indicated above, this chapter aims to extend SoV using the $Q$-operator method to the irreducible character of the symplectic group, $\chi_\lambda$, which is the Schur polynomial for the root system of type $C$. The only other study of SoV methods for root systems other than type $A$ that we are aware of is for the open quantum $\rm{sl}(2,\mathbb{R})$ spin chain \cite{DerkachovKM03}.

The key hurdle faced when applying SoV to type $A$ polynomials is the inversion of the method. The inverse of the separating operator has not been found for type $A$ polynomials in general (a notable exception is the Schur function, see \cite{KuzSkl06} for more details), and another method has to be used to reverse the process, known as the `lifting operator'. For the symplectic character, it has not been possible to find an equivalent lifting operator, but the inverse separating operator is easy to find, presumably because of the symplectic character's relationship to the Schur function. This raises some questions about the invertibility of the method for more general type $C$ polynomials.

The key ingredient of the $Q$-operator method is the construction of the separating operator $\mathcal S$ whose action on a polynomial $P_\lambda$ is proportional to a product of single variable polynomials,
\be
\label{eq:SPq}
\big(\mathcal S P_\lambda\big)(x_1,\ldots ,x_L) =P_\lambda(1,\ldots,1)\prod_{i=1}^L q_\lambda (x_i).
\ee
For the case of the symplectic character, as with Schur functions, $\mathcal S$ is invertible. The structure of the method is given in the next section.

\section{The $Q$-operator method}
\label{sec:method}
We consider a quantum integrable system with a commuting set of Hamiltonians, given by differential operators $H_i$ ($i\in\{1,\ldots,L\}$). These have eigenstates equal to the polynomial $P_\lambda$:
\be
\label{eq:hamil}
H_j P_\lambda (x_1,\ldots ,x_L)=h_j(\lambda )P_\lambda (x_1,\ldots ,x_L),
\ee
where the $h_j$ are the corresponding eigenvalues, which depend on the multi-degree $\lambda=(\lambda_1,\ldots,\lambda_L)$ of the polynomial.

The aim of the $Q$-operator method is to find a version of the spectral problem \eqref{eq:hamil} that involves a related factorised polynomial in place of $P_\lambda$. Acting on both sides of \eqref{eq:hamil} with $\mathcal S$,
\[ \mathcal SH_j\mathcal S^{-1}\ \big(\mathcal SP_\lambda\big) (x_1,\ldots ,x_L)=h_j(\lambda )\ \left(\mathcal SP_\lambda\right) (x_1,\ldots ,x_L),\]
and using $\eqref{eq:SPq}$,
\[ \left (\mathcal SH_j\mathcal S^{-1}\right ) \prod_{i=1}^L q_\lambda (x_i)=h_j(\lambda )\prod_{i=1}^L q_\lambda (x_i).\]

The main results of this chapter are the explicit construction of the operators $\mathcal S$ and $\mathcal S^{-1}$ for the symplectic characters, as well as a `factorised Hamiltonian' $\widetilde H_j$, which acts in the same way as $\mathcal SH_j\mathcal S^{-1}$ on the factorised polynomial. $\widetilde H_j$ is not uniquely defined, and can be constructed using $q_\lambda$ as shown in the next section.

\subsection{Factorised Hamiltonian}
\label{sec:fachamil}
The operator $\widetilde H_j$ can be constructed from a single variable operator in the following way. If there exists a differential operator in $x_i$, denoted $W_i$, such that
\be
\label{eq:Wq0}
W_i q_\lambda (x_i)=0,\qquad [W_i\ , q_\lambda (x_k)]=0\quad (k\neq i),
\ee
then we construct the operator
\[ W_{i,j}=W_i+h_j(\lambda ) .\]
Noting that $W_{i,j}$ commutes with any function of $x_k$ where $k\neq i$, it is easy to see that
\be
\label{eq:Wqhq}
W_{i,j} \prod_{k=1}^L q_\lambda (x_k)=h_j(\lambda )\prod_{k=1}^L q_\lambda (x_k),\qquad \forall i,
\ee
and therefore any linear combination of the form
\[\widetilde H_j=\sum_{i=1}^L c_i W_{i,j},\qquad\qquad \sum_{i=1}^L c_i=1,\]
will also satisfy \eqref{eq:Wqhq}.

\subsection{Factorisation of the separating operator $\mathcal S$}
We will show that $\mathcal S$ can be written as a product of operators,
\be
\label{eq:SQQ}
\mathcal S= \big(\rho _0Q_{z_1}\ldots Q_{z_L}\big)\big|_{z_1=x_1,\ldots,z_L=x_L},
\ee
where the operator $Q_{z_i}$ is an integral operator that acts as
\be
\label{eq:QPqP}
\big(Q_{z_i}P_\lambda\big)(x_1,\ldots ,x_L)=q_\lambda (z_i) P_\lambda (x_1,\ldots ,x_L),
\ee
and $\rho_0$ sends $f(x_1,\ldots,x_L)\rightarrow f(1,\ldots,1)$. Furthermore, the $Q_{z_i}$ have the important properties
\[ \left[Q_{z_i},Q_{z_j}\right ]=0,\qquad \left[Q_{z_i},H_j\right ]=0,\qquad\forall i,j.\]
Having found the operator $Q_{z_i}$, we introduce new operators $\mathcal A_i$ for which
\[ \big(\rho_{i-1}Q_{z_i}\big)\big|_{z_i=x_i}=\mathcal A_i \rho _i.\]
Here, $\rho_j$ sends $f(x_1,\ldots,x_L)\rightarrow f(x_1,\ldots,x_j,1,\ldots,1)$. These new operators act as
\be
\label{eq:APqP}
\big(\mathcal A_iP_\lambda\big) (x_1,\ldots ,x_i,1,\ldots ,1)=q_\lambda (x_i)P_\lambda (x_1,\ldots ,x_{i-1},1\ldots ,1),
\ee
and \eqref{eq:SQQ} becomes the factorisation
\[ \mathcal S=\mathcal A_1\ldots\mathcal A_L.\]

\subsection{Summary}
Given a particular family of polynomials, the $Q$-operator method can be condensed into five steps:
\vspace{.3cm}\\
\begin{minipage}{\textwidth}
\begin{enumerate*}
\item{Specify $H_j$ and $h_j(\lambda)$ in \eqref{eq:hamil},}
\item{Specify $W_i$ satisfying \eqref{eq:Wq0}, and construct the factorised Hamiltonian,}
\item{Construct $\mathcal S^{-1}$,}
\item{Find $Q_z$ such that \eqref{eq:QPqP} is satisfied,}
\item{Find $\mathcal A_i$ such that \eqref{eq:APqP} is satisfied, and construct $\mathcal S$.}
\end{enumerate*}
\end{minipage}\vspace{.3cm}

The inverse separating operator is given before the construction of $\mathcal S$ for two reasons; firstly, for the symplectic character it turns out to be simpler in form, and secondly, the details relate to those of the differential operators $H$ and $W$ used in the first two steps.

As previously stated, this chapter focuses on the case where $P_\lambda$ is equal to $\chi_\lambda$, the symplectic character, which will be defined in \secref{symp}. For this case, we closely follow \cite{KuzSkl06}, in which the factorisation of the Schur functions is treated. The technical detail in the symplectic case however is more elaborate.

\secref{symp} describes the symplectic character and some of its properties. The sections following proceed in the order of the steps above. \secref{invS} also describes the inverse of the operator $\mathcal S_k=\mathcal A_1\ldots\mathcal A_k$, which satisfies
\[ \mathcal S_k^{-1}\prod_{i=1}^k q_\lambda (x_i)=\frac{\chi_\lambda (x_1,\ldots ,x_k,1,\ldots ,1)}{\chi_\lambda (1,\ldots ,1)}.\]

\section{Properties of the symplectic character}
\label{sec:symp}

The polynomials we will consider, defined in \eqref{eq:sympchar}, are the irreducible characters of the symplectic group. These form a basis of $\mathbb C[x_1^{\pm},\ldots,x_L^{\pm}]^{\mathcal W_{\rm B}}$, the ring of Laurent polynomials with complex coefficients invariant under the action of the Weyl group of type $B_L$, i.e., polynomials symmetric in $x_i$ and invariant under $x_i\rightarrow x_i^{-1}$. Each character is labelled by a partition $\lambda=(\lambda_1,\ldots,\lambda_L)$, with $\lambda_i\geq\lambda_{i+1}$. We refer to them as symplectic characters \cite{Weyl39,FultonH91}, and for the purposes of this chapter define them as follows,
\begin{defn}
\[ \chi _\lambda (x_1,\ldots ,x_L)=\frac{a_\mu (x_1,\ldots ,x_L)}{a_\delta (x_1,\ldots ,x_L)},\]
where
\[
a_\mu (x_1,\ldots ,x_L)=\dt{x_i^{\mu _j}-x_i^{-\mu _j}},
\]
$\delta$ is defined as the partition $(L,\ldots ,L-i+1,\ldots ,2,1)$, and we use $\mu$ to denote the partition $\lambda+\delta$, so $\mu_i=\lambda_i+L-i+1$. The subscript $L$ on the determinant denotes that the matrix is size $L\times L$.
\end{defn}

The polynomials $a_\mu$ are Laurent polynomials in the ring $\mathbb C[x_1^{\pm},\ldots,x_L^{\pm}]$ that are antisymmetric in $(x_1,\ldots,x_L)$ and under $x_i\rightarrow x_i^{-1}$. These antisymmetries mean that $a_\mu(x_1,\ldots,x_L)=0$ whenever $x_i=x_j$ or $x_i=1$. For the definition of $\mathcal A_i$, we will need an expression for $\chi_\lambda$ when some of the arguments are set to 1, so we must find an alternate definition for $\chi_\lambda$ in this limit. We therefore define a `truncated' version of $a_\mu$:
\begin{defn}
\be
\label{eq:atrunc}
a^{(k)}_\mu (x_1,\ldots ,x_{k-1})=\dt{\begin{array}{c}
\left[ x_i^{\mu _j}-x_i^{-\mu _j} \right ]_{i < k} \\[5mm]
\left[\mu _j^{2(L-i)+1}\right ]_{i\geq k}
\end{array}}.
\ee
\end{defn}

\begin{lm}
\label{lm:kprop}
For $i=k,\ldots,L$, set $x_i=\e^{\varepsilon u_i}$ with $u_i\in\mathbb R$. Then
\[
a^{(k)}_\mu(x_1,\ldots,x_{k-1}) \sim c_k(\varepsilon) a_\mu(x_1,\ldots,x_L),
\]
in the limit as $\varepsilon\rightarrow 0$. The prefactor $c_k(\varepsilon)$ does not depend on $\mu$.
\end{lm}

\begin{proof}
As an example, we first prove the above for the case where $k=L$. In this case the parameter $u$ can be set to $1$. In the $L$th row of the determinant $a_\mu(x_1,\ldots,x_{L-1},\e^{\varepsilon})$, the $j$th element is
\begin{align*}
\e^{\varepsilon \mu _j}-\e^{-\varepsilon \mu _j}&=2\sinh{(\varepsilon \mu _j)}\\
&=2\varepsilon \mu _j+ {\rm h.o.t.}
\end{align*}
Taking this to first order in $\varepsilon$, we can factor $2\varepsilon $ out of the determinant, leaving $\mu_j$ in the bottom row. Then
\[a^{(L)}_\mu(x_1,\ldots,x_{L-1})=\lim_{\varepsilon\rightarrow 0}\frac1{2\varepsilon }a_\mu(x_1,\ldots,x_{L-1},\e^\varepsilon),\]
so $c_L(\varepsilon)=1/2\varepsilon $.

In the case of general $k$, the parameters $u_i$ allow us to take multiple arguments to $1$ along distinct trajectories. In the rows from $k$ to $L$ of $a_\mu(x_1,\ldots,x_{k-1},\e^{\varepsilon u_k},\ldots,\e^{\varepsilon u_L})$, we expand up to order $\varepsilon^{2(L-k)+1}$,
\[
2\sinh{(\varepsilon u_i \mu _j)}\approx \sum_{m=0}^{L-k}\frac{2(\varepsilon u_i \mu _j)^{2m+1}}{(2m+1)!}=\sum_{m=k}^{L}\frac{2(\varepsilon u_i \mu _j)^{2(L-m)+1}}{(2(L-m)+1)!}.
\]
The resulting matrix can be shown to be a product of two matrices. We use $\textbf{1}_n$ for the identity matrix of size $n$, and $\textbf{0}$ for a matrix of zeroes whose size is determined from context:
\begin{align*}
&\dt{\begin{array}{c}\left[ x_i^{\mu_j}-x_i^{-\mu_j}\right ] _{i<k}\vspace{.1cm}\\[4mm] \left[ \sum_{m=k}^L \frac{2(\varepsilon u_i\mu_j )^{2(L-m)+1}}{(2(L-m)+1)!}\right ] _{i\geq k}\end{array}}\\[4mm]
&=\det\nolimits_L{\left (\left[\begin{array}{cc}\textbf{1}_{k-1} & \textbf{0}\vspace{.1cm}\\[4mm] \textbf{0} & \left[ \frac{2(\varepsilon u_i)^{2(L-j)+1}}{(2(L-j)+1)!} \right ]_{i,j\geq k}\end{array}\right ]\left[\begin{array}{c}\left[ x_i^{\mu_j}-x_i^{-\mu_j}\right ] _{i<k}\\[4mm] \left[ \mu_j^{2(L-i)+1}\right ] _{i\geq k}\end{array}\right ]\right )}\\[4mm]
&=\frac{2^{L-k+1}\varepsilon^{(L-k+1)^2}}{\prod_{m=k}^{L}(2(L-m)+1)!}\ a^{(1)}_{(u_k,\ldots ,u_L)}\ a^{(k)}_\mu (x_1,\ldots ,x_{k-1}).
\end{align*}
This leads to the result
\begin{align}
\label{eq:aprop}
& a^{(k)}_\mu (x_1,\ldots ,x_{k-1})\\
&\qquad =\lim_{\varepsilon\rightarrow 0}\frac{\varepsilon^{-(L-k+1)^2}\prod_{m=0}^{L-k}(2m+1)!}{2^{L-k+1} a^{(1)}_{(u_k,\ldots ,u_L)}}a_\mu (x_1,\ldots, x_{k-1},\e^{\varepsilon u_k},\ldots ,\e^{\varepsilon u_L}),\nn
\end{align}
so the proportionality factor $c_k(\varepsilon)$ is
\[c_k(\varepsilon)=\frac{\varepsilon^{-(L-k+1)^2}\prod_{m=0}^{L-k}(2m+1)!}{2^{L-k+1} a^{(1)}_{(u_k,\ldots ,u_L)}}.\qedhere\]
\end{proof}
We note that there is an alternative inductive proof similar to that used in \cite{KuzSkl06}, which we give in \apref{altproof}.
\begin{cor}
Because $c_k(\varepsilon)$ is independent of $\mu$, at the point $x_{k+1}=\ldots=x_L=1$ we have an alternate definition of $\chi$,
\[
\chi _\lambda (x_1,\ldots ,x_k,1,\ldots ,1)=\frac{ a^{(k+1)}_\mu (x_1,\ldots ,x_k)}{a^{(k+1)}_\delta (x_1,\ldots ,x_k)}.
\]
\end{cor}
This new definition allows us to directly calculate the symplectic character when some of its arguments are set to $1$.

\subsection{Factorised forms and homogeneous identities}
\label{sec:factor}
This section will list some useful identities and factorised expressions for $a_\mu$ and $\chi_\lambda$. First recall that the denominator of a Schur function is given by the Vandermonde determinant, which has the product form
\be
\label{eq:vdm}
\dt{x_i^{L-j}}=\prod_{1\leq m<n\leq L}(x_m-x_n).
\ee
The Weyl denominator formula for type $C$ gives us an analogous identity for the symplectic character \cite{FultonH91},
\be
\label{eq:ad}a_\delta (x_1,\ldots ,x_L)=\prod _{i=1}^L x_i^{i-L} \prod _{1\leq i<j\leq L}(x_i-x_j)\prod _{1\leq i\leq j\leq L}(x_i-x_j^{-1}).
\ee
We can also use \eqref{eq:vdm} to show
\begin{align}
\label{eq:am1}
a^{(1)}_\mu &=\dt{\mu _i^{2(L-j)+1}} =\prod_{i=1}^L\mu _i\prod_{1\leq i<j\leq L}(\mu _i^2-\mu _j^2).
\end{align}
This result and simple row expansion of \eqref{eq:atrunc} immediately leads to the following identity,
\be
\label{eq:am2}
a^{(2)}_\mu (z)=\sum_{k=1}^L(-1)^{k-1}\left (\prod_{i\neq k}\mu_i\prod_{\stack{1\leq i<j\leq L}{i,j\neq k}}(\mu_i^2 -\mu_j^2 )\right )(z^{\mu_k}-z^{-\mu_k}),
\ee
which will be useful later. Furthermore, from \eqref{eq:ad} and \eqref{eq:aprop}, it is easily shown that
\begin{align}
a^{(k)}_\delta (x_1,\ldots ,x_{k-1})&=\prod _{i=0}^{L-k}(2i+1)!\prod _{i=1}^{k-1}x_i^{i-L}(x_i-1)^{2(L-k+1)}\nonumber \\ &\times\prod _{1\leq i<j\leq k-1}(x_i-x_j)\prod _{1\leq i\leq j\leq k-1}(x_i-x_j^{-1}).
\label{eq:adk}
\end{align}
In particular,
\be
\label{eq:ad1}
a^{(1)}_\delta=\prod_{i=1}^{L-1}(2i+1)!\ ,
\ee
which, along with \eqref{eq:am1}, leads to
\[
\chi_\lambda(1,\ldots,1)=\prod_{i=1}^L\frac{\mu_i}{(2i-1)!}\prod_{1\leq i<j\leq L}(\mu _i^2-\mu _j^2),
\]
which is Weyl's dimension formula for the symplectic group \cite{Weyl39,FultonH91}.

\section{Hamiltonians and eigenvalues}
\label{sec:Hamil}
We now construct the system of mutually commuting Hamiltonians $H_j$ that satisfy the eigenvalue equation \eqref{eq:hamil} in the symplectic case, i.e.,
\be
\label{eq:Hchi}
H_j \chi_\lambda (\mathbf{x})=h_j(\lambda )\chi_\lambda (\mathbf{x}),
\ee
where we have used a bold $\mathbf{x}$ to refer to a list of variables $x_1,\ldots,x_L$.

Let $D_x=x\de{x}$, and recall the definition of the usual elementary symmetric function $e_j$,
\[
e_j(x_1,\ldots ,x_L)=\sum_{1\leq r_1<\ldots <r_j\leq L}x_{r_1}\ldots x_{r_j}.
\]
\begin{lm}
\label{lm:Hamil}
The Hamiltonians $H_j$, for $j=1,\ldots,L$, are given by
\[ H_jf_\lambda (\mathbf x)=(a_\delta (\mathbf x) )^{-1}e_j(D_{x_1}^2,\ldots ,D_{x_L}^2)\ a_\delta (\mathbf x)f_\lambda (\mathbf x),\]
with corresponding eigenvalues
\[ h_j(\lambda )=e_j(\mu _1^2,\ldots ,\mu _L^2),\]
recalling that $\mu=\lambda+\delta$.
\end{lm}
\begin{proof}
The proof that these Hamiltonians satisfy \eqref{eq:Hchi} is equivalent to the proof of
\[ e_j(D_{x_1}^2,\ldots ,D_{x_L}^2)\ a_\mu (\mathbf{x})=e_j(\mu _1^2,\ldots ,\mu _L^2)\ a_\mu (\mathbf{x}).\]
We set $v_i^j=x_i^{\mu_j}-x_i^{-\mu_j}$ and note that $D_{x_i}^2 v_i^j=\mu_j^2\ v_i^j$.
Writing $a_\mu$ as
\[ a_\mu (x_1,\ldots ,x_L)=\sum_{\sigma\in S_L}(-1)^\sigma v_1^{\sigma_1}\ldots v_L^{\sigma_L},\]
we then have
\begin{align*}
e_j(D_{x_1}^2,\ldots ,D_{x_L}^2)\ a_\mu &=\sum_{1\leq r_1<\ldots <r_j\leq L}\left (\sum_{\sigma\in S_L}D_{x_{r_1}}^2\ldots D_{x_{r_j}}^2(-1)^\sigma v_1^{\sigma_1}\ldots v_L^{\sigma_L}\right )\\
&=\sum_{\sigma\in S_L}\left(\sum_{1\leq r_1<\ldots <r_j\leq L}\mu_{\sigma_{r_1}}^2\ldots \mu_{\sigma_{r_j}}^2\right) (-1)^\sigma v_1^{\sigma_1}\ldots v_L^{\sigma_L}.
\end{align*}
Due to symmetry, the inner sum is independent of $\sigma$, and equal to $e_j(\mu _1^2,\ldots ,\mu _L^2)$. This can be taken out of the outer sum, which is equal to $a_\mu$.
\end{proof}

\section{Differential equation for $q_\lambda$}
\label{sec:diff}

In this section we describe the single variable polynomial $q_\lambda(z)$, and define the differential operator $W_i$ that satisfies \eqref{eq:Wq0}.
\begin{defn}
The polynomial $q_\lambda$ is given by
\[
q_\lambda (z)=\frac{\chi_\lambda(z,1,\ldots,1)}{\chi_\lambda(1,\ldots,1)}.
\]
\end{defn}
It will be useful to introduce $\phi_\mu(z_1,\ldots,z_k)$, defined as
\[
\phi_\mu (z_1,\ldots,z_k)=a^{(k+1)}_\mu(z_1,\ldots,z_k)/a^{(1)}_\mu,
\]
and in particular
\begin{align}
\label{eq:phi1}
\phi_\mu (z)&=\sum_{j=1}^L\frac{z^{\mu_j}-z^{-\mu_j}}{\mu_j\prod_{i\neq j}(\mu _j^2-\mu _i^2)},
\end{align}
where we have used the identity \eqref{eq:am2}. Recalling that $\mu=\lambda+\delta$, we can write $q_\lambda (z)$ as
\be
\label{eq:qphi}
q_\lambda (z)=\phi_\mu (z)/\phi_\delta(z).
\ee
\begin{lm}
The polynomial $\phi_\mu(z)$ satisfies the differential equation
\be
\label{eq:DEphi}
\prod_{n=1}^L (D_z^2-\mu _n^2)\phi_\mu (z)=0.
\ee
\end{lm}
\begin{proof}
By definition,
\[ D_z^2 \phi_\mu (z)=\sum_{j=1}^L \mu_j^2\, \frac{(z^{\mu_j}-z^{-\mu_j})}{\mu_j\prod_{i\neq j}(\mu _j^2-\mu _i^2)},\]
so
the $n$th term of the sum in $\phi_\mu$ is reduced to $0$ by the $n$th factor in the product.
\end{proof}

\begin{lm}
The differential equation $Wq_\lambda(z)=0$ is satisfied when
\[
W=\prod_{n=1}^L (Z^2-\mu_n^2),
\]
where
\[ Z=D_z +\frac{(Lz^2+2Lz-2z+L)}{(z^2-1)}.\]
\end{lm}
\begin{proof}
Equation \eqref{eq:DEphi} can be written as
\be
\label{eq:diff2}
\prod_{n=1}^L (D_z^2-\mu _n^2)\ \phi_\delta(z) q_\lambda (z)=0.
\ee
Using \eqref{eq:adk} and \eqref{eq:ad1}, we obtain
\[\phi_\delta(z) =\frac{(z+1)(z-1)^{2L-1}}{z^L(2L-1)!},\]
and applying the derivative,
\begin{align*}
D_z \phi_\delta(z) &=\frac{(z-1)^{2L-2}(Lz^2+2Lz-2z+L)}{z^L(2L-1)!}\\
&=\phi_\delta(z) \frac{(Lz^2+2Lz-2z+L)}{(z^2-1)}.
\end{align*}
Since $D_z$ obeys the same product rule as the usual derivative, we have
\begin{align*}
D_z\ \phi_\delta(z)f(z)&=\phi_\delta(z)\left (D_z+\frac{(Lz^2+2Lz-2z+L)}{(z^2-1)}\right )f(z)\\
&=\phi_\delta(z)Zf(z).
\end{align*}
Substituting this result back into \eqref{eq:diff2} we obtain
\[
0
=\prod_{n=1}^L (Z^2-\mu _n^2)\ q_\lambda (z),
\]
for $z\neq\pm 1$.
\end{proof}

\section{Inverse separating operator}
\label{sec:invS}
The inverse of the operator $\mathcal S=\mathcal A_1\ldots\mathcal A_L$ must satisfy
\[ \mathcal S^{-1}\prod _{i=1}^L q_\lambda (x_i)=\frac{\chi_\lambda (x_1,\ldots ,x_L)}{\chi_\lambda (1,\ldots ,1)},\]
for all $\lambda$.
\begin{prop}
\label{prop:SL}
$\mathcal S^{-1}$ is given by the differential operator
\[ \left(\mathcal S^{-1} f\right)(x_1,\ldots,x_L) =\\ (-1)^{\frac{L(L-1)}{2}} \phi_\delta (x_1,\ldots ,x_L)^{-1}\ K_L\ \prod_{i=1}^L \phi_\delta (x_i)\ f(x_1,\ldots,x_L),\]
where
\[ K_L=\dt{ D_{x_i}^{2(L-j)}}.\]
\end{prop}
\begin{proof}
Acting with the operator $\mathcal S^{-1}$ on the product of $q_\lambda(x_i)$ results, using \eqref{eq:qphi}, in $K_L$ acting on a product of $\phi_\mu(x_i)$. Taking one factor of this product into each row of the determinant $K_L$,
\[ \dt{ D_{x_i}^{2(L-j)} \phi_\mu(x_i) } = \dt{ \sum _{m=1}^L\frac{\mu _m^{2(L-j)+1}(x_i^{\mu _m}-x_i^{-\mu _m})}{\mu _m^2\prod_{n\neq m}(\mu _m^2-\mu _n^2)}},\]
where we have used the explicit expression for $\phi_\mu(x_i)$ given in \eqref{eq:phi1}. This can be expressed as the product of two matrices,
\begin{align*}
\dt{\sum _{m=1}^L\frac{\mu _m^{2(L-j)+1}(x_i^{\mu _m}-x_i^{-\mu _m})}{\mu _m^2\prod_{n\neq m}(\mu _m^2-\mu _n^2)}}&=\dt{\mu_i^{2(L-j)+1}}\dt{\frac{x_i^{\mu_j}-x_i^{-\mu_j}}{\mu_j^2\prod_{n\neq j}(\mu _j^2-\mu _n^2)}}\\
&=\frac{\dt{\mu_i^{2(L-j)+1}}\dt{x_i^{\mu_j}-x_i^{-\mu_j}}}{\prod_{m=1}^L\left (\mu_m^2\prod_{n\neq m}(\mu _m^2-\mu _n^2)\right )}.
\end{align*}
The first determinant in the numerator is just $a^{(1)}_\mu$, see \eqref{eq:am1}, and the second is the definition of $a_\mu(x_1,\ldots,x_L)$. Since the denominator is equal to $(-1)^{\frac{L(L-1)}{2}}(a^{(1)}_\mu)^2$, we finally obtain
\[ K_L\prod_{i=1}^L\phi_\delta (x_i) q_\lambda (x_i)=(-1)^{\frac{L(L-1)}{2}} \phi_\mu (x_1,\ldots ,x_L),\]
from which it immediately follows that $\mathcal S^{-1}$ satisfies the required property.
\end{proof}
We can also find the inverse of the operator $\mathcal S_k=\mathcal A_1\ldots\mathcal A_k$, which must satisfy
\[ \mathcal S_k^{-1}\prod_{i=1}^k q_\lambda (x_i)=\frac{\chi_\lambda (x_1,\ldots ,x_k,1,\ldots ,1)}{\chi_\lambda (1,\ldots ,1)}.\]
This will be useful for calculating the asymptotics of $\chi_\lambda$ with all but a few of its arguments set to one.
\begin{prop}
\label{prop:Sk}
$\mathcal S_k^{-1}$ is given by the differential operator
\[ \left(\mathcal S_k^{-1}f\right)(x_1,\ldots,x_k)=(-1)^{\frac{k(k-1)}{2}}\phi_\delta(x_1,\ldots ,x_k)^{-1} \dt[k]{D_{x_i}^{2(k-j)}}\prod_{i=1}^k \phi_\delta(x_i) f(x_1,\ldots,x_k).\]
\end{prop}
The power of $-1$ here has been corrected from that which appears in the published version of \cite{dGP11}.
\begin{proof}
The proof of this is equivalent to the proof of
\be
\label{eq:redSk}
\dt[k]{D_{x_i}^{2(k-j)}}\prod_{i=1}^k \phi_\mu (x_i)=(-1)^{\frac{k(k-1)}{2}}\phi_\mu(x_1,\ldots ,x_k).
\ee
On the LHS, we insert one factor $\phi_m(x_i)$ into row $i$ of the determinant, so that each element of the matrix becomes
\[ D_{x_i}^{2(k-j)}\phi_\mu (x_i)=\sum_{m=1}^L\mu _m^{2(k-j)}\ \frac{(x_i^{\mu_m}-x_i^{-\mu_m})}{\mu _m\prod_{n\neq m}(\mu _m^2-\mu _n^2)}.\]
Using the proof of \propref{SL} as a guide, we rewrite the RHS as
\[ (-1)^{\frac{k(k-1)}{2}}\frac{ a_\mu^{(1)} a_\mu^{(k+1)}(x_1,\ldots ,x_k)}{\left ( a_\mu^{(1)}\right )^2},\]
and use the product formula for the denominator while using the determinant formula for the numerator, to produce
\begin{align}
\label{eq:messdetk}&(-1)^{\frac{k(k-1)}{2}}\frac{\dt{\begin{array}{c}\left[x_i^{\mu_j}-x_i^{-\mu_j}\right ]_{i\leq k}\\[4mm] \left[\mu _j^{2(L-i)+1}\right ]_{i>k}\end{array}}\dt{\mu _i^{2(L-j)+1}}}{\left (\prod_{m}\mu _m\prod_{n>m}(\mu _m^2-\mu _n^2)\right )^2}\nn\\[4mm]
&=(-1)^{\frac{k(k-1)+L(L-1)}{2}}\dt{\begin{array}{c}\left[x_i^{\mu_j}-x_i^{-\mu_j}\right ]_{i\leq k}\\[4mm] \left[\mu _j^{2(L-i)+1}\right ]_{i>k}\end{array}}\dt{\frac{\mu _i^{2(L-j)+1}}{\mu _i^2\prod_{n\neq i}(\mu _i^2-\mu _n^2)}}\nn\\[4mm]
&=(-1)^{\frac{k(k-1)+L(L-1)}{2}}\dt{\begin{array}{c}\left[\sum_{m=1}^L \frac{\mu _m^{2(L-j)}(x_i^{\mu_m}-x_i^{-\mu_m})}{\mu _m\prod_{n\neq m}(\mu _m^2-\mu _n^2)}\right ]_{i\leq k}\\[4mm] \left[\sum_{m=1}^L \frac{\mu _m^{4L-2(i+j)+2}}{\mu _m^2\prod_{n\neq m}(\mu _m^2-\mu _n^2)}\right ]_{i>k}\end{array}}.
\end{align}
The elements in rows $k$ to $L$ can be written, with $\eta=i+j$, as
\begin{align*}
&\frac1{\prod_{r<s}(\mu _r^2-\mu _s^2)}\sum_{m=1}^L (-1)^{m-1}\mu _m^{4L-2\eta}\prod_{\stack{1\leq r<s\leq L}{r,s\neq m}}(\mu _r^2-\mu _s^2)\\
&=\frac{\dt{\begin{array}{cc}\left[\mu_i^{4L-2\eta}\right ]_{j=1} & \left[\mu_i^{2(L-j)}\right ]_{j\geq 2}\end{array}}}{\dt{\mu _i^{2(L-j)}}},
\end{align*}
which is $0$ when $\eta>L+1$, and $1$ when $\eta=L+1$. This means that \eqref{eq:messdetk} can be expressed as
\[ (-1)^{\frac{k(k-1)+L(L-1)}{2}}\dt{\begin{array}{cc}\left[*\right ]_{\stack{i\leq k}{j\leq L-k}} &\left[\sum_{m=1}^L\frac{\mu_m^{2(L-j)}(x_i^{\mu_m}-x_i^{-\mu_m})}{\mu_m\prod_{n\neq m}(\mu _m^2-\mu _n^2)}\right ]_{\stack{i\leq k}{j\geq L-k+1}}\\[6mm] \text{\bf A} &\textbf{0}_{\stack{i\geq k+1}{j\geq L-k+1}}\end{array}},\]
where `$*$' is an entry that does not contribute to the determinant, and \textbf{A} is a matrix with $1$'s on the backwards diagonal, $0$'s below and `$*$'s above. Then the expression can be reduced to
\begin{align*}
&\dt[k]{\sum_{m=1}^L \frac{\mu _m^{2(k-j)}(x_i^{\mu_m}-x_i^{-\mu_m})}{\mu _m\prod_{n\neq m}(\mu _m^2-\mu _n^2)}},
\end{align*}
which is exactly the LHS of \eqref{eq:redSk}.
\end{proof}

\section{The integral operator $Q_z$}
\label{sec:Q}
In this section we will construct the operator $Q_z$ satisfying
\be
\label{eq:Qchi}\left[Q_z\chi_\lambda\right ](x_1,\ldots ,x_L)=q_\lambda (z)\,\chi_\lambda (x_1,\ldots ,x_L).
\ee
In order to construct $Q_z$ as an integral operator, we will need to define an appropriate domain of integration, which we will do first. The integration variables are $t_1,\ldots,t_{L-1}$, $y_1,\ldots,y_L$, and $w$. The integration variables interlace the $x$'s as $x_i\leq y_i/t_i\leq x_{i+1}$, for $i=1,\ldots,L-1$.

\begin{defn}
\label{def:P}
For any Laurent series $f(t)=\sum_{m\in\mathbb{Z}} c_m t^m$ with no constant term, i.e., $c_0=0$, the domain $P$ of the integral over $t$ is defined as
\begin{align*}\int _P \frac{\dd t}{t} f(t)&:=\int _0^1\frac{\dd t}{t} \sum _{m>0} c_m t^m -\int _1^\infty\frac{\dd t}{t} \sum _{m<0} c_m t^m
= \sum_{m\in\mathbb{Z}} \frac{c_m}{m}.
\end{align*}
For $1\leq i\leq L-1$, we also define
\[\iint_{\mathcal{D}_i}\dd Y_i=\int _P\frac{\dd t_i}{t_i}\int _{t_ix_i}^{t_ix_{i+1}}\frac{\dd y_i}{y_i},\]
and we will need two domains for the integrals over $y_L$, 
\[
x_L<\frac{y_L}{ t_1\ldots t_{L-1}}<\infty, \qquad \text{and}\qquad
0<\frac{y_L}{t_1\ldots t_{L-1}}<x_1.
\]
\end{defn}

With these definitions we can now write down explicitly the operator $Q_z$ that satisfies \eqref{eq:Qchi}.
\begin{prop}
\label{prop:Q}
$Q_z$ is given by
\begin{align}
&\left[Q_z f\right ](x_1,\ldots ,x_L)=\nn\\
&\quad \frac{1}{\phi_\delta (z)a_\delta (x_1,\ldots ,x_L)}\int _1^z\frac{\dd w}{w}\int_{\mathbf D}\dd \mathbf Y\ a_\delta (y_1,\ldots ,y_L)f(y_1,\ldots ,y_L),\nn
\end{align}
where
\begin{align*}
\int_{\mathbf D} \dd\mathbf Y&=\iint _{\mathcal{D}_1}\dd Y_1\cdots\iint _{\mathcal{D}_{L-1}}\dd Y_{L-1}\left[\int_{t_1\ldots t_{L-1}x_{L}}^\infty \dd y_L\ \delta\left (y_L-wx_L\prod_{l=1}^{L-1}\frac{x_l t_l^2}{y_l}\right )\right .\\
&\quad\left .+\int _0^{t_1\ldots t_{L-1}x_1} \dd y_L\ \delta\left (y_L-\frac{x_L}{w}\prod_{l=1}^{L-1}\frac{x_l t_l^2}{y_l}\right )\right ].
\end{align*}
\end{prop}
\begin{proof}
The LHS of \eqref{eq:Qchi} becomes
\[\frac{1}{\phi_\delta (z)a_\delta (x_1,\ldots ,x_L)}\int _1^z\frac{\dd w}{w}\int_{\mathbf D} \dd \mathbf Y\ a_\mu (y_1,\ldots ,y_L),\]
while the RHS can be written as
\[\frac{ \phi_\mu (z)}{\phi_\delta(z)} \frac{a_\mu (x_1,\ldots ,x_L)}{a_\delta (x_1,\ldots ,x_L)}.\]
It is therefore enough to prove that
\begin{align}
\label{eq:Qred}
&\int _1^z\frac{\dd w}{w}\int_{\mathbf D}\dd\mathbf Y\ a_\mu (y_1,\ldots ,y_L)=\phi_\mu (z)\ a_\mu (x_1,\ldots ,x_L).
\end{align}
To do this, we first use the following three ingredients:
\begin{enumerate}
\item[i.] We expand $a_\mu(\mathbf y)$ along the bottom row, producing
\be \sum_{r=1}^L (-1)^{L+r}(y_L^{\mu_r}-y_L^{-\mu_r})\dt[L-1]{y_i^{\mu_j}-y_i^{-\mu_j}}_{\genfrac{}{}{0pt}{}{i\neq L}{j\neq r}}.
\label{eq:aexp}
\ee
For each term in this sum, the integrals over $y_L$ can be performed easily, resulting in
\be
\label{eq:wprod}
y_L^{\mu_r}-y_L^{-\mu_r} \rightarrow
(w^{\mu_r}+w^{-\mu_r})\left [\left (x_L\prod _{l=1}^{L-1}\frac{x_l t_l^2}{y_l}\right )^{\mu_r}-\left (\frac1{x_L}\prod _{l=1}^{L-1}\frac{y_l}{x_l t_l^2}\right )^{\mu_r}\right ].
\ee
\item[ii.] The integral over $w$ becomes elementary:
\be
\label{eq:wint}
\int _1^z\frac{\dd w}{w}(w^{\mu_r}+w^{-\mu_r})=\frac{1}{\mu_r} \left(z^{\mu_r}-z^{-\mu_r}\right).
\ee
\item[iii.]
Combining \eqref{eq:aexp} and \eqref{eq:wprod} we find the further simplification
\begin{align*}
&\left [\left (x_L\prod _{l=1}^{L-1}\frac{x_l t_l^2}{y_l}\right )^{\mu_r}-\left (\frac1{x_L}\prod _{l=1}^{L-1}\frac{y_l}{x_l t_l^2}\right )^{\mu_r}\right ]\ \dt[L-1]{y_i^{\mu_j}-y_i^{-\mu_j}}_{\genfrac{}{}{0pt}{}{i\neq L}{j\neq r}}\\
&=\prod _{l=1}^{L}x_l^{\mu_r}\dt[L-1]{t_i^{2\mu_r}(y_i^{\mu_j -\mu_r}-y_i^{-\mu_j -\mu_r})}_{\genfrac{}{}{0pt}{}{i\neq L}{j\neq r}}\\
&\qquad\qquad -\prod _{l=1}^{L}x_l^{-\mu_r}\dt[L-1]{t_i^{-2\mu_r}(y_i^{\mu_j +\mu_r}-y_i^{\mu_r -\mu_j})}_{\genfrac{}{}{0pt}{}{i\neq L}{j\neq r}},
\end{align*}
after which the remaining integrals over $Y_i$ can be inserted into each row of the determinants, e.g.,
\begin{align*}
&\iint _{\mathcal{D}_1}\dd Y_1\cdots\iint _{\mathcal{D}_{L-1}}\dd Y_{L-1} \dt[L-1]{t_i^{2\mu_r}(y_i^{\mu_j -\mu_r}-y_i^{-\mu_j -\mu_r})}_{\genfrac{}{}{0pt}{}{i\neq L}{j\neq r}}\\
=&\dt[L-1]{\iint _{\mathcal{D}_{i}}\dd Y_{i}\ t_i^{2\mu_r}(y_i^{\mu_j -\mu_r}-y_i^{-\mu_j -\mu_r})}_{\genfrac{}{}{0pt}{}{i\neq L}{j\neq r}} \\
=&\dt[L-1]{\frac1{\mu_j^2 -\mu_r^2}(x_{i+1}^{\mu_j -\mu_r}-x_i^{\mu_j -\mu_r}-x_{i+1}^{-\mu_j -\mu_r}+x_i^{-\mu_j -\mu_r})}_{\genfrac{}{}{0pt}{}{i\neq L}{j\neq r}}\\
=&\frac{(-1)^{L-1}}{\prod _{j\neq r}(\mu_r^2 -\mu_j^2 )} \dt[L-1]{x_{i+1}^{-\mu_r}(x_{i+1}^{\mu_j}-x_{i+1}^{-\mu_j})-x_i^{-\mu_r}(x_i^{\mu_j}-x_i^{-\mu_j})}_{\genfrac{}{}{0pt}{}{i\neq L}{j\neq r}}.
\end{align*}
\end{enumerate}

Using these ingredients, we finally find the following expression for the LHS of \eqref{eq:Qred},
\begin{multline}
\label{eq:Qred2}
\sum_{r=1}^L \frac{z^{\mu_r}-z^{-\mu_r}}{\mu_r \prod _{j\neq r}(\mu_r^2 -\mu_j^2 )} \times\\
(-1)^{r-1} \left (\prod _{l=1}^{L}x_l^{\mu_r}\dt[L-1]{x_{i+1}^{-\mu_r}(x_{i+1}^{\mu_j}-x_{i+1}^{-\mu_j})-x_i^{-\mu_r}(x_i^{\mu_j}-x_i^{-\mu_j})}_{\genfrac{}{}{0pt}{}{i\neq L}{j\neq r}}\right.\\
\left. -\prod _{l=1}^{L}x_l^{-\mu_r}\dt[L-1]{x_{i+1}^{\mu_r}(x_{i+1}^{\mu_j}-x_{i+1}^{-\mu_j})-x_i^{\mu_r}(x_i^{\mu_j}-x_i^{-\mu_j})}_{\genfrac{}{}{0pt}{}{i\neq L}{j\neq r}}\right).
\end{multline}
At this point it has become clear that if $(-1)^{r-1}$ times the difference of products is independent of $r$ (so that we can factor it out of the sum), the remaining sum will be precisely equal to $\phi_\mu(z)$. Hence, it remains to show that for each $r$,
\begin{align*}
a_\mu (\mathbf x)& =(-1)^{r-1}\left (\prod _{l=1}^{L}x_l^{\mu_r}\dt[L-1]{x_{i+1}^{-\mu_r}(x_{i+1}^{\mu_j}-x_{i+1}^{-\mu_j})-x_i^{-\mu_r}(x_i^{\mu_j}-x_i^{-\mu_j})}_{\genfrac{}{}{0pt}{}{i\neq L}{j\neq r}}\right .\\
&\quad\left .-\prod _{l=1}^{L}x_l^{-\mu_r}\dt[L-1]{x_{i+1}^{\mu_r}(x_{i+1}^{\mu_j}-x_{i+1}^{-\mu_j})-x_i^{\mu_r}(x_i^{\mu_j}-x_i^{-\mu_j})}_{\genfrac{}{}{0pt}{}{i\neq L}{j\neq r}}\right ).
\end{align*}

This can be achieved by increasing the size of the matrices in the determinants by one row and one column, at $i=1$ and $j=r$, cancelling the factor of $(-1)^{r-1}$. The new column has entries of $0$ except at $i=1$, which is 1, and the new row has entries $x_1^{-\mu_r}(x_1^{\mu_j}-x_1^{-\mu_j})$ for the first determinant and $x_1^{\mu_r}(x_1^{\mu_j}-x_1^{-\mu_j})$ for the second. We then use row reduction, adding the $1$st row to the $2$nd row, then the $2$nd to the $3$rd, and so on. For ease of display, we place the additional column on the end, producing an extra factor of $(-1)^{L-r}$ that will be cancelled at the very end when we permute the column back to $j=r$. Thus we obtain
\begin{multline*}
(-1)^{L-r} \left( \prod_{l=1}^L x_l^{\mu_r}\det\nolimits_L\Bigg[{\begin{array}{ll}
\Big[x_i^{-\mu_r}(x_i^{\mu_j}-x_i^{-\mu_j})\Big ]_{j\neq r} & \Big[1\Big ]
\end{array}}\Bigg ] \right.\\
\left.-
\prod_{l=1}^L x_l^{-\mu_r}\det\nolimits_L\Bigg[{\begin{array}{ll}
\Big[x_i^{\mu_r}(x_i^{\mu_j}-x_i^{-\mu_j})\Big ]_{j\neq r} & \Big[1\Big ]
\end{array}}\Bigg ]\right),
\end{multline*}
where $[1]$ denotes a column of 1's. The factors outside the determinants are then inserted into their respective rows, resulting in
\[
=(-1)^{L-r} \left(\det\nolimits_L\Bigg[{\begin{array}{ll}
\Big[x_i^{\mu_j}-x_i^{-\mu_j}\Big ]_{j\neq r} & \Big[x_i^{\mu_r}\Big ]
\end{array}}\Bigg ]-\det\nolimits_L\Bigg[{\begin{array}{lll}
\Big[x_i^{\mu_j}-x_i^{-\mu_j}\Big ]_{j\neq r} & \Big[x_i^{-\mu_r}\Big ]
\end{array}}\Bigg ]\right).
\]
These two determinants can be combined by simply performing the subtraction in the last column. Permuting this column back to $j=r$, this finally results in
\[\dt{x_i^{\mu_j}-x_i^{-\mu_j}}=a_\mu (x_1,\ldots ,x_L).\qedhere\]
\end{proof}

\section{The integral operator $A_k$}
\label{sec:A}
The next and final step is to obtain the operator $\mathcal{A}_k$, which satisfies
\begin{align}
\label{eq:QA}\big(\rho _{k-1}Q_{z}f(\mathbf{x})\big)\big|_{z=x_k} =\mathcal A_k\rho _k f(\mathbf{x}),
\end{align}
where
\[\rho _kf(x_1,\ldots ,x_L)=f(x_1,\ldots ,x_k,1,\ldots ,1).\]
\begin{prop}
\label{prop:A}
Relation \eqref{eq:QA} is satisfied by
\begin{align*}
\mathcal{A}_k&=\lim_{\varepsilon\rightarrow 0}\frac{1}{2\varepsilon} \frac1{\phi_\delta (x_k) a^{(k)}_\delta (x_1,\ldots ,x_{k-1})}\int _1^{x_k}\frac{\dd w}{w}\int_{\mathbf D^{(\varepsilon)}} \dd \mathbf Y^{(\varepsilon)}\  a^{(k+1)}_\delta(y_1,\ldots ,y_k),
\end{align*}
where
\begin{align*} \int_{\mathbf D^{(\varepsilon)}} \dd\mathbf Y^{(\varepsilon)}&=\iint _{\mathcal D_1}\dd Y_1\cdots\iint _{\mathcal D_{k-1}}\dd Y_{k-1}\left[\int _0^{t_1\ldots t_{k-1}x_1} \dd y_k\ \delta\left (y_k-\frac{\e^\varepsilon}{w}\prod_{l=1}^{k-1}\frac{t_l^2 x_l}{y_l}\right )\right .\\
&\left .+\int_{t_1\ldots t_{k-1}\e^\varepsilon}^\infty \dd y_k\ \delta\left (y_k-w \e^\varepsilon\prod_{l=1}^{k-1}\frac{t _l^2 x_l}{y_l}\right )\right ],
\end{align*}
and
\[\iint_{\mathcal D_{k-1}}\dd Y_{k-1}=\int _P\frac{\dd t_{k-1}}{ t_{k-1}}\int_{ t_{k-1}x_{k-1}}^{ t_{k-1}\e^\varepsilon}\frac{\dd y_{k-1}}{y_{k-1}}.\]
We remind the reader that the domain of integration $P$ is given in \defref{P}.
\end{prop}
\begin{proof}
The $\chi_\lambda$ form a linear basis of $\mathbb C[y_1^\pm,\ldots ,y_L^\pm ]^{\mathcal W_{\rm B}}$, so it suffices to show that \eqref{eq:QA} holds for $f=\chi_\lambda$, and the general case will follow by linearity. We thus want to prove that
$\mathcal{A}_k \rho_k\chi_\lambda(\mathbf{x})=q_\lambda(x_k)\rho_{k-1}\chi_\lambda(\mathbf{x})$, i.e.,
\begin{align}
\label{eq:Aam}
\lim_{\varepsilon\rightarrow 0}\frac1{2\varepsilon}\int _1^{x_k}\frac{\dd w}{w}\int_{\mathbf D^{(\varepsilon)}} \dd \mathbf Y^{(\varepsilon)}\  a^{(k+1)}_\mu(y_1,\ldots ,y_k)=\phi_\mu(x_k)a^{(k)}_\mu(x_1,\ldots,x_{k-1}),
\end{align}
using the fact that $a^{(k+1)}_\delta\rho_k\chi_\lambda=a^{(k+1)}_\mu$.

This proof is very similar to that of \propref{Q}, but with a few added subtleties. The differences in the proof are outlined here, and we give the details in \apref{Ak}.

As in step i.~of the proof of \propref{Q}, the determinant $a^{(k+1)}_\mu(x_1,\ldots,x_k)$ is expanded along a row, but this time along the $k$th row instead of the last. Also, the integrals over $y_k$ take the place of that over $y_L$ in this step. The simplification in step iii.~is also very similar. The products outside the determinant now run from $1$ to $k-1$, and contain an extra factor of $\e^{\pm\varepsilon\mu_r}$. The $k-1$ integrals over $Y_1,\ldots, Y_{k-1}$ are inserted into the first $k-1$ rows of each determinant.

After these integrals are performed, the sizes of the matrices are increased as before. This time we choose the entries of the new row proportional to $\sinh{(\varepsilon\mu_j)}$ so that the matrices simplify after row reduction. After this, the prefactors in front of the two determinants can be cancelled, and the determinants can be combined.

The final step is to approximate to first order in $\varepsilon$, and then perform row reduction on the rows from $k$ to $L$, in order to remove the dependence on $\mu_r$ from each row. The result is a factor $a^{(k)}_\mu(x_1,\ldots,x_{k-1})$, which can be factored out of the remaining sum over $r$. This last sum is simply equal to $\phi_\mu(x_k)$, and the proof is complete.
\end{proof}

\chapter*{Conclusion}
\addcontentsline{toc}{chapter}{Conclusion}

Two physical quantities of interest for the Temperley--Lieb loop model at the special point $n=1$ have been calculated in this thesis, the groundstate eigenvector and its normalisation in \chapref{2BSoln} and the boundary-to-boundary correlation function in \chapref{Flow}. In addition, a key step has been laid down for the asymptotic analysis of these results in \chapref{Branching}.

We have shown in \chapref{2BSoln} that the ground state eigenvector equation for the TL($1$) loop model with open boundaries is equivalent to the $q$-deformed Knizhnik--Zamolodchikov equation. We have used this connection to obtain an explicit description of key components of the ground state $\ket\Psi$ for finite system sizes and without resorting to the Bethe ansatz. The derivation of the normalisation $Z$ of the TL($1$) ground state is presented in \secref{normalisation}, with the result expressed as a product of four symplectic characters $\chi$ (also known as Schur polynomials of the type $C$ root system). Our results for two open boundaries contain as special cases those of reflecting and mixed boundary conditions, which have been considered before \cite{DF05,ZJ07,dGP07}. The general result for the open boundary case has proved to be less tractable than those corresponding to these other boundary conditions, as it is not possible to express the eigenvector components as factorised operators acting on a `highest weight' component. Though we were able to find symmetries and relations that the eigenvector components must satisfy, in some cases the relations did not contain enough information to specify the components exactly.

Our result for the normalisation leads in \chapref{Flow} to an explicit expression for a finite size correlation function $F$ of the open TL($1$) loop model. In a two-dimensional percolation model on a finite width lattice, this quantity is the average number of percolating clusters passing in between two chosen points. This observable also falls within a class of parafermionic correlation functions that can be shown to be discretely holomorphic, and hence are precursors to analytic correlators in the conformally invariant continuum limit. We stress furthermore that our result is exact for systems of finite width, and not asymptotic, which is unusual for systems that are not free fermionic. Our main results \eqref{eq:Xval} and \eqref{eq:Yval} are expressed in terms of the same symplectic character $\chi$ as appeared in the normalisation. Analogous results have been obtained for cylindrical boundary conditions for both site and bond percolation \cite{NienJdF10}.

Naturally we desire to study the asymptotic limit of these results, as this limit corresponds to the thermodynamic limit of the system. In the case of the correlation function, we wish to use the asymptotic analysis to check our results against the continuum limit, which was calculated non-rigorously using conformal field theory by Cardy \cite{Cardy00}. To calculate the asymptotic limit of $Z$ and $F$, an asymptotic analysis of the symplectic character $\chi$ must first be conducted. We are particularly interested in the role the two boundaries play in the limit as $L$ goes to infinity, so we would especially like to study the asymptotics of the symplectic character when all $z_i$ are set to $1$ but $\zeta_0$ and $\zeta_L$ are kept general.

In \chapref{Branching} we have set out the $Q$-operator method of separation of variables for the symplectic character, as initiated by Sklyanin \cite{Skl88} and as applied to type $A$ Jack polynomials in \cite{KuzMS03,KuzSkl06}. The steps of the separation method are laid out in \secref{method}. The separation of variables for the symplectic character follows a process similar to that of the type $A$ Schur polynomial (more commonly known as the Schur function). However, some of the intermediate steps are technically much more involved. In particular, the $Q$- and $\mathcal A$-operators contain a double integration in each variable, whereas the corresponding operators for the Schur function contain only one. This is related to the fact that the Hamiltonian for the symplectic character contains double derivatives.

The problem of asymptotics for the symplectic character can now be regarded as a problem of asymptotics for the separated polynomial, which is a product of single-variable polynomials $q_\lambda$. As mentioned earlier, we are particularly interested in the asymptotic limit of $\chi_\lambda$ when all but $k$ variables are set to 1, and the operator $\mathcal S_k$ of \propref{Sk} and its inverse are useful here. The differential equation satisfied by $q_\lambda$, given in \secref{diff}, is expected to be helpful in determining the asymptotics of $q_\lambda$. This is certainly worthy of further investigation.

\chapter*{Outlook}
\addcontentsline{toc}{chapter}{Outlook}
The XXZ quantum spin chain with open boundaries at both ends is closely related to the TL($1$) loop model at the anisotropy point $\Delta=-1/2$, as discussed in the Introduction. We hope and expect that the results in this thesis will lead to the calculation of finite size correlation functions for the XXZ spin chain. The symplectic character can also be interpreted as a Toda lattice wave function, an observation which hints at an interesting link between the quantum integrable TL($1$) model and classical integrable models. Other incarnations of the TL($1$) model to which our results may be applied include the conformally invariant stochastic raise and peel model \cite{PearceRdGN02,Pyat04} and supersymmetric lattice models with boundaries \cite{YangF04,dGNPR05}.

In the case of the stochastic raise and peel model, the ground state is a stationary state distribution. In this setting an expression for the normalisation is important, so that the stationary state is a properly normalised probability distribution.

One way forward would be to express certain linear combinations of the components of the ground state eigenvector in terms of multiple contour integrals such as was done for reflecting boundary conditions \cite{DFZJ07,ZJDF08}. However, the success of this approach seems to be related to the existence of factorised expressions for the ground state components, which exist for reflecting and mixed boundary conditions \cite{dGP07}. In the case of open boundaries, the lack of a pseudovacuum leads to a lack of such convenient factorised expressions, so we anticipate some fundamental difficulties with this approach. We further hope to make a connection between our solutions and Macdonald--Koornwinder polynomials of type $(C^\vee,C)$ for specialised parameters \cite{Kasa08}, as well as with those in the form of Jackson integrals for $q$KZ equations on tensor product spaces, see \cite{TaraV94,Mimachi96} for the case of type $A$.

There is another, independent reason for computing the normalisation, which is in the context of the Razumov--Stroganov conjecture \cite{RazStr04,BatdGN01,dG05}. The original version of this conjecture was proved in \cite{CantiniS11}, but the more general versions remain unproved. This conjecture states that there is an intriguing relation between the TL($1$) ground state $\ket{\Psi_L}$ and the combinatorics of a fully packed loop (FPL) model on finite geometries, as well as other combinatorial objects such as alternating sign matrices and symmetric plane partitions~\cite{Bres99}. In particular it states that the ground state normalisation is equal to the statistical mechanical partition function of an FPL model on a finite patch of the square lattice. Such a link has been made for the ground state of the model with \textit{identified} open boundaries, see \cite{dGR04}, but for the model considered here, which has two genuinely open boundaries, it is not known which FPL geometry (if any) gives rise to a partition function equal to the normalisation of $\ket{\Psi_L}$ as given in \eqref{eq:homonorm}. Understanding the underlying combinatorics for the general case of two boundaries will therefore lead to a deepening of our understanding of the Razumov--Stroganov correspondence, as well as to possible generalisations of symmetric plane partitions and alternating sign matrices.

The method of separation of variables for the symplectic character is a small part of a much larger theory of polynomials. An obvious extension of this work is to generalise the method to Jack polynomials of type $C$ and $BC$, or even Koornwinder polynomials. It is not expected that the inverse separating operator will be easy to construct in the more general case, as it is still an unsolved problem in the case of type $A$ Jack polynomials. However, another method for reversing the SoV process was used in \cite{KuzMS03}, using a so-called lifting operator, and it is hoped that an analogue of this will prove useful for the more general $C$ and/or $BC$ polynomials.


\appendix
\addtocontents{toc}{\contentsline {chapter}{Appendices}{}}
\numberwithin{equation}{chapter}
\numberwithin{defn}{chapter}
\chapter{The Hecke algebra}
\label{ap:hecke}
$R$-matrices are fundamental objects in the study of integrable lattice models \cite{BaseilhacK05}. In a lattice model of width $L$, $\check R_i$ describes the possible states at site $i$. The $R$-matrix defined in \eqref{eq:R_z} is just one example of a more general $R$-matrix,
\[ \check R_i(z)=a(z)+b(z)T_i,\qquad 1\leq i\leq L-1,\]
where the $T_i$ are algebraic objects, and $a(z)$ and $b(z)$ are coefficient functions. The $T_i$ satisfy certain relations such that when $a(z)$ and $b(z)$ are chosen appropriately, the Yang--Baxter equation,
\[\check R_i(z)\check R_{i+1}(zw) \check R_i(w)=\check R_{i+1}(w)\check R_i(zw)\check R_{i+1}(z),\]
is satisfied.

If the model in question is periodic, the site $i=L$ is identified with $i=1$, and the Yang--Baxter equation is also valid for $\check R_{L-1}$ and $\check R_1$.

If the model has non-diagonal boundary conditions, $K$-matrices are introduced to describe the possible states at one or both of the boundaries. A more general form of a $K$-matrix is
\[ \check K_i(z)=c(z)+d(z)T_i,\qquad i=0,L.\]
Again, $c(z)$ and $d(z)$ are coefficient functions and the algebraic object $T_i$ is chosen so that the reflection equation,
\begin{align*}
\check K_0(z)\check R_1(zw)\check K_0(w)\check R_1(w/z)&=\check R_1(w/z)\check K_0(w)\check R_1(zw)\check K_0(z),\qquad \text{for}\; i=0,\\
\check K_L(z)\check R_{L-1}(zw)\check K_L(w)\check R_{L-1}(w/z)&=\check R_{L-1}(w/z)\check K_L(w)\check R_{L-1}(zw)\check K_L(z),\qquad \text{for}\; i=L,
\end{align*}
is satisfied.

The relations satisfied by the $T_i$ are those of the Hecke algebra \cite{MartinWL00,Sahi07}. The Hecke algebra is related to the symmetric group, and it has a number of versions relating to the Dynkin diagram `type' classifications.

\begin{defn}
\label{def:HA}
The Hecke algebra of type $A_L$ is generated by $\{T_i\ |\ 1\leq i\leq L-1\}$, and defined by the relations
\[ (T_i-q)(T_i+q^{-1})=0,\qquad\qquad T_iT_{i+1}T_i=T_{i+1}T_iT_{i+1},\]
with the $T_i$ commuting otherwise. The latter of these relations is commonly known as the braid relation.
\end{defn}

The most relevant presentation for the Temperley--Lieb loop model is given by the simple transformation
\be
\label{eq:Hrep1}
e_i=T_i-q,
\ee
so that
\[ e_i^2=-(q+q^{-1})e_i,\qquad\qquad e_ie_{i+1}e_i-e_i=e_{i+1}e_ie_{i+1}-e_{i+1}. \]

By setting $e_ie_{i+1}e_i-e_i=0$, we obtain the $sl_2$ quotient known as the (zero-boundary) Temperley--Lieb algebra given in \defref{0BTL}, with $q=\e^{\ii\gamma}$. A representation of the Temperley--Lieb algebra on the tensor product space $\otimes_{i=1}^L \mathbb C^2$ is
\[ e_i \quad\mapsto\quad 1^{\otimes i-1}\otimes\left[
\begin{array}{cccc}
0 & 0 & 0 & 0 \\
0 & 1 & q & 0 \\
0 & q^{-1} & 1 & 0 \\
0 & 0 & 0 & 0 \\
\end{array}
\right]\otimes 1^{\otimes L-i-1}, \]
and is used to build the $R$-matrix of the six vertex model, $\check R_{6v}$.

By taking $sl_k$ quotients of the Hecke algebra, one can build modules of rank $k-1$. Such higher rank modules are discussed in \cite{ShiU07,Doikou05}. For example, one can construct a representation on the tensor product space $\otimes_{i=1}^L \mathbb C^k$, which can be used to build an $R$-matrix generalising $\check R_{6v}$ \cite{BelliardR08,KulishR83}.

In addition, systems with boundaries can be dealt with by including additional generators in the algebra, with boundary-type braid relations. We give one example here.
\begin{defn}
\label{def:HB}
The Hecke algebra of type $B_L$ has the generators and relations of \defref{HA}, along with the additional generator $T_0$ that satisfies
\[
(T_0-\e^{\ii\omega_0})(T_0-\e^{-\ii\omega_0})=0,\qquad\qquad T_0T_1T_0T_1=T_1T_0T_1T_0. \]
\end{defn}
The presentation defined by \eqref{eq:Hrep1} can be extended to the type $B$ case by defining
\[ e_0=\frac{T_0-\e^{\ii\omega_0}}{2\ii\sin{(\omega_0+\gamma)}}, \]
so that
\[ e_0^2=\frac{-\sin{\omega_0}}{\sin{(\omega_0+\gamma)}}e_0,\qquad\qquad e_0e_1e_0e_1-e_0e_1=e_1e_0e_1e_0-e_1e_0. \]
Again, the $sl_2$ quotient is obtained by setting both sides of this last relation to $0$, and this produces the one-boundary Temperley--Lieb algebra from \defref{1BTL}.

Similarly, the Hecke algebra of type $BC_L$ is defined by the generators and relations of type $B_L$, along with the additional generator $T_L$, which satisfies similar relations to $T_0$, with $T_{L-1}$ taking the place of $T_1$. However, the type $BC_L$ algebra is no longer finite-dimensional, as elements of the form $(T_0T_1\ldots T_L)^n$ cannot be reduced. The presentation defined in \eqref{eq:Hrep1} can again be extended, and the $sl_2$ quotient produces the two-boundary Temperley--Lieb algebra from \defref{2BTL} \cite{dGN09}.

Finally, it is worthwhile to point out that the operators $a_i$ and $s_i$ defined in \eqref{eq:adef} and \eqref{eq:sdef} are generators of the Hecke algebra of type $BC$ \cite{dGP07}. They are related to the $T_i$ by
\begin{align*}
a_i&=q-T_i, & s_i&=q^{-1}+T_i,\qquad\qquad 1\leq i\leq L-1,\\
a_k&=\frac{\e^{\ii\omega_k}-T_k}{2\ii\sin{(\omega_k+\gamma)}}, & s_k&=\frac{T_k-\e^{-\ii\omega_k}}{2\ii\sin{(\omega_k+\gamma)}},\qquad\qquad k=0,L,
\end{align*}
and satisfy
\[ a_is_i=s_ia_i=0, \qquad 0\leq i\leq L,\]
along with the relations (we list them here for the $a_i$; the relations satisfied by the $s_i$ are identical)
\begin{align*}
a_i^2&=(q+q^{-1})a_i,\qquad\qquad 1\leq i\leq L-1,\\
a_k^2&=\frac{\sin{\omega_k}}{\sin{(\omega_k+\gamma)}}a_k,\qquad\qquad k=0,L,
\end{align*}
and
\begin{align*}
a_ia_{i+1}a_i-a_i&=a_{i+1}a_ia_{i+1}-a_{i+1},\qquad\qquad 1\leq i\leq L-1,\\
a_0a_1a_0a_1-a_0a_1&=a_1a_0a_1a_0-a_1a_0,\\
a_La_{L-1}a_La_{L-1}-a_La_{L-1}&=a_{L-1}a_La_{L-1}a_L-a_{L-1}a_L.
\end{align*}

For the zero-boundary and one-boundary $q$KZ equations (corresponding to the type $A$ and type $B$ Hecke algebras respectively), the highest weight element of the solution vector can be found explicitly by fixing its degree and considering its symmetries. All the remaining elements can then be written in terms of operators acting on the highest weight element. These operators are Baxterised versions of the Hecke algebra generators $s_i$ \cite{dGP07}, and have the form
\[ h(u)=s_i-\frac{q^{u-1}-q^{1-u}}{q^u-q^{-u}}. \]
Because the type $BC$ Hecke algebra is infinite-dimensional and lacks a highest weight element, this approach only partially works for the $q$KZ equation with two boundaries.

\chapter{Proofs of the transfer matrix recursions}
\label{ap:recursion}
\section{Proof of \propref{Trecur}: The bulk recursion}
\label{ap:bulkrecur}
In the following we will use the shorthand notation
\[
a(z)=\frac{[q/z]}{[qz]},\qquad b(z)=-\frac{[z]}{[qz]}.
\]
We first define $\alpha'_L=\varphi_i\alpha_{L-2}$ to be the link pattern of length $L$ with a small link connecting sites $i$ and $i+1$ inserted into the link pattern $\alpha_{L-2}$. Restricting our focus to the action of the transfer matrix on the sites $i$ and $i+1$, we have
\[
T_L(w;z_{i+1}=qz_i)\ket{\alpha'_L}=\quad\raisebox{-40pt}{\includegraphics[height=100pt]{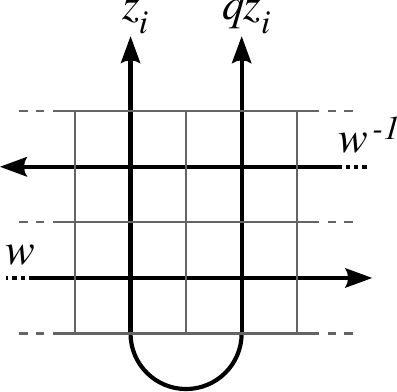}}.
\]
As each $R$-operator consists of two terms, the action of $T_L$ on sites $i$ and $i+1$ produces sixteen terms,
\begin{align*}
&a(z_iw)a(qz_iw)a(w/z_i)a(w/qz_i)\quad\raisebox{-22pt}{\includegraphics[height=50pt]{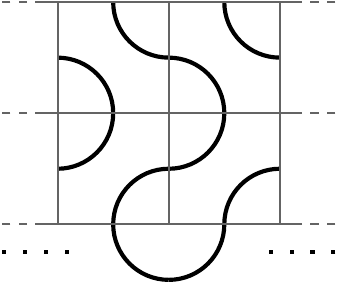}}\,+\\
&a(z_iw)a(qz_iw)a(w/z_i)b(w/qz_i)\quad\raisebox{-22pt}{\includegraphics[height=50pt]{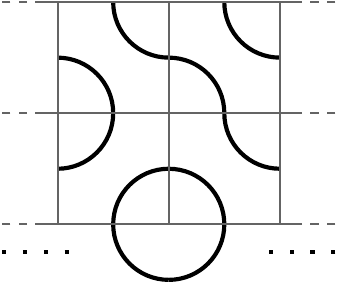}}\,+\,\ldots\,.
\end{align*}
Some of these pictures are equivalent with respect to their external connectivities. In total there are five different kinds of connectivities. For instance, one of the connectivities has
\begin{align*}
&a(z_iw)b(qz_iw)a(w/z_i)b(w/qz_i)\quad\raisebox{-22pt}{\includegraphics[height=50pt]{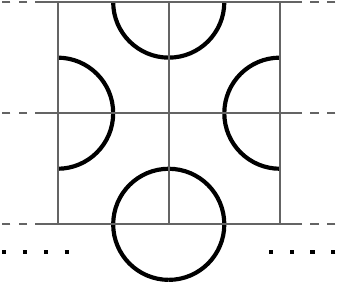}}\,+\\
&a(z_iw)b(qz_iw)b(w/z_i)b(w/qz_i)\quad\raisebox{-22pt}{\includegraphics[height=50pt]{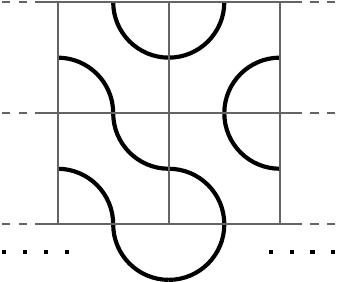}}\,+\\
&a(z_iw)b(qz_iw)a(w/z_i)a(w/qz_i)\quad\raisebox{-22pt}{\includegraphics[height=50pt]{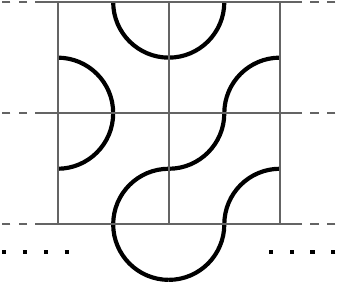}}.
\end{align*}
The closed loop in the first diagram is erased at the expense of a factor $-(q+q^{-1})$, after which the coefficients of the three diagrams sum to $0$. Using the fact that $a(qu)a(u)+b(qu)b(u)-{(q+q^{-1})}a(qu)b(u)=0$, it is easy to show that this happens for four of the five total kinds of connectivities, and that we are then left with
\begin{align*}
&T_L(w;z_{i+1}=qz_i)\ket{\alpha'_L}\\
&\qquad=a(z_iw)b(qz_iw)b(w/z_i)a(w/qz_i)\quad\raisebox{-38pt}{\includegraphics[height=80pt]{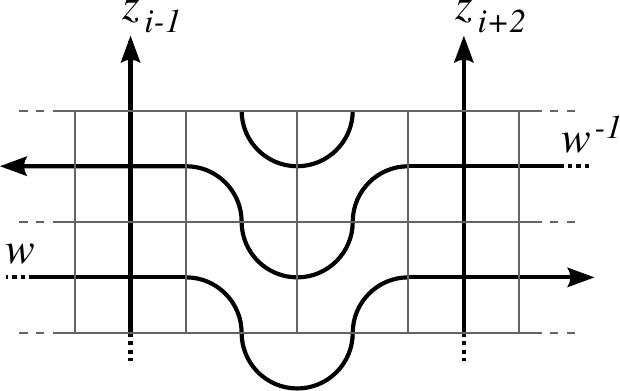}}\\
&\qquad=a(z_iw)b(qz_iw)b(w/z_i)a(w/qz_i)\ \varphi_i\quad\raisebox{-38pt}{\includegraphics[height=80pt]{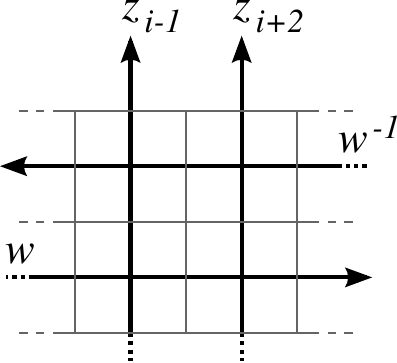}}\\
&\qquad=\frac{[q/z_iw][q^2z_i/w]}{[q^2z_iw][qw/z_i]}\ \varphi_i\, T_{L-2}(w;\hat z_i,\hat z_{i+1})\ket{\alpha_{L-2}}.
\end{align*}

\section{The left boundary recursion \eqref{eq:0boundrecur}}
\label{ap:boundrecur}
Here we will describe the recursion of the transfer matrix at the left boundary. The recursion at the right boundary is similar.

Setting $z_1=q\zeta_0$, the entries of the eigenvector are zero unless they correspond to link patterns with a little link from position 1 to the left boundary. We have the transfer matrix acting on this little link,
\[
T_L\circ\varphi_0\Big|_{z_1=q\zeta_0}=\quad\raisebox{-38pt}{\includegraphics[height=70pt]{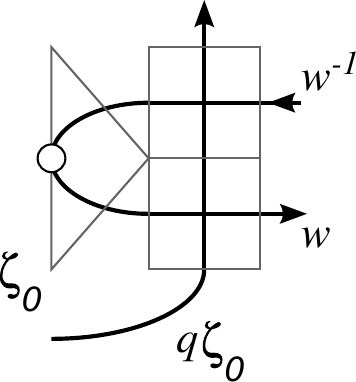}}.
\]

There are $8$ possible configurations in the above picture, which can be grouped into 3 kinds of connectivities. The following is a summary of the connectivities along with their weights:
\begin{align*}
\raisebox{-40pt}[0pt][0pt]{\includegraphics[height=50pt]{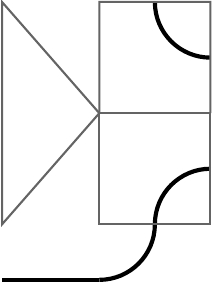}}\hspace{2cm}
&\frac{[q/q\zeta_0w]}{[q^2\zeta_0w]}\left(\frac{[q^2\zeta_0/w]}{[qw/q\zeta_0]}+\frac{[q\zeta_0/w]}{[qw/q\zeta_0]}\right)\left(\frac{k(q/w,\zeta_0)}{k(w/q,\zeta_0)}+\frac{[1/q][q^2/w^2]}{k(w/q,\zeta_0)}\right)\\
&\quad+\frac{[1/q\zeta_0w]}{[q^2\zeta_0w]}\frac{[q^2\zeta_0/w]}{[qw/q\zeta_0]}\frac{k(q/w,\zeta_0)}{k(w/q,\zeta_0)}\\[2mm]
&=\frac{[1/\zeta_0w]}{[q^2\zeta_0w]}+\frac{[1/q\zeta_0w][q^2\zeta_0/w][qw\zeta_0/q][qw/q\zeta_0]}{[q^2\zeta_0w][w/\zeta_0][q^2/w\zeta_0][q^2\zeta_0/w]}\\[2mm]
&=\frac{[1/\zeta_0w]}{[q^2\zeta_0w]}+\frac{[w\zeta_0]}{[q^2\zeta_0w]}=0,\\[8mm]
\raisebox{-30pt}[0pt][0pt]{\includegraphics[height=40pt]{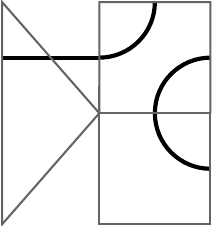}}\hspace{2cm}
&\frac{[1/q\zeta_0w]}{[q^2\zeta_0w]}\frac{[q\zeta_0/w]}{[qw/q\zeta_0]}\left(\frac{k(q/w,\zeta_0)}{k(w/q,\zeta_0)}+\frac{[1/q][q^2/w^2]}{k(w/q,\zeta_0)}\right)\\[2mm]
&=\frac{[q\zeta_0/w][1/q\zeta_0w]}{[w/\zeta_0][q^2\zeta_0w]}\\[2mm]
&=\frac{k(q/w,q\zeta_0)}{k(w/q,q\zeta_0)},\\[8mm]
\raisebox{-40pt}[0pt][0pt]{\includegraphics[height=50pt]{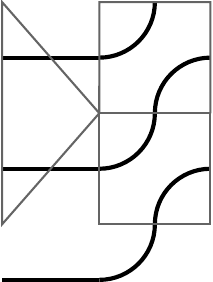}}\hspace{2cm}
&\frac{[1/q\zeta_0w]}{[q^2\zeta_0w]}\frac{[q^2\zeta_0/w]}{[qw/q\zeta_0]}\frac{[1/q][q^2/w^2]}{k(w/q,\zeta_0)}\\[2mm]
&=\frac{[1/q\zeta_0w][q^2\zeta_0/w][1/q][q^2/w^2]}{[q^2\zeta_0w][w/\zeta_0][q^2\zeta_0/w][q^2/w\zeta_0]}\\[2mm]
&=\frac{[1/q][q^2/w^2]}{[w/\zeta_0][q^2\zeta_0w]}\\[2mm]
&=\frac{[1/q][q^2/w^2]}{k(w/q,q\zeta_0)}.
\end{align*}
The second and third of these can be seen as the two contributions to a left boundary $K$-matrix, with an inserted loop connecting to the left boundary. The weights agree, and the whole sum can therefore be written as $\varphi_0\circ \check K_0(w,q\zeta_0)$, in a system of size $L-1$. This implies
\[
T_L(\zeta_0;z_1,\ldots)\circ\varphi_0\Big|_{z_1=q\zeta_0}=\varphi_0\circ T_{L-1}(q\zeta_0;z_2,\ldots).
\]

\chapter{Symmetries of $Z$ and the proportionality factors $p_i$, $r_0$, and $r_L$}
\label{ap:propsym}
Recall the ground state eigenvector normalisation,
\[
Z_L=\bra{\Psi}\Psi\rangle,
\]
which is equal to the sum over the components of $\ket\Psi$. Acting with the left eigenvector of the transfer matrix, $\bra\Psi$, on both sides of the $q$KZ equation \eqref{eq:qKZTL_TypeC2}, and using the fact that $\pi_i$ commutes with $\bra\Psi$,
\begin{align*}
\pi_iZ_L&=\bra{\Psi}\check R_i(z_i/z_{i+1})\ket{\Psi}\\
&=\sum_\alpha\psi_{\alpha} \bra\Psi \check R_i(z_i/z_{i+1})\ket\alpha\\
&=\sum_\alpha\psi_{\alpha}\left(\frac{[qz_{i+1}/z_i]}{[qz_i/z_{i+1}]} \bra\Psi\alpha\rangle -\frac{[z_i/z_{i+1}]}{[qz_i/z_{i+1}]}\bra\Psi e_i\ket\alpha\right).
\end{align*}
Since $e_i\ket\alpha=\ket{\alpha'}$ for some link pattern $\alpha'$, and $\bra\Psi\alpha\rangle=1$ for all $\alpha$, this becomes
\begin{align*}
\pi_iZ_L&=\sum_\alpha\psi_{\alpha}\left(\frac{[qz_{i+1}/z_i]}{[qz_i/z_{i+1}]}-\frac{[z_i/z_{i+1}]}{[qz_i/z_{i+1}]}\right)\\
&=\sum_\alpha\psi_{\alpha}=Z_L.
\end{align*}
Similar arguments can be made to show that $\pi_0 Z_L=Z_L$ and $\pi_L Z_L=Z_L$. We therefore know that $Z_L$ remains unchanged under any permutation of the variables $z_i$. Recalling that
\[
\sum_\alpha\psi^L_{\varphi_i\alpha}(z_{i+1}=qz_i)\ \varphi_i\ket\alpha=p_i(z_i;\ldots,\hat z_i,\hat z_{i+1},\ldots)\sum_\alpha\psi^{L-2}_{\alpha}(\hat z_i,\hat z_{i+1})\ \varphi_i\ket\alpha,
\]
we have
\be
Z_L(z_{i+1}=qz_i)=p_i(z_i;\ldots,\hat z_i,\hat z_{i+1},\ldots)Z_{L-2}(\hat z_i,\hat z_{i+1}).
\label{eq:Zrecur}
\ee
Since $Z$ is symmetric in the $z_i$, $p_i$ becomes $p_j$ under the transformation $z_i\leftrightarrow z_j$. We henceforth drop the index $i$ from $p_i$. Further, since the LHS is symmetric in everything except $z_i$, as is $Z_{L-2}$, $p(z_i;\ldots)$ must also be symmetric in all its variables except for $z_i$.

Completely analogously we can derive boundary recursions for the normalisation, which are
\be
\begin{split}
Z_L(z_1=q\zeta_0,\ldots,z_L;\zeta_0,\zeta_L) &= r_0(z_2,\ldots,z_{L};\zeta_0) Z_{L-1}(z_2,\ldots,z_{L};q\zeta_0,\zeta_L),\\
Z_L(z_1,\ldots,z_L=\zeta_L/q;\zeta_0,\zeta_L) &= r_L(z_1,\ldots,z_{L-1};\zeta_L) Z_{L-1}(z_1,\ldots,z_{L-1};\zeta_0,\zeta_L/q).
\label{eq:Zrecurbound}
\end{split}
\ee
As before, this shows that the proportionality factors $r_0$ and $r_L$ must be symmetric in all their variables except for $\zeta_0$ and $\zeta_L$ respectively.

\chapter{Solving the $q$KZ equation for $L=3$}
\label{ap:L=3}
Here we will solve the $q$KZ equation \eqref{eq:qKZea} with $L=3$. For this example we will take $q=\e^{2\pi \ii/3}$ and $s^4=1$, but we will leave $b$ generic. We use the notation
\begin{align*}
\psi_1&=\psi_{(((}\hspace{3cm}\psi_5=\psi_{)((}\\
\psi_2&=\psi_{(()}\hspace{3cm}\psi_6=\psi_{)()}\\
\psi_3&=\psi_{()(}\hspace{3cm}\psi_7=\psi_{))(}\\
\psi_4&=\psi_{())}\hspace{3cm}\psi_8=\psi_{)))},
\end{align*}
and we recall the definition of $s_i$ in \eqref{eq:s}
\[
s_i =-1-a_i.
\]

Considering in turn each $i$ and $\alpha$, the $q$KZ equation,
\[
\sum_\alpha \psi_\alpha \left(e_i\ket\alpha\right)=-\sum_\alpha \left(a_i\psi_\alpha\right) \ket\alpha,
\]
implies the following system of $32$ equations:
\begin{equation}
\left.\begin{array}{r@{\;}c@{\;}l}
a_0\psi_\alpha & = & 0 \\
s_0\psi_{\alpha+4} &= & \psi_\alpha
\end{array}
\right\}
\qquad \alpha=1,\ldots,4
\end{equation}
\begin{align}
a_1\psi_\alpha &=0 & &\hspace{-1cm} \alpha=1,2,5,\ldots,8\nn\\
s_1\psi_3&=\psi_1+\psi_2+b\psi_5+\psi_7 & & \label{eq:s1}\\
s_1\psi_4&=\psi_6+\psi_8 & & \nn
\end{align}
\begin{align}
a_2\psi_\alpha &=0 & &\hspace{-1cm} \alpha=1,3,4,5,7,8\nn\\
s_2\psi_6&=\psi_4+\psi_5+\psi_7+\psi_8 & & \label{eq:s2}\\
s_2\psi_2&=\psi_1+\psi_3 & & \nn
\end{align}
\begin{equation}
\begin{array}{r@{\;}c@{\;}ll}
a_3\psi_{2\alpha} &=& 0\qquad \qquad & \alpha=1,\ldots,4 \\
s_3\psi_{2\alpha-1} &=& \psi_{2\alpha} & \alpha=1,3\\
s_3\psi_{2\alpha-1} &=& b\psi_{2\alpha} & \alpha=2,4
\end{array}
\end{equation}
The relations where the action of the projector $a_i$ is zero entails certain symmetry restrictions on the components. For instance,
\begin{align*}
\psi_1&=\prod_{1\leq i<j\leq 3}k(z_j,z_i)\,\prod_{i=1}^3 k(z_i,\zeta_0)\,f_1(z_1,z_2,z_3),\\
\psi_8&=\prod_{1\leq i<j\leq 3}k(1/sz_i,1/sz_j)\,\prod_{i=1}^3 k(1/sz_i,1/s\zeta_L)\,f_8(sz_1,sz_2,sz_3),
\end{align*}
where the $f_\alpha(z_1,z_2,z_3)$ are symmetric functions invariant under $z_i \rightarrow 1/z_i$.

Using the system of equations we may obtain the components $\psi_\alpha$ in terms of $\psi_1$ and $\psi_8$. Assuming we know $\psi_1$, we find $\psi_2=s_3\psi_1$ , then $\psi_3=s_2\psi_2-\psi_1$, and $\psi_4=b^{-1} s_3\psi_3$. Then we can find $\psi_8$ from
\begin{align}
\psi_8&=s_2\psi_6-\psi_4-\psi_5-\psi_7,
\label{eq:psi8}
\end{align}
by using $\psi_6=s_1\psi_4-\psi_8$ and applying $s_3$ on both sides of \eqref{eq:psi8},
\[
-\psi_8=s_3s_2(s_1\psi_4-\psi_8)+\psi_4-\psi_6-b\psi_8,
\]
which implies
\[
(b-1)\psi_8=(s_3s_2s_1-s_1+1)\psi_4.
\]
The expressions for $\psi_2$, $\psi_3$, $\psi_4$ as well as $\psi_8$ can be neatly rewritten in a factorised form as in \cite{dGP07}. However, if $b=1$, the component $\psi_8$ cannot be determined this way.

In a similar way, given $\psi_8$ we can find $\psi_4$, then $\psi_6$, and $\psi_2$. Since we can express $\psi_4$ and $\psi_2$ in two different ways, these components have to satisfy certain consistency conditions. Now we can find the remaining two components,
\begin{align*}
(b-1)\psi_5&=s_1\psi_3-\psi_1-\psi_2-s_2\psi_6+\psi_4+\psi_8,\\
\psi_7&=s_1\psi_3-\psi_1-\psi_2-b\psi_5.
\end{align*}
Again, if $b=1$, $\psi_5$ and $\psi_7$ cannot be found separately in this way. However, their sum can be determined. Assuming we find an expression for one of these components (say, by solving $s_0\psi_5=\psi_1$ for $\psi_5$) that satisfies the appropriate degree, the boundary recursions in \lmref{comprecur}, and the symmetries imposed by \eqref{eq:s1} and \eqref{eq:s2}, the entire system can be shown to be consistent.

At $b=1$, as $L$ becomes larger there are more and more components like $\psi_5$ and $\psi_7$, which cannot be calculated directly from the extremal components. However, as above, certain sums of these components can be found, and consistency conditions can be checked. Using the recursions discussed in \secref{estaterecur}, recursive conditions can also be put on these components.

\chapter{Inductive proof of \lmref{kprop}}
\label{ap:altproof}
\begin{proof}
This proof shows that $a^{(k)}_\mu(x_1,\ldots,x_{k-1})$ is proportional to $a^{(k+1)}_\mu(x_1,\ldots,x_k)$ in the limit as $x_k\rightarrow 1$, and relies on the proof for $k=L$ given in \secref{symp}.

We take $\lim_{\varepsilon\rightarrow 0}a^{(k+1)}_\mu(x_1,\ldots,x_{k-1},\e^\varepsilon)$. In the $k$th row of the determinant (given in \eqref{eq:atrunc}), the $j$th element is
\[
\e^{\varepsilon\mu _j}-\e^{-\varepsilon\mu _j}=2\sinh{(\varepsilon\mu _j)}=2\sum_{n=0}^\infty \frac{(\varepsilon\mu _j)^{2n+1}}{(2n+1)!}.
\]
We use row reduction with the rows below the $k$th to remove the terms up to and including $n=L-(k+1)$. The remainder of the series is
\[ 2\frac{(\varepsilon\mu _j)^{2(L-k)+1}}{(2(L-k)+1)!}+ {\rm h.o.t.},\]
so $ a^{(k)}_\mu(x_1,\ldots,x_{k-1})=\lim_{\varepsilon\rightarrow 0}\frac{(2(L-k)+1)!}{2\varepsilon^{2(L-k)+1}} a^{(k+1)}_\mu(x_1,\ldots,x_{k-1},\e^\varepsilon)$. Then by induction we have that
\[ a^{(k)}_\mu(x_1,\ldots,x_{k-1})\propto\lim_{x_k,\ldots,x_L\rightarrow 1}a_\mu(x_1,\ldots,x_L).\qedhere\]
\end{proof}

\chapter{Proof of \propref{A}}
\label{ap:Ak}
\begin{proof}
The proof of equation \eqref{eq:Aam} is similar to that of \propref{Q}:
\begin{itemize}
\item[i.]
The first step is to expand the determinant $a^{(k+1)}_\mu$ over the $k$th row, giving
\be
\label{eq:sumA} \sum_{r=1}^L (-1)^{k+r}(y_k^{\mu_r}-y_k^{-\mu_r})\dt[L-1]{\begin{array}{c}
\left[y_i^{\mu_j}-y_i^{-\mu_j}\right ]_{i<k}\\[4mm]
\left[\mu_j^{2(L-i)+1}\right ]_{i>k}
\end{array}}_{j\neq r}.\ee
We can perform the integrals over $y_k$ in \eqref{eq:Aam}, and, as before, the dependence on $w$ factors out.
\item[ii.] The integral over $w$ has the same evaluation as \eqref{eq:wint}.
\item[iii.] The remaining factor is combined with the determinant,
\begin{align*}
&\left[\left (\e^\varepsilon\prod_{l=1}^{k-1}\frac{t _l^2 x_l}{y_l}\right )^{\mu_r}-\left (\e^{-\varepsilon}\prod_{l=1}^{k-1}\frac{y_l}{t_l^2 x_l}\right )^{\mu_r}\right ]\dt[L-1]{\begin{array}{c}
\left[y_i^{\mu_j}-y_i^{-\mu_j}\right ]_{i<k}\\[4mm]
\left[\mu_j^{2(L-i)-1}\right ]_{k\leq i<L}
\end{array}}_{j\neq r}\\
&=\e^{\varepsilon\mu_r}\prod_{l=1}^{k-1}x_l^{\mu_r}\dt[L-1]{\begin{array}{c}
\left[t_i^{2\mu_r}(y_i^{\mu_j -\mu_r}-y_i^{-\mu_j -\mu_r})\right ]_{i<k}\\[4mm]
\left[\mu_j^{2(L-i)-1}\right ]_{k\leq i<L}
\end{array}}_{j\neq r}\\
&\quad -\e^{-\varepsilon\mu_r}\prod_{l=1}^{k-1}x_l^{-\mu_r}\dt[L-1]{\begin{array}{c}
\left[t_i^{-2\mu_r}(y_i^{\mu_j +\mu_r}-y_i^{-\mu_j +\mu_r})\right ]_{i<k}\\[4mm]
\left[\mu_j^{2(L-i)-1}\right ]_{k\leq i<L}
\end{array}}_{j\neq r}.
\end{align*}
The $k-1$ integrals over $Y_i$ in \eqref{eq:Aam} can be inserted into the first $k-1$ rows of the determinants, and evaluated as before. The two determinants then become
\begin{align*}
&\e^{\pm\varepsilon\mu_r} \prod_{l=1}^{k-1}x_l^{\pm\mu_r}\dt[L-1]{\begin{array}{c}
\left[\frac1{\mu_j^2 -\mu_r^2}(x_{i+1}^{\mp\mu_r}(x_{i+1}^{\mu_j}-x_{i+1}^{-\mu_j})-x_i^{\mp\mu_r}(x_i^{\mu_j}-x_i^{-\mu_j}))\right ]_{i<k-1}\\[4mm]
\frac1{\mu_j^2 -\mu_r^2}(2\e^{\mp\varepsilon\mu_r}\sinh{(\varepsilon\mu_j)}-x_{k-1}^{\mp\mu_r}(x_{k-1}^{\mu_j}-x_{k-1}^{-\mu_j}))\\[4mm]
\left[\mu_j^{2(L-i)-1}\right ]_{k-1<i<L}
\end{array}}_{j\neq r}\\[4mm]
&=\frac{(-1)^{L-1}\e^{\pm\varepsilon\mu_r} \prod_{l=1}^{k-1}x_l^{\pm\mu_r}}{\prod_{j\neq r}(\mu_r^2 -\mu_j^2 )}\\
&\qquad\times\dt[L-1]{\begin{array}{c}
\left[x_{i+1}^{\mp\mu_r}(x_{i+1}^{\mu_j}-x_{i+1}^{-\mu_j})-x_i^{\mp\mu_r}(x_i^{\mu_j}-x_i^{-\mu_j})\right ]_{i<k-1}\\[4mm]
2\e^{\mp\varepsilon\mu_r}\sinh{(\varepsilon\mu_j)}-x_{k-1}^{\mp\mu_r}(x_{k-1}^{\mu_j}-x_{k-1}^{-\mu_j})\\[4mm]
\left[(\mu_j^2 -\mu_r^2 )\mu_j^{2(L-i)-1}\right ]_{k-1<i<L}
\end{array}}_{j\neq r}.
\end{align*}
Again, as in \eqref{eq:Qred2}, we will extract the ingredients needed for $\phi_\mu(x_k)$ and show that the rest is independent of $r$ so that it can be factored out of the sum in \eqref{eq:sumA}.
\end{itemize}

As in \secref{Q}, we increase the size of each matrix by adding a row at $i=L$ and a column at $j=r$, introducing a factor of $(-1)^{L+r}$. The new column has entries equal to $0$ except at $i=L$, which equals $1$, and the new row has entries of $2\e^{\mp\varepsilon\mu_r}\sinh{(\varepsilon\mu_j)}$. We depict column $r$ at the end:
\begin{align*}
&(-1)^{r+1}\e^{\pm\varepsilon\mu_r} \prod_{l=1}^{k-1}x_l^{\pm\mu_r}\\
&\qquad\times\dt{\begin{array}{cc}
\left[(x_{i+1}^{\mp\mu_r}(x_{i+1}^{\mu_j}-x_{i+1}^{-\mu_j})-x_i^{\mp\mu_r}(x_i^{\mu_j}-x_i^{-\mu_j}))\right ]_{i<k-1, j\neq r} & \mathbf{0}\\[4mm]
\left[2\e^{\mp\varepsilon\mu_r}\sinh{(\varepsilon\mu_j)}-x_{k-1}^{\mp\mu_r}(x_{k-1}^{\mu_j}-x_{k-1}^{-\mu_j})\right ]_{j\neq r} & 0\\[4mm]
\left[(\mu _j^2-\mu _r^2)\mu _j^{2(L-i)-1}\right ]_{k-1<i<L, j\neq r} & \mathbf{0}\\[4mm]
\left[2\e^{\mp\varepsilon\mu_r}\sinh{(\varepsilon\mu_j)}\right ]_{j\neq r} & 1
\end{array}}.
\end{align*}

Now we use row reduction, subtracting the $L$th row from the $k$th and then adding the $(k-1)$th to the $k$th, the $(k-2)$th to the $(k-1)$th, etc. We then multiply the first $k$ rows by $-x_i^{\pm\mu_r}$, and the $L$th by $\e^{\pm\varepsilon\mu_r}$,
\[
(-1)^{k-1}\dt{\begin{array}{cc}
\left[(x_i^{\mu_j}-x_i^{-\mu_j})\right ]_{i<k, j\neq r} & \left[x_i^{\pm\mu_r}\right]_{i<k}\\[4mm]
\left[(\mu _j^2-\mu _r^2)\mu _j^{2(L-i)-1}\right ]_{k\leq i<L, j\neq r} & \mathbf{0}\\[4mm]
\left[2\sinh{(\varepsilon\mu_j)}\right ]_{j\neq r} & \e^{\pm\varepsilon\mu_r}
\end{array}}.
\]
At this point we are able to combine the two determinants by subtracting the $r$th column of the second from the $r$th column of the first, obtaining
\[
\dt{\begin{array}{cc}
\left[(x_i^{\mu_j}-x_i^{-\mu_j})\right ]_{i<k, j\neq r} & \left[(x_i^{\mu_r}-x_i^{-\mu_r})\right ]_{i<k}\\[4mm]
\left[(\mu _j^2-\mu _r^2)\mu _j^{2(L-i)-1}\right ]_{k\leq i<L, j\neq r} & \mathbf{0}\\[4mm]
\left[2\sinh{(\varepsilon\mu_j)}\right ]_{j\neq r} & 2\sinh{(\varepsilon\mu_r)}
\end{array}}.
\]
Now we approximate to first order in $\varepsilon$, resulting in $2\varepsilon\mu_j$ in the bottom row. We can factor out $2\varepsilon$, and then use row reduction once again: To each row in turn from $i=L-1$ to $i=k$ we add $\mu_r^2$ times the row below, and we are finally left with
\begin{align*}
2\varepsilon\dt{\begin{array}{c}
\left[(x_i^{\mu_j}-x_i^{-\mu_j})\right ]_{i<k}\\[4mm]
\left[\mu _j^{2(L-i)+1}\right ]_{i\geq k}
\end{array}}=2\varepsilon  a_\mu^{(k)}(x_1,\ldots ,x_{k-1}).
\end{align*}
This then factors out of the sum in \eqref{eq:sumA}, which is equal to $\phi_\mu(x_k)$. Putting everything together, the factors of $(-1)$ cancel out, as does the factor of $2\varepsilon$, and we are left with
\[
a^{(k)}_\mu (x_1,\ldots ,x_{k-1})\phi_\mu (x_k),
\]
which is the RHS of \eqref{eq:Aam}.
\end{proof}
\end{document}